\documentclass[12pt]{article}
\pdfoutput=1

\usepackage{amsmath}
\usepackage{amstext,amssymb,amsfonts,amsthm}

\usepackage[pdftex]{graphicx}
\usepackage{booktabs}
\usepackage{xcolor}

\advance\voffset by -1.5cm
\advance\hoffset by -1.25cm
\newcommand{\pmat}[1]{\begin{pmatrix} #1 \end{pmatrix}}

\newcommand{\JJ}{\ensuremath{\mathcal{J}}}
\newcommand{\EE}{\ensuremath{\mathcal{E}}}
\newcommand{\cI}{\ensuremath{\mathcal{I}}}
\newcommand{\one}{\ensuremath{\mathbf{1}}}
\newcommand{\dra}{\rangle\!\rangle}

\newcommand{\polyRing}[2]{\ensuremath{\dfrac{\mathbb{C}[y_1,y_2]}{\left\langle #1, #2 \right\rangle}}}

\numberwithin{equation}{section}

\usepackage[unicode]{hyperref}
\hypersetup{colorlinks=true, breaklinks=true, linkcolor=black, menucolor=black, urlcolor=black!20!green!70!blue,citecolor=black!20!green!70!blue}

\begin{document}

\vspace*{-1.5cm}
\thispagestyle{empty}
\begin{flushright}
AEI-2010-092 
\end{flushright}
\vspace*{2.5cm}

\begin{center}
{\Large
{\bf D-branes and matrix factorisations in supersymmetric coset models}}
\vspace{2.5cm}

{\large Nicolas Behr$^{1}$}
\footnotetext[1]{{\tt E-mail: Nicolas.Behr@aei.mpg.de}}
{and \large Stefan Fredenhagen$^{2}$}
\footnotetext[2]{{\tt E-mail: Stefan.Fredenhagen@aei.mpg.de}}

\vspace*{0.5cm}

Max-Planck-Institut f{\"u}r Gravitationsphysik,
Albert-Einstein-Institut\\
D-14424 Golm, Germany\\
\vspace*{3cm}

{\bf Abstract}
\end{center}
\sloppy Matrix factorisations describe B-type boundary conditions in
$\mathcal{N}=2$ supersymmetric Landau-Ginzburg models. At the
infrared fixed point, they correspond to superconformal boundary
states. We investigate the relation between boundary states and matrix
factorisations in the Grassmannian Kazama-Suzuki coset models.  For
the first non-minimal series, i.e.\ for the models of type $SU
(3)_{k}/U (2)$, we identify matrix factorisations for a subset of the
maximally symmetric boundary states. This set provides a basis for the
RR charge lattice, and can be used to generate (presumably all) other
boundary states by tachyon condensation.

\newpage
\setcounter{tocdepth}{2}
\tableofcontents
\newpage

\section{Introduction}

\fussy In this article we want to study the relation between two different
and complementary descriptions of B-type boundary conditions in
$\mathcal{N}= (2,2)$ supersymmetric two-dimensional field theories:
the description in terms of matrix factorisations of a superpotential,
and the description in terms of boundary states. Such field theories
arise as world-sheet theories of open strings which end on B-type
D-branes. To motivate our investigation let us look at the moduli
space of string theory compactified on a six-dimensional Calabi-Yau
manifold. This moduli space is in general very complicated and
consists of different phases~\cite{Witten:1993yc}. In a large volume
regime we have a description in terms of a non-linear sigma-model on
the background geometry, and we can use geometric tools. At some other
region of the moduli space we might have a description in terms of
Landau-Ginzburg models governed by some holomorphic superpotential
$W$. At special points of the moduli space, the superconformal field
theory that is described by the Landau-Ginzburg model is in fact a
rational conformal field theory (CFT), which means that it has a large
chiral symmetry algebra that turns the theory solvable. A typical
example is the Gepner point in moduli space.

When we discuss D-branes in such backgrounds, a natural question
to ask is how they behave when the closed string moduli are
deformed. We shall focus in this paper on B-type D-branes. In the
aforementioned regimes, one has different descriptions for the branes. In
the geometric regime they are described by holomorphic submanifolds,
or more generally by complexes of coherent sheaves (see e.g.\
\cite{Aspinwall:2004jr}). In the Landau-Ginzburg models the B-type
boundary conditions are described by factorisations of the
superpotential in terms of matrices (see e.g.\
\cite{Jockers:2007ng}). The connection between these descriptions has
been clarified in~\cite{Herbst:2008jq} using gauged linear
sigma models as a description in the whole moduli space.

It is less clear how to connect the description in terms of matrix
factorisations to the formulation of B-type boundary conditions at the
points where we have a rational conformal field theory
description. Such a connection would be desirable to have, because
both descriptions have their advantages. Matrix factorisations easily
allow to discuss the dependence on the moduli, whereas the rational
CFT description is only available at one point. On the other hand, in
the Landau-Ginzburg formulation, one can only access few data
directly, namely topological data such as RR charges, but not e.g.\
the mass of the brane, whereas in the rational CFT we know the
couplings of all fields to the brane.

We are looking for some dictionary between matrix factorisations
and rational boundary states, not only for the case of Calabi-Yau
backgrounds, but for the general situation where a supersymmetric
rational CFT admits a Landau-Ginzburg description. Setting up such a
dictionary is a highly non-trivial problem. To get from the
Landau-Ginzburg formulation to the CFT description one has to follow a
renormalisation group flow to the infrared, but these flows are
usually not under good control. Only some 'topological' data is
protected under renormalisation.

The other problem we have to face is that on the CFT side, our tools
only allow us to construct rational boundary states, i.e.\ boundary
states which preserve the chiral symmetry algebra. In general, this
will only be a subset of all superconformal boundary states. Therefore
we should not expect to find a simple prescription of how to obtain a
boundary state from any matrix factorisation. More realistically, one
can hope to find answers to the following two questions: Can we
determine a matrix factorisation from a given boundary state? Can we
understand on the matrix factorisation side what distinguishes the
'rational' boundary conditions from the rest?

One approach to these questions is to study the relation of matrix
factorisations and rational boundary states in a large class of
models, and to look for general patterns. Up to now, most comparisons
have been performed in minimal
models~\cite{Brunner:2003dc,Kapustin:2003rc,Brunner:2005pq,Keller:2006tf}.
Minimal models are very special in the sense that we only have a
finite number of elementary boundary states and matrix factorisations
that have to be matched. Also products of minimal models have been
considered~\cite{Brunner:2005fv,Enger:2005jk}. Here one encounters for
the first time the situation that the rational boundary states only
present a subset of all boundary states.

A more general class of rational $\mathcal{N}= (2,2)$ supersymmetric
rational CFTs is provided by the Kazama-Suzuki
models~\cite{Kazama:1989qp}, which are based on a coset construction
$G/H$. Not all of these models, however, have a description as a
Landau-Ginzburg theory. A subclass with this property is given by
those models where the group $G$ is simply laced, the corresponding
level is $1$, and $G/H$ is a Hermitian symmetric
space~\cite{Lerche:1989uy}. A two-parameter family of such models is
given by the Grassmannian cosets, where $G=SU (n+k)$ and $H=S (U
(n)\times U (k))$. For $n=1$ one recovers the minimal models. The
first non-minimal family of Grassmannian models is given by $n=2$. For
these models we want to extend the connection between the coset and
the Landau-Ginzburg description to the case when B-type boundary
conditions are present.\footnote{For A-type boundary conditions in the
$SU (3)/U (2)$ models, the relation between rational boundary states
and Landau-Ginzburg solitons has been investigated
in~\cite{Nozaki:2001mz}.}  
\medskip

In the Grassmannian models $SU (3)_{k}/U (2)$, we explicitly identify
the matrix factorisations that correspond to a set of B-type boundary
states that form a basis of the Ramond-Ramond charge lattice. To do
the identifications between matrix factorisations and boundary states
we compare the open string spectra, the RR charges and also
information on boundary renormalisation group flows. We expect to find
all other boundary states and matrix factorisations from tachyon
condensation of the basic ones. We illustrate and confirm this idea,
and construct matrix factorisations that correspond to another subset
of boundary states. For low levels ($k=1,2$), this means that we can
identify matrix factorisations for all rational boundary states, for
higher levels we believe that by performing more tachyon condensations
we would eventually identify all remaining factorisations.  
\medskip

The paper is organised as follows: In section~2 we shall discuss the
Grassmannian Kazama-Suzuki models, in particular their field content
and their B-type boundary states. For the model $SU(3)/U(2)$ we then
go more into detail and evaluate the spectra of the boundary theories
and the RR charges. In section~3 the Landau-Ginzburg description is
introduced. First we review the identification of the superpotential
that corresponds to the Grassmannian cosets, then we study
factorisations of the superpotentials for the $SU(3)/U(2)$ series. A
number of basic factorisations is given and the corresponding RR
charges are determined. Section~4 then deals with the comparison
between CFT and LG description. By analysing spectra and RR charges,
it is shown how to identify some of the boundary states with matrix
factorisations. We then discuss boundary renormalisation group flows
and tachyon condensation. On the one hand, we can use these to compare
the CFT and LG description, on the other hand we can use them to find
factorisations for the remaining boundary states. This is exemplified
for another family of boundary states. For low levels, where our
models are equivalent to minimal models, we compare in section~5 our
findings to results in the literature. In the concluding section~6 we
discuss some open problems and possible routes to solve them. Two
appendices contain the details of the calculations that form the basis
of our identifications between CFT and LG description.

\section{Kazama-Suzuki models}

Kazama and Suzuki~\cite{Kazama:1988uz,Kazama:1989qp} constructed a
large class of rational CFTs with $\mathcal{N}=2$ superconformal
symmetry as coset models of the form
\begin{equation}
\frac{G_{k}\times SO (2d)_{1}}{H}\ .
\end{equation}
Here, $G$ is a simple, compact Lie group, $k$ the corresponding level,
$2d$ is the difference of the dimensions of $G$ and of the regularly
embedded subgroup $H$ (which we take to have the same rank as $G$). To
have $\mathcal{N}=2$ supersymmetry, $G/H$ has to be K{\"a}hler and hence
the difference of dimensions, $2d$, is even.

Of particular interest are the models where $G$ is simply laced, the
level is $k=1$, and $G/H$ is a Hermitian symmetric space. In this
case, the CFTs have a description as Landau-Ginzburg
models~\cite{Lerche:1989uy}. These theories have been
classified~\cite{Kazama:1989qp}, and a prominent family of such models
is provided by the Grassmannians, where $G=SU (n+k)$ and $H=S (U
(n)\times U (k))$.

\subsection{Grassmannians: the bulk theory}
\label{sec:CFT_bulk}
 
The Grassmannian cosets are of the form
\begin{equation}\label{eq:cosetDef}
\frac{SU (n+k)_{1}\times SO (2nk)_{1}}{SU (n)_{k+1}\times
SU(k)_{n+1}\times U (1)} 
\cong \frac{SU (n+1)_{k}\times SO (2n)_{1}}{SU (n)_{k+1}\times U
(1)}\ .
\end{equation}
with central charge $c=\frac{3nk}{k+n+1}$.  The equivalence used here
is known as level-rank
duality~\cite{Kazama:1989qp,Lerche:1989uy,Gepner:1988wi,Fuchs:1993at,Blau:1995np,Naculich:1997ic,Xu:2001yg,Ali:2002vd}. 

We shall most of the time work in the formulation on the right hand
side. The 'embedding' homomorphism of the denominator group into the
numerator group is
\begin{equation}
i (h,\zeta) = \begin{pmatrix}
h\zeta & 0\\
0& \zeta^{-n}
\end{pmatrix} \in SU (n+1) \ ,
\end{equation}
where $h\in SU (n)$ is a $n\times n$-matrix, and $\zeta \in U (1)$ is
a phase. Note that this is not a one-to-one mapping, because $i
(\xi^{-1}\mathbf{1},\xi)=\mathbf{1}$ for $\xi^{n}=1$. This just means
that the denominator group only becomes a subgroup of the numerator
group after taking a $\mathbb{Z}_{n}$ quotient,
\begin{equation}
U (n) = \big(SU (n)\times U (1) \big)/\mathbb{Z}_{n}\ .
\end{equation}
This will become important shortly when we discuss selection and
identification rules.

The sectors of the theory are labelled by quadruples
$(\Lambda ,\Sigma ;\lambda ,\mu)$, where $\Lambda$ is a dominant weight
of $su (n+1)_{k}$, $\lambda$ is a dominant weight of $su (n)_{k+1}$,
$\mu$ is an integer labelling a $u(1)$-representation, and finally
$\Sigma$ labels a dominant weight of $so (2n)_{1}$, so it labels
either the trivial representation $0$, the vector ($v$), the spinor ($s$)
or the anti-spinor $(\bar{s})$ representation. Representations with $\Sigma
=0,v$ belong to the Neveu-Schwarz sector, $\Sigma =s,\bar{s}$ belong
to the Ramond sector.

As usual, the representation labels are restricted by selection rules,
and we have an equivalence relation on the allowed labels given by
identification
rules~\cite{Lerche:1989uy,Moore:1989yh,Gepner:1989jq}. The appearance
of selection and identification 
rules is connected to the existence of a non-trivial common center $Z$
of the numerator and denominator theory, or better the preimage
$Z=i^{-1} (Z_{G})$ of the center of the numerator group $G=SU
(n+1)$. Here, $Z_{SU (n+1)}=\{\eta\mathbf{1}|\eta^{n+1}=1 \}$, so that
$Z=\{(\xi^{-1}\mathbf{1},\xi \eta) | \xi^{n}=1,\eta^{n+1}=1 \}$. This
is a cyclic group $\mathbb{Z}_{n (n+1)}$ with generator $(e^{-2\pi
i/n}\mathbf{1},e^{2\pi i/n}e^{2\pi i/ (n+1)})$. 

Corresponding to the center $Z$, there is a cyclic simple current
group $G_{\text{id}}$ that acts on the
weights~\cite{Schellekens:1989am,Schellekens:1989uf}. It is generated
by the simple current $J_{0}=(J_{n+1},v;J_{n},k+n)$, where
$J_{n+1}=k\omega_{1}$ generates the simple current group of
$su(n+1)_{k}$, and $J_{n}= (k+1)\omega_{1}$ generates the simple
current group of $su(n)_{k+1}$ (here, we denote for both $su (n)$ and
$su (n+1)$ the first fundamental weight by $\omega_{1}$). In the $u
(1)$-part, the simple current acts as $\mu\to \mu +k+n$. Since
$J_{0}^{n (n+1)}$ should act as the identity, the $u(1)$ labels $\mu$
should be periodically identified with period $n (n+1) (k+n)$. This
means that the $u(1)$ Heisenberg algebra can be enlarged to $u(1)_{n
(n+1) (k+n)}$.

The simple current group $G_{\text{id}}$ acts without fixed-points on
the quadruples of weights and generates the identification rules. On
the other hand, the selection rules are encoded in the requirement
that the monodromy charges of the numerator and denominator parts
should be equal,
\begin{equation}\label{eq:mCharges}
Q_{J_{n+1}} (\Lambda) +Q_{v} (\Sigma)\overset{!}{=}Q_{J_{n}} (\lambda)+Q_{k+n}
(\mu)\ .
\end{equation} 
The monodromy charges are defined as usual as differences of
conformal weights, $Q_{J} (\phi)=h_{J}+h_{\phi}-h_{J\phi} \mod 1$.

The sectors of the theory are labelled by equivalence classes
$[\Lambda ,\Sigma ;\lambda ,\mu]$ of allowed labels. An important subset
of representations of the coset algebra is the set of chiral primary
states. It can be shown~\cite{Gepner:1988wi} that in the Grassmannian models a
chiral primary can be represented as
\begin{equation}\label{chiral_primary}
[\Lambda ,0;P_{n}\Lambda ,P_{U}\Lambda] \ .
\end{equation}
Here $P_{n}$ and $P_{U}$ are the projection matrices that map $su
(n+1)$ weights to $su(n)$ and $u(1)$ weights, respectively. In terms
of Dynkin labels they are explicitly given as
\begin{align}
P_{n} (\Lambda_{1},\dotsc ,\Lambda_{n}) & = (\Lambda_{1},\dotsc ,\Lambda
_{n-1}) & P_{U} (\Lambda_{1},\dotsc ,\Lambda_{n}) & =
\Lambda_{1}+2\Lambda_{2}+\dotsb +n\Lambda_{n} \ . 
\end{align}
The above statement about the form of the chiral primaries makes it
easy to obtain the number of chiral primary states -- it is just given
by the number of dominant highest weights of $su (n+1)_{k}$, i.e.
\begin{equation}
\text{number of chiral primaries}\ = \binom{k+n}{n} \ .
\end{equation}
\smallskip

\noindent Up to now we have only discussed representation theoretic
aspects. When we want to consider a conformal field theory (without
boundaries for the moment), we have to specify the spectrum, which we
shall take to be of (almost) diagonal form,
\begin{equation}
\mathcal{H} = \bigoplus_{[\Lambda ,\Sigma ;\lambda ,\mu]}
\mathcal{H}_{[\Lambda ,\Sigma ;\lambda ,\mu ]} \otimes \mathcal{H}_{[\Lambda
,\Sigma^{+} ;\lambda ,\mu ]} \ .
\end{equation}
Two comments are in order. The most natural thing would be to consider
the charge conjugated spectrum. It turns out, however, that the
diagonal spectrum is the one that is related to the Landau-Ginzburg
models that we shall discuss later. Of course, we can use the mirror
automorphism to map one spectrum into the other, but then we would
also map B-type boundary conditions to A-type, and if we want to
relate B-type conditions in the coset model to B-type in the
Landau-Ginzburg theory, it is the diagonal spectrum that we have to choose.
The other comment concerns the small deviation from the diagonal
theory, namely the charge conjugation on the $so (2n)_{1}$
representation. This is the right choice to obtain the Landau-Ginzburg
theories with the standard potentials that we introduce later. If we
twist the spectrum by applying the outer automorphism that exchanges
spinor and anti-spinor, we obtain the theory where we add a quadratic
term $z^{2}$ to the superpotential.

\subsection{Boundary conditions}

We now want to discuss the theory on a world-sheet with a
boundary,\footnote{Boundary conditions in (non-minimal) Kazama-Suzuki
models have been discussed before
in~\cite{Lerche:2000iv,Nozaki:2001mz,Ishikawa:2003kk}.} which we take
to be the upper half plane. At the real axis, we impose B-type gluing
conditions for the energy momentum tensor $T$, the current $J$ and the
supercurrents $G^{\pm}$,
\begin{align}\label{Btypegluing}
T (z) & = \bar{T} (\bar{z}) & J (z) & = \bar{J} (\bar{z}) &
G^{\pm} (z) & = \eta \bar{G}^{\pm} (\bar{z})
\end{align}
at $z=\bar{z}$. Here, $\eta$ is a sign corresponding to the choice of
a spin structure. The sign of $\eta$ does of course not affect the
gluing conditions for the fields of the bosonic subalgebra of the
$\mathcal{N}=2$ superconformal algebra. 

In general, the classification and construction of boundary states
with the above gluing conditions is a difficult and unsolved
problem. We need to restrict our focus on highly symmetric boundary
conditions, which satisfy gluing conditions on more fields of our
chiral symmetry algebra. Denoting by $W (z)$ any chiral field of the
coset algebra, we can impose the gluing condition~\cite{Cardy:1989ir}
\begin{equation}
W (z) = \omega (\bar{W}) (\bar{z})\quad  \text{at}\ z=\bar{z}\ .
\end{equation}
Here, $\omega$ is an automorphism of the coset algebra.  The coset
algebra contains the bosonic subalgebra of the $\mathcal{N}=2$
superconformal algebra, so the gluings we choose for the coset theory
should be consistent with the B-type gluing conditions.

The classification of automorphisms of coset algebras is not known,
but there is a particularly nice class of automorphisms that we can
use. An automorphism of this class is induced by an automorphism
$\omega_{G}$ of the group $G$ that can be restricted to an
automorphism $\omega_{H}$ of $H$, in the sense that $i (\omega_{H}
(h))=\omega_{G} (i (h))$ for all $h\in H$. In~\cite{Ishikawa:2003kk}
the automorphisms of this type have been classified, and it is also
analysed which automorphisms correspond to B-type gluing
conditions. In the Grassmannian models, only the trivial automorphism
is possible.

This, however, still means that we have to deal with twisted boundary
conditions, because we chose a diagonal bulk spectrum which is twisted
(by conjugation) with respect to the standard theory with charge
conjugated spectrum. In particular this means that only those sectors
of the bulk theory can couple to the branes which are invariant under
charge conjugation.

Our discussion leads to the conclusion that
only those bulk fields can couple to the boundary that belong to
$\mathcal{H}_{[\Lambda ,\Sigma ;\lambda ,\mu]}\otimes
\mathcal{H}_{[\Lambda ,\Sigma^{+};\lambda ,\mu]}$ satisfying
\begin{equation}\label{invariant_sectors}
[\Lambda ,\Sigma ;\lambda ,\mu ] = [\Lambda^{+} ,\Sigma ;\lambda^{+}
,-\mu]\ .
\end{equation}
Note that because of our choice of the spectrum, the $so
(2n)_{1}$-label $\Sigma$ appears without conjugation on the right hand
side. 

To analyse the condition~\eqref{invariant_sectors}, we have to take
into account that only the equivalence classes of labels have
to agree. Let us denote the quadruples by $\alpha$ and the
automorphism appearing on the right hand side
of~\eqref{invariant_sectors} by $C$. Solving $[\alpha]=[C (\alpha)]$
then means to find all equivalence classes $[\alpha]$ such that 
\begin{equation}\label{invariant_alphas}
\alpha = J C (\alpha)
\end{equation}
for some simple current $J$ of the identification group
$G_{\text{id}}$. If $\alpha$ is a solution to the above equation, then
of course $J'\alpha$ is also a solution, but possibly for a different
$J$. In our case, commuting the charge conjugation with the action of
a simple current just inverts the current, so that we get
\begin{equation}\label{twisted_gluing}
J'\alpha = J' J C (\alpha) = J'JJ' C (J'\alpha) \ .
\end{equation}
Hence, $J'\alpha$ satisfies~\eqref{invariant_alphas} if $J$ is replaced
by $J'JJ'$. In other words we only have to
investigate~\eqref{invariant_alphas} for one representative $J$ of each
orbit $\mathcal{C}_{J}=\{J'JJ'|J'\in G_{\text{id}} \}$. In our case
where $G_{\text{id}}$ is just a cyclic group of even order, there are
two orbits: one generated by $1$ (containing the even powers of
$J_{0}$) and one generated by $J_{0}$ (consisting of the odd powers of
$J_{0}$). So we are led to consider solutions to the condition
\begin{equation}\label{Ishibashi_condition}
(\Lambda ,\Sigma ;\lambda ,\mu) = (\Lambda^{+} ,\Sigma ;\lambda^{+}
,-\mu )
\end{equation} 
and solutions of
\begin{equation}
(\Lambda ,\Sigma ;\lambda ,\mu) = (J_{n+1}\Lambda^{+},v\Sigma
;J_{n}\lambda^{+},-\mu+k+n) \ .
\end{equation}
As is obvious from the condition on the $so (2n)_{1}$-label, the
latter equation does not have a solution, so the only sectors that
couple to the boundary correspond to solutions of the first
condition. On the set of labels that satisfy this condition, we still
have the action of a subgroup of the identification group; it is clear
from the discussion above and~\eqref{twisted_gluing} that apart from
the identity only the element $J_{0}^{n (n+1)/2}$ maps this set to
itself. We can use this identification to set the $U (1)$-label to
$\mu=0$, since the other solution, namely $\mu=\pm \frac{n (n+1)}{2}
(k+n+1)$, is mapped to $\mu =0$ by $J_{0}^{n (n+1)/2}$.

There is one further issue that we have to take into account, namely
that some sectors are forbidden by selection rules. As we have said,
the selection rule is encoded in the monodromy charges
(\ref{eq:mCharges}). For a self-conjugate representation
$\lambda=\lambda^{+}$ of $su (n)$, the monodromy charge with respect
to the generating simple current $J_{(n)}$ is either zero (if $n$ is
odd) or given by $\frac{1}{2}\lambda_{n/2}$ (for even
$n$).\footnote{similarly for $su (n+1)$} For the $so (2n)_{1}$
representation $\Sigma$, the monodromy charge is $0$ for $\Sigma =0,v$
and $\frac{1}{2}$ for $\Sigma =s,\bar{s}$. So for given $\lambda$ and
$\Lambda$, the selection rules restrict the choice of $\Sigma$ to two
values.  \smallskip

In each allowed sector that couples to the brane, we can construct
(twisted) Ishibashi states~\cite{Ishibashi:1988kg}. The set of
Ishibashi states $|\Lambda ,\Sigma ;\lambda ,0\dra$ is labelled by
self-conjugate labels $\Lambda =\Lambda^{+}$, $\lambda =\lambda^{+}$
and an $so (2n)_{1}$-label $\Sigma$ (that is constrained by the
selection rule).  The task is now to find the right linear
combinations that form the boundary states. The problem of
constructing twisted boundary states in coset models has been analysed
in~\cite{Ishikawa:2001zu,Ishikawa:2002wx,Fredenhagen:2003xf,Ishikawa:2003kk}
(see also~\cite{Maldacena:2001ky,Gawedzki:2001ye,Elitzur:2001qd}). In
the case at hand, we are in a standard situation where the set of
Ishibashi labels is just given by a tuple of twisted Ishibashi labels
of the constituent models, acted upon by an identification group
without fixed-points. In this case the Ansatz of factorised boundary
states~\cite{Ishikawa:2001zu} works, i.e.\ we take the coefficients of
the twisted boundary states of the constituent theories, and multiply
them,
\begin{equation}\label{boundarystates}
|L,S;l\rangle = \mathcal{N}
\sum_{(\Lambda,\Sigma;\lambda,0)\in\mathcal{V}} 
\frac{\psi_{L\Lambda}^{(n+1)}
S^{(so)}_{S\Sigma}\bar{\psi}_{l\lambda}^{(n)}}{\sqrt{S^{(n+1)}_{0\Lambda}
S^{(so)}_{0\Sigma}S^{(n)}_{0\lambda}}} |\Lambda ,\Sigma ;\lambda
,0\dra \ .
\end{equation}
Here, $S^{(n)}$ is the modular S-matrix of $su (n)_{k+1}$,
$\psi^{(n)}$ is its twisted S-matrix (similarly for $n+1$). $S^{(so)}$
is the modular S-matrix of $so (2n)_{1}$, and $\mathcal{V}$ denotes
the set of labels $(\Lambda ,\Sigma ;\lambda,0)$ with
$\Lambda=\Lambda^{+},\lambda =\lambda^{+}$ and which in addition
satisfy the selection rules. The normalisation $\mathcal{N}$ will be
determined shortly.

The label $S$ is a usual $so (2n)_{1}$-representation. The labels
$L,l$ denote representations of the twisted affine algebras
$A_{n}^{(2)}$ and $A_{n-1}^{(2)}$, respectively. Let us for a moment
concentrate just on the numerator part, $A_{n}^{(2)}$. The label $L$
can be represented as a tuple $(L_{1},\dotsc ,L_{\lfloor
\frac{n+1}{2}\rfloor})$ with the condition that
$2\sum_{i=1}^{n/2}L_{i}\leq k$ for $n$ even, and
$L_{1}+\sum_{i=2}^{(n+1)/2}L_{i}\leq k$ for $n$ odd. Also for $n$ odd,
there is a simple current like action on the label, $L\mapsto
\mathcal{J}L$, that replaces $L_{1}$ by
$(\mathcal{J}L)_{1}=k-L_{1}-2\sum_{i=2}^{(n+1)/2}L_{i}$. The twisted
S-matrix satisfies
\begin{equation}
\psi_{\mathcal{J}\!L\, \Lambda }^{(n+1)} = \psi_{L\, \Lambda}^{(n+1)}
(-1)^{\Lambda_{(n+1)/2}} \ .
\end{equation}
The discussion for the denominator part $su (n)_{k+1}$ is similar. 

The selection rules on the Ishibashi states induce identifications of
labels of boundary states, namely we have that
\begin{equation}
|L,S;l\rangle  = |\mathcal{J}L,vS;\mathcal{J}l \rangle \ ,
\end{equation}
where it is understood that $\mathcal{J}$ acts trivially on $L$ when
$n$ is even, and trivially on $l$ when $n$ is odd.
\smallskip

Having identified the set of Ishibashi states and boundary
states, we can now determine the spectra. This will then also fix
the normalisation constant $\mathcal{N}$.

For the closed string overlap amplitude between two boundary states,
or equivalently the one-loop open string partition function, we have
($q=e^{2\pi i\tau}, \tilde{q}=e^{-2\pi i/\tau}$)
\begin{align}
\langle L_{1},S_{1};l_{1}| & \tilde{q}^{\frac{1}{2}
(L_{0}+\bar{L}_{0}-\frac{c}{12})}|L_{2},S_{2};l_{2}\rangle \nonumber \\
& = \mathcal{N}^{2} \bigg( \frac{n (n+1)}{k+n+1}\bigg)^{1/2} 
\sum_{(\Lambda ,\Sigma ;\lambda ,0)\in \mathcal{V}}\ 
\sum_{[\Lambda',\Sigma ';\lambda',\mu ']}\nonumber\\
& \quad\quad   \frac{1}{2}\bigg(
\frac{\bar{\psi}^{(n+1)}_{L_{1}\Lambda}\psi^{(n+1)}_{L_{2}\Lambda}
S^{(n+1)}_{\Lambda'\Lambda}}{S_{0\Lambda}^{(n+1)}}
\frac{\psi^{(n)}_{l_{1}\lambda}\bar{\psi}^{(n)}_{l_{2}\lambda}
\bar{S}^{(n)}_{\lambda'\lambda}}{S_{0\lambda}^{(n)}}
\frac{\bar{S}^{so}_{S_{1}\Sigma} S^{so}_{S_{2}\Sigma}S^{so}_{\Sigma '
\Sigma}}{S^{so}_{0\Sigma}} \nonumber\\
& \quad \quad\quad  + \big( (L_{1},S_{1},l_{1})\to
(\mathcal{J}L_{1},vS_{1},\mathcal{J}l_{1})\big)\bigg) \chi_{[\Lambda
',\Sigma ';\lambda ',\mu ']} (q)\\
& = \sum_{[\Lambda ',\Sigma ';\lambda ',\mu ']} \Big( n^{(n+1)}_{\Lambda'
L_{2}}{}^{L_{1}} n^{(n)}_{\lambda 'l_{2}}{}^{l_{1}} N^{so}_{\Sigma
'S_{2}}{}^{S_{1}} \nonumber\\
& \quad \quad + \big( (L_{1},S_{1},l_{1})\to
(\mathcal{J}L_{1},vS_{1},\mathcal{J}l_{1})\big) \Big) \chi_{[\Lambda
',\Sigma ';\lambda ',\mu ']} (q)\ .
\label{spectrum}
\end{align}
The sum over the orbit of $(\mathcal{J},v;\mathcal{J})$ has been
introduced to take care of the selection rules for Ishibashi
states. The factor $(n (n+1) (k+n+1))^{-1/2}$ comes from the modular
transformation of the $u(1)$-part (see~\eqref{app:SmatrixU1}), the
factor $n (n+1)$ comes from the relation of the coset modular S-matrix
to the product of the S-matrices of the constituent models. In the
last step we have used the Verlinde formula and its twisted version to
get the (twisted) fusion rules $n^{(n+1)}$, $n^{(n)}$ and
$N^{so}$. The normalisation factor has been set to $\mathcal{N}^{4}=4
(k+n+1)/ (n (n+1))$ in (\ref{spectrum}) such that the vacuum state has
multiplicity one in the self-spectra.  
\smallskip

The boundary states that we have introduced are consistent with the
B-type gluing conditions for the supercurrents with either sign for
$\eta$ in~\eqref{Btypegluing}. By restricting to boundary state labels
$S=0,v$, we fix one sign of $\eta$, i.e.\ we fix the
spin-structure. From now on, we only allow $S$ to be either of the
two values. On the other hand, changing the $so$-label from $0$ to $v$
and vice versa means to exchange brane and anti-brane (the RR part of
the boundary state changes sign). In the following we shall use the
notation 
\begin{equation}
|L,l\rangle \equiv |L,0;l\rangle\quad \text{and}\quad 
\overline{|L,l}\rangle \equiv |L,v;l\rangle \ .
\end{equation}
The identification rule on the boundary states is then
\begin{equation}
|L,l\rangle = \overline{|\mathcal{J}L,\mathcal{J}l}\rangle \ .
\end{equation}

\noindent We are particularly interested in the chiral primary fields that
appear in the open string spectrum, because their multiplicities can
be compared to the computations in the Landau-Ginzburg models. Chiral
primaries are of the form~\eqref{chiral_primary}, so in the overlap of
$|L_{1},l_{1}\rangle$ and $|L_{2},l_{2}\rangle$ we find a chiral
primary state $(\Lambda ,0;P_{n}\Lambda ,P_{U}\Lambda)$ with
multiplicity $n^{(n+1)}_{\Lambda L_{2}}{^{L_{1}}}n^{(n)}_{P_{n}\Lambda
\, l_{2}}{}^{l_{1}}$. The number of chiral primaries $(\Lambda
,0;P_{n}\Lambda ,P_{U}\Lambda)$ in the spectrum minus the number of
superpartners $(\Lambda ,v;P_{n}\Lambda ,P_{U}\Lambda)$ of chiral
primaries defines the intersection index between two boundary states,
\begin{equation}
I (L_{1},l_{1}|L_{2},l_{2}) = \sum_{\Lambda } \Big(n^{(n+1)}_{\Lambda
L_{2}}{}^{L_{1}}n^{(n)}_{P_{n}\Lambda \, l_{2}}{}^{l_{1}} -
n^{(n+1)}_{\Lambda L_{2}}{}^{\mathcal{J}L_{1}}n^{(n)}_{P_{n}\Lambda \,
l_{2}}{}^{\mathcal{J}l_{1}} \Big) \ .
\end{equation}
The intersection index carries information about the RR charges of the
D-branes, and it is conserved in dynamical processes like tachyon condensation.

This ends our discussion of B-type boundary states in the Grassmannian
series. We have identified the maximally symmetric boundary states
$|L,l\rangle$, and determined the spectra in terms of twisted fusion
rules that can be found in~\cite{Gaberdiel:2002qa}. In the following
sections we shall concentrate on the case $n=2$ and work out the
explicit formulae.

\subsection{The \texorpdfstring{$SU (3)/U (2)$}{SU (3)/U (2)} series}

In the Kazama-Suzuki model based on $SU (3)/U (2)$, the sectors are
labelled by quadruples $(\Lambda ,\Sigma ;\lambda ,\mu)$ where $\Lambda
= (\Lambda_{1},\Lambda_{2})$ with $\Lambda_{1}+\Lambda_{2}\leq k$ is a
dominant weight of $su (3)_{k}$, $\Sigma$ labels a representation of $so
(4)_{1}$, $\lambda\in \{0,\dotsc ,k+1 \}$ labels a dominant weight of
$su (2)_{k+1}$ and $\mu$ is a $6 (k+3)$-periodic integer labelling
representations of $u (1)_{6 (k+3)}$. The selection rule for a
quadruple reads
\begin{equation}
\frac{\Lambda_{1}+2\Lambda_{2}}{3} + \frac{|\Sigma|}{2}
-\frac{\lambda}{2}+\frac{\mu}{6} \in \mathbb{Z} \ ,
\end{equation}
where $|\Sigma |$ is defined to be $1$ for $\Sigma =s,\bar{s}$ and $0$
for $\Sigma =0,v$. The simple current 
\begin{equation}
J_{0}= ((k,0),v;k+1,k+3)
\end{equation}
that generates the identification group $G_{\text{id}}$ leads to the
following identification of labels,
\begin{equation}
((\Lambda_{1},\Lambda_{2}),\Sigma ;\lambda ,\mu ) \sim
((k-\Lambda_{1}-\Lambda_{2},\Lambda_{1}),v\Sigma ;k+1-\lambda ,\mu +k+3)\ .
\end{equation}
The order of the identification group is $6$, so out of the total
number 
\begin{equation}
N_{\text{tot}} = \frac{(k+1) (k+2)}{2}\cdot 4 \cdot (k+2) \cdot 6
(k+3) = 12 (k+1) (k+2)^{2} (k+3) 
\end{equation}
of quadruples, only $N_{\text{tot}}/36$ label
allowed and inequivalent representations. 

The conformal weight $h$ and the $U (1)$-charge $q$ (with respect to
the $U (1)$ of the superconformal algebra) of a representation labelled
by $(\Lambda ,\Sigma ;\lambda ,\mu )$ are given by
\begin{align}
h & = \frac{1}{2 (k+3)} \Big((\Lambda ,\Lambda +2\rho) - \frac{\lambda
(\lambda +2)}{2} -\frac{\mu ^{2}}{6} \Big) + h_{\Sigma} \mod 1\\
q & = -q_{\Sigma} + \frac{\mu }{k+3} \mod 2 \ .
\end{align}
Here, $\rho$ denotes the Weyl vector of $su(3)$, $h_{\Sigma}$ and
$q_{\Sigma}$ are the contributions from the $so (4)_{1}$-part, they
are given as
\begin{align}
h_{0}&=0 & h_{v}&=\frac{1}{2} & h_{s}&=\frac{1}{4} & h_{\bar{s}}& 
= \frac{1}{4}\\
q_{0}&=0 & q_{v}&=1 & q_{s}&=1 & q_{\bar{s}} & = 0 \ .
\end{align}

\noindent The chiral primary states are labelled by
$((\Lambda_{1},\Lambda_{2}),0;\Lambda_{1},\Lambda_{1}+2\Lambda_{2})$.
They have $U (1)$-charge $q=\frac{\Lambda_{1}+2\Lambda_{2}}{k+3}$ and
conformal weight $h=\frac{1}{2}q$. In total there are $(k+1) (k+2)/2$
chiral primaries. The set of chiral primaries has a ring structure, and
we shall discuss this chiral ring when we discuss the connection to
the Landau-Ginzburg models in section~\ref{sec:LG_bulk} .
\smallskip

An important property of the superconformal algebra is the existence
of a spectral flow. The spectral flow automorphism extends to the
coset algebra, and the action of a flow by half a unit on a
representation $(\Lambda ,\Sigma ;\lambda ,\mu)$ is given by
\begin{equation}
(\Lambda ,\Sigma ;\lambda ,\mu ) \mapsto (\Lambda ,s\times \Sigma
;\lambda ,\mu +3)\ ,
\end{equation}
so it is generated by the simple current $(0,s;0,3)$ (for a general
Grassmannian model, $3$ is replaced by $\frac{n
(n+1)}{2}$)~\cite{Lerche:1989uy,Hosono:1990yj}. The flow by half a
unit maps the Ramond sector to the Neveu-Schwarz sector and vice versa.

In the $SU (3)/U (2)$ Grassmannian model, the boundary label $L$ and
$l$ are just integers ranging from $L=0,\dotsc ,\lfloor
\frac{k}{2}\rfloor$ and $l=0,\dotsc ,k+1$. The identification is 
\begin{equation}
|L,l\rangle = \overline{|L,k+1-l}\rangle \ .
\end{equation}
The explicit formula for the boundary states can be found in
Appendix~\ref{sec:boundarystates}. For the denominator part $su
(2)_{k+1}$, charge conjugation is trivial, so the relevant fusion
rules that appear in the open string spectra are the ordinary
untwisted ones that we denote by $N^{(k+1)}_{\lambda
l_{2}}{}^{l_{1}}$. The twisted fusion rules for the numerator theory
$su (3)_{k}$ have been explicitly computed in~\cite{Gaberdiel:2002qa},
their expressions involve either the fusion rules of $su (2)$ at level
$2k+4$ or (for odd $k$) at level $(k-1)/2$. For our purposes, however,
it is convenient to write them in terms of $su (2)$ fusion rules at
level $k+1$,
\begin{equation}\label{twistedfusion}
n_{\Lambda L_{2}}{}^{L_{1}} = \sum_{\gamma} b^{\Lambda}_{\gamma} \big(
N^{(k+1)}_{\gamma \ L_{2}}{}^{L_{1}} - N^{(k+1)}_{k+1-\gamma
\ L_{2}}{}^{L_{1}}\big) \ .
\end{equation}
Here $\Lambda = (\Lambda_{1},\Lambda_{2})$ is a dominant weight of $su
(3)_{k}$, $\gamma$ denotes a dominant weight of $su (2)$ and
$b^{\Lambda}_{\gamma}$ is the branching rule of the regular embedding of $su
(2)\subset su (3)$ with embedding index $x=1$. This expression for the
twisted fusion rules appears to be new (although closely related to
the results of~\cite{Gaberdiel:2002qa}) and is proved in
appendix~\ref{sec:twisted_su3_fusion}. 

The open string spectrum is now obtained by specialising the
formula~\eqref{spectrum} for the spectrum in a general Grassmannian
model to the case of $SU(3)/U(2)$. For the intersection index, we find 
\begin{align}
I (L_{1},l_{1}|L_{2},l_{2}) & = \sum_{\Lambda  =(\Lambda_{1},\Lambda_{2})}
n_{\Lambda L_{2}}{}^{L_{1}} \Big( N^{(k+1)}_{\Lambda_{1}l_{2}}{}^{l_{1}}
- N^{(k+1)}_{\Lambda_{1}l_{2}}{}^{k+1-l_{1}}\Big)\nonumber \\
& = \sum_{\Lambda ,\gamma } b^{\Lambda}_{\gamma}
\Big(N^{(k+1)}_{\gamma L_{2}}{}^{L_{1}} - N^{(k+1)}_{k+1-\gamma\,
L_{2}}{}^{L_{1}} \Big) \Big(N^{(k+1)}_{\Lambda_{1}l_{2}}{}^{l_{1}} -
N^{(k+1)}_{k+1-\Lambda_{1}\, l_{2}}{}^{l_{1}} \Big) \ .
\end{align}
We observe that the labels $L_{i}$ and $l_{i}$ enter the formula in a
similar, but not symmetric way. Some explicit results for the
spectra of chiral primaries are collected in
appendix~\ref{sec:CFTspectra}.

\subsection{RR charges and g-factors}

D-branes can be charged under RR fields. B-type D-branes can only
couple to RR ground states that have opposite $U (1)$-charge for the
left and right-movers. In our case where we consider a diagonal bulk
spectrum, the B-type condition thus only allows a coupling to
RR ground states with vanishing $U (1)$-charge. 

Let us first look at the left-movers. Ramond ground states
are obtained from chiral primary states by the
application of spectral flow by half a unit, so the set of Ramond ground
states is given by
\begin{equation}
\text{RGS} =
\{[(\Lambda_{1},\Lambda_{2}),s;\Lambda_{1},\Lambda_{1}+2\Lambda_{2}+3]
\} \ .
\end{equation}
The $U (1)$-charge is given by
$q=-1+\frac{\Lambda_{1}+2\Lambda_{2}+3}{k+3}$, so the uncharged Ramond
ground states correspond to labels satisfying
$\Lambda_{1}+2\Lambda_{2}=k$. We are now looking for representatives
of these states that have a symmetric $su(3)$-weight. Applying
$J_{0}^{5}=J_{0}^{-1}$ to the labels, we obtain the following form of
the set of uncharged Ramond ground states,
\begin{equation}
\text{RGS}_{0} = \{[(\Lambda_{2},\Lambda_{2}),\bar{s};2\Lambda_{2}+1,0] \}
\ .
\end{equation}
Combining such Ramond ground states from left- and right-movers, we
obtain the RR ground states that can couple to our B-type branes. The
RR charges of the brane described by a boundary state $|L,l\rangle$
are then given by the coefficients in front of the corresponding RR
ground states in~(\ref{boundarystates}). The charge
$\text{ch}_{j}(|L,l\rangle)$ with respect to the RR ground state with
symmetric $su(3)$ weight $(j,j)$ is given by
\begin{equation}
\text{ch}_{j} (|L,l\rangle) = \mathcal{N}\frac{
\psi^{(3)}_{L\, (j,j)} S^{so}_{0\bar{s}} S^{(2)}_{l\, 2j+1}
}{\sqrt{S^{(3)}_{(0,0) (j,j)}S^{so}_{0\bar{s}} S^{(2)}_{0\, 2j+1}}} \ .
\end{equation}
Employing the explicit formulae for the (twisted) S-matrices (see
appendix~\ref{sec:boundarystates}), we get
\begin{equation}\label{RRcharge}
\text{ch}_{j} (|L,l\rangle) = \frac{1}{\sqrt{2}}\frac{\sin
\big(\frac{2\pi (L+1) (j+1)}{k+3} \big) \sin \big(
\frac{\pi (l+1) (2j+2)}{k+3} \big)}{\sin \big(\frac{\pi
(j+1)}{k+3} \big) \sin \big(\frac{2\pi (j+1)}{k+3}\big)}\ .
\end{equation}
As there are only $\lfloor \frac{k}{2}\rfloor +1$ uncharged Ramond
ground states, it is clear that the charge vectors of the boundary
states are not linearly independent. A basis is for example given by
the charge vectors of the boundary states $|L,0\rangle$; it is
straightforward to verify that
\begin{equation}\label{RRcharges}
\text{ch}_{j} (|L,l\rangle) = \sum_{L'=0}^{\lfloor \frac{k}{2}\rfloor}
\big(N^{(k+1)}_{L L'}{}^{l} - N^{(k+1)}_{L L'}{}^{k+1-l} \big) \text{ch}_{j}
(|L',0\rangle)
\end{equation} 
for all $j=0,\dotsc ,\lfloor \frac{k}{2}\rfloor$. Let us briefly
remark that this fits nicely with an analysis of the dynamics of such
branes in the limit of large level $k$ along the lines
of~\cite{Fredenhagen:2001kw,Alekseev:2002rj,Fredenhagen:thesis}. In
this limit, the branes are labelled by a representation $L$ of the
invariant subgroup $SU (2)\subset SU (3)$ and a representation $l$ of
the numerator group $SU (2)$. The dynamics at large level $k$ suggest
that the charge of the branes $(L,l)$ is measured by the
representation $L\otimes l$ of the diagonally embedded $SU (2)$. This
matches precisely with the charge formula in~\eqref{RRcharges}.
\smallskip

Another useful information on the D-branes is provided by their mass,
or in the CFT language, the g-factor of the boundary condition. It is
given by the coefficient of the boundary state $|L,l\rangle$ in front
of the vacuum state, which -- up to an overall normalisation -- is given by
\begin{equation}
\tilde{g}_{L,l} = \sin \left(\frac{2\pi (L+1)}{k+3} \right) \sin
\left(\frac{\pi (l+1)}{k+3} \right) \ .
\end{equation}
We chose the notation $\tilde{g}$ to emphasise that this is an
unnormalised g-factor. The g-factor has the symmetry
\begin{equation}
\tilde{g}_{L,2L'+1} = \tilde{g}_{L',2L+1} \ ,
\end{equation}
and also, because of the identification rule,
$\tilde{g}_{L,l}=\tilde{g}_{L,k+1-l}$ (brane and anti-brane have of
course the same g-factor). For odd $k$, there is in addition the
symmetry $\tilde{g}_{L,l}=\tilde{g}_{\frac{k-1}{2}-L,l}$. 
For odd $k$, the smallest g-factor (corresponding to the lightest
D-brane) is carried by $|0,0\rangle$ and $|\frac{k-1}{2},0\rangle$
(and their-anti-branes). For even $k$, the lightest D-brane
corresponds to $|\frac{k}{2},0\rangle$ and its
anti-brane. 
\smallskip

This concludes our presentation of the CFT results on boundary states
in Grassmannian Kazama-Suzuki models. We shall now turn towards the
Landau-Ginzburg description.

\section{Landau-Ginzburg theory}

In this section we shall discuss the description of B-type boundary
conditions in Landau-Ginzburg models that correspond to Grassmannian
coset models. We shall first introduce the bulk models in section~3.1,
and then discuss the concept of matrix factorisations in section~3.2.
Sections~3.3 and~3.4 then analyse factorisations in the $SU (3)/U (2)$ model.

\subsection{Landau-Ginzburg description of Kazama-Suzuki models}
\label{sec:LG_bulk}

A Landau-Ginzburg theory is a theory of chiral scalar superfields
$\Phi_{i}$ with action (in superspace notation)
\begin{equation}
\mathcal{S}_{LG} = \int d^2 z d^4 \theta K(\Phi,\bar\Phi) + \int d^2 z
\Big( d^2 \theta W(\Phi) +c.c.\Big)\ ,
\end{equation}
where $K(\Phi,\bar\Phi)$ denotes the K{\"a}hler potential and $W(\Phi)$ is
the superpotential. This theory is in general not scale invariant, and
one can study its behaviour under renormalisation group (RG) flow.
Due to non-renormalisation theorems, the superpotential is not
renormalised~\cite{Vafa:1988uu,Howe:1989qr}, but only the D-term
involving the K{\"a}hler potential. In this way, one can obtain some
information on the behaviour of the theory in the infrared.

In the course of the RG flow, the fields $\Phi_{i}$ undergo
wavefunction renormalisation, so they are rescaled during the
flow, and in that sense there is a change in the superpotential. In
the infrared, where one expects a scale-invariant theory, the
superpotential therefore has to be quasi-homogeneous,
\begin{equation}
W (e^{i\lambda q_{i}}\Phi_{i}) = e^{2i\lambda} W (\Phi_{i}) \ ,
\end{equation}
where the fields can have different weights $q_{i}$ under
scaling. The infrared fixed-points of Landau-Ginzburg theories are
therefore characterised by such quasi-homogeneous superpotentials.
The central charges of the fixed-point theories are completely
determined by the weights $q_{i}$ (see e.g.\ \cite{Vafa:1988uu}),
\begin{equation}
c=\sum_{i}3 (1-q_{i})\ .
\end{equation}
The superpotential now determines the ring of chiral primary
operators, the chiral ring
\begin{equation}
R=\frac{\mathbb{C}[x_{1},\dotsc ,x_{n}]}{\langle \partial_{i}W
\rangle} \ .
\end{equation}
It is this chiral ring that we can compare to the chiral ring in
the superconformal coset models to get the identification of the
theories. 

From the CFT side, the multiplication in the chiral ring is given by
the non-singular term in the operator product expansion (OPE) of two
chiral primary operators, which again has to be chiral primary.  The
OPEs consist of the fusion rules that essentially govern the
representation theoretic constraints on the operator products, and
some structure constants, which in general are rather difficult to
compute. To obtain the ring structure, one is however allowed to
rescale the chiral primary fields to have simpler coefficients. In the
case of the Grassmannian coset models $SU (n+1)/U (n)$,
Gepner has shown~\cite{Gepner:1991gr} that the structure constants
involved in the definition of the chiral ring can be set to $1$, so
that the chiral ring structure is given by the appropriate truncation
of the fusion rules to chiral primary fields.

That being said, we can now review how to obtain the corresponding
chiral rings. As we have discussed in section~\ref{sec:CFT_bulk} (see
eq.\ \eqref{chiral_primary}), the chiral primary
fields are labelled by representations of $su (n+1)$. These
representations can all be generated by tensor products from the
fundamental representations that we denote by $y_{1},\dotsc
,y_{n}$. Any representation $\Lambda =(\Lambda_{1},\dotsc ,\Lambda_{n})$ can
be written as a polynomial $U_{\Lambda } (y_{i})$ in the
$y_{i}$. These polynomials are given by Giambelli's formula 
\begin{equation}\label{Giambelli}
U_{\Lambda} (y) = \det \big(y_{a_{i}+i-j} \big)_{1\leq i,j\leq
|\Lambda |}\ .
\end{equation}
Here, $|\Lambda |=\Lambda_{1}+\dotsb +\Lambda_{n}$, and the integers
$a_{i}$ describe the decomposition of $\Lambda$ in terms of the
fundamental weights $\omega_{i}$, $\Lambda =\sum_{j=1}^{|\Lambda
|}\omega_{a_{j}}$, with $1\leq a_{1}\leq \dotsb \leq a_{|\Lambda
|}\leq n$. In~\eqref{Giambelli} we have set $y_{j}=1$ for $j\leq 0$ or
$j\geq n+1$.

Let us denote the chiral primary fields corresponding to the
fundamental representations of $su (n+1)$ also by $y_{i}$.  The chiral
primary field corresponding to a representation $\Lambda$ can then be
written as a polynomial $\tilde{U}_{\Lambda} (y_{i})$ in the chiral
primary fields $y_{i}$. The polynomial $\tilde{U}_{\Lambda}$ is in
general different from $U_{\Lambda}$, because when we describe the
chiral ring, we have to truncate the fusion to chiral primary fields.
The chiral primary labelled by $\Lambda$ has $U (1)$-charge
$q_{\Lambda}=\frac{\sum_{i}i\Lambda_{i}}{k+n+1}$, hence in the
polynomial $U_{\Lambda} (y_{i})$, only the term that under the
transformation $y_{i}\mapsto y_{i}\lambda^{i}$ scales with
$\lambda^{\sum_{j}j\Lambda_{j}}$ corresponds to a chiral primary
field. In other words, to obtain $\tilde{U}_{\Lambda}$ we truncate
$U_{\Lambda}$ to the term with the highest $U (1)$ charge, 
\begin{equation}\label{truncpoly}
\tilde{U}_{\Lambda} (y_{i}) = \lim_{\lambda \to \infty}
\lambda^{-\sum_{j}j\Lambda_{j}} U_{\Lambda} (\lambda^{i}y_{i}) \ .
\end{equation}  
Until now, the level $k$ did not enter. The polynomial expressions do
not change when we consider fusion in the affine theory instead of
tensor products. Of course there is a truncation in that we have to
set to zero some of the polynomials, namely those that lie in the
fusion ideal (the ideal that one has to divide out from the
representation ring to obtain the fusion ring). For $su (n+1)_{k}$, a
basis for this fusion ideal is given by $\{(k+i,0,\dotsc ,0)|
i=1,\dotsc,n\}$~\cite{Bouwknegt:2002bq}.  Dividing out the
corresponding polynomials $\tilde{U}$ results in the chiral ring.

Let us see how this works in detail. From the $su (n+1)$ tensor product
rules, we see that the polynomials
$U_{(\Lambda_{1},0,\dotsc ,0)} (y)$ satisfy the recursion relation
\begin{equation}
U_{(\Lambda_{1},0,\dotsc ,0)} (y) = \sum_{j=1}^{n+1}
(-1)^{j-1}y_{j}U_{(\Lambda_{1}-j,0,\dotsc ,0)} (y) \ ,
\end{equation}
where $y_{n+1}\equiv 1$, $\Lambda_{1}\geq 0$, $U_{(0,0)}=1$, and polynomials
$U_{\Lambda}$ with negative Dynkin indices are set to zero.
For the generating function
\begin{equation}
F_{n;1} (y_{1},\dotsc ,y_{n};t) = \sum_{\Lambda_{1}=0}^{\infty}
U_{(\Lambda_{1},0,\dotsc ,0)} (y)t^{\Lambda_{1}}
\end{equation}
this implies the relation
\begin{align}
F_{n;1} (y,t) & = 1 + \sum_{\Lambda_{1}>0}^{\infty} U_{(\Lambda_{1},0,\dotsc
,0)} (y) t^{\Lambda_{1}} \nonumber\\
& = 1 + (y_{1}t -y_{2}t^{2}+\dotsb + (-1)^{n}t^{n+1}) F_{n;1} (y,t)\ .
\end{align}
We conclude that the generating function is given by
\begin{equation}
F_{n;1} (y_{1},\dotsc ,y_{n};t) =
\Big(1-ty_{1}+t^{2}y_{2}- \dotsb + (-t)^{n}y_{n}+ (-t)^{n+1}\Big)^{-1} \ .
\end{equation} 

\noindent The polynomials $\tilde{U}$ are obtained from the limiting procedure
in~\eqref{truncpoly}, so their generating function is
\begin{align}
\tilde{F}_{n;1} (y_{1},\dotsc , y_{n-1};t) & = \sum_{\Lambda_{1}=0}^{\infty}
\tilde{U}_{(\Lambda_{1},0,\dotsc , 0)} (y)t^{\Lambda_{1}}\nonumber\\
& = \lim_{\lambda \to \infty} F_{n;1} (\lambda y_{1},\dotsc ,
\lambda^{n-1}y_{n-1};\lambda^{-1}t) \nonumber\\
& = \Big( 1- t y_{1}+t^{2}y_{2} +\dotsb + (-t)^{n}y_{n}\Big)^{-1} \ .
\label{genfuncforUtilde}
\end{align}
For fixed $k$ and $n$, the polynomials $\tilde{U}_{(k+i,0,\dotsc , 0)}$
for $i=1,\dotsc , n$ generate the ideal that has to be divided out from
the polynomial ring $\mathbb{C}[y_{1},\dotsc ,y_{n}]$ to obtain the
chiral ring. The polynomials $\tilde{U}$ can be obtained from a 
potential $W_{k,n}$ as 
\begin{equation}\label{Ufrompotential}
\tilde{U}_{(k+i,0,\dotsc ,0)} (y_{1},\dotsc , y_{n}) = (-1)^{n-i}
\frac{\partial}{\partial y_{n+1-i}} W_{k,n} (y_{1},\dotsc , y_{n})\ ,
\end{equation}
where the generating function for the potentials $W_{k,n}$ is given by 
\begin{align}
w_{n} (y_{1},\dotsc , y_{n};t) & = \sum_{k=-n}^{\infty} W_{k,n}
(y_{1},\dotsc , y_{n}) t^{k+n+1} \nonumber\\
& = - \log \Big( 1-ty_{1}+ \dotsb + (-t)^{n}y_{n}\Big)\ .
\label{genfuncforW}
\end{align}
The relation~\eqref{Ufrompotential} can be easily verified by
differentiating~\eqref{genfuncforW} with respect to $y_{i}$ and
comparing the result to~\eqref{genfuncforUtilde}.
In this way one arrives at an expression for the superpotential
$W_{k,n}$ of the Landau-Ginzburg model that corresponds to the $SU
(n+1)/U (n)$ Kazama-Suzuki model~\cite{Gepner:1991gr}.

There is a coordinate change that makes the expression for the
superpotential simpler. If we write the $y_{i}$ as the elementary
symmetric polynomials in some auxiliary variables $x_{j}$,
$y_{i}=\sum_{j_{1}<\dotsb <j_{i}}x_{j_{1}}\dotsb x_{j_{i}}$, the
generating function becomes
\begin{align}
w_{n} (x_{1},\dotsc , x_{n};t) 
& = -\log \prod_{i=1}^{n} (1-tx_{i})\nonumber\\
& = \sum_{k=-n}^{\infty} \frac{1}{k+n+1}\big(x_{1}^{k+n+1}+\dotsb
+x_{n}^{k+n+1} \big)t^{k+n+1} \ .
\end{align}
Note however that the transformation to the variables $x_{i}$ is
non-linear, so considering the Landau-Ginzburg model with chiral
superfields corresponding to the $x_{i}$ will lead to a different
theory.\footnote{In fact, this would result in the tensor product 
of $n$ minimal
models.}

By expanding the generating function one can obtain explicit
expressions for the superpotential in terms of the variables
$y_{i}$. For the case of $SU (3)/U (2)$ ($n=2$) the result is
\begin{equation}\label{superpotsu3}
W_{k,2} (y_{1},y_{2}) = \sum_{i=0}^{\lfloor \frac{k+3}{2}
\rfloor}y_{1}^{k+3-2i}y_{2}^{i}
(-1)^{i}\frac{1}{k+3-i}\binom{k+3-i}{i} \ .
\end{equation}

\noindent We have now obtained an expression for the superpotential. For the
precise dictionary between chiral primary fields in the CFT, which are
labelled by weights $\Lambda = (\Lambda_{1},\dotsc ,\Lambda_{n})$, and
the corresponding expressions in the Landau-Ginzburg models, we still
need to determine the polynomials $\tilde{U}$. There are different
ways to proceed -- we shall use the technique of generating functions to
get the result for the case of $SU (3)/U(2)$. The generalised
Chebyshev polynomials $U_{\Lambda} (y_{1},y_{2})$ have the generating
function~\cite[eq.(13.241)]{FrancescoCFT}
\begin{align}
F_{2} (y_{1},y_{2};t_{1},t_{2}) & =
\sum_{\Lambda_{1},\Lambda_{2}=0}^{\infty}
U_{(\Lambda_{1},\Lambda_{2})} (y_{1},y_{2}) t_{1}^{\Lambda_{1}}
t_{2}^{\Lambda_{2}} \nonumber \\
& = \frac{1-t_{1}t_{2}}{(1-t_{1}y_{1}+t_{1}^{2}y_{2}-t_{1}^{3})
(1-t_{2}y_{2}+t_{2}^{2}y_{1}-t_{2}^{3})} \ . 
\end{align}
The truncated polynomials $\tilde{U}_{\Lambda} (y_{1},y_{2})$
(see~\eqref{truncpoly}) that describe the elements of the chiral ring
then have the generating function
\begin{align}
\tilde{F}_{2}
(y_{1},y_{2}; t_{1},t_{2}) & =
\sum_{\Lambda_{1},\Lambda_{2}=0}^{\infty}
\tilde{U}_{(\Lambda_{1},\Lambda_{2})} (y_{1},y_{2}) t_{1}^{\Lambda_{1}}
t_{2}^{\Lambda_{2}} \nonumber \\
& = \lim_{\lambda \to \infty} F (\lambda
y_{1},\lambda^{2}y_{2};\lambda^{-1}t_{1},\lambda^{-2}t_{2})\nonumber\\
& = \frac{1}{(1-t_{1}y_{1}+t_{1}^{2}y_{2}) (1-t_{2}y_{2})}\ .
\end{align}
This is similar to the generating function $F_{1}$ for the usual
Chebyshev polynomials of the second kind\footnote{Our convention for
these polynomials is taken from~\cite{FrancescoCFT}; it is related to
the more common convention (used e.g.\ in~\cite{Gradshteyn:book}) by
$U_{\text{here}} (x)=U_{\text{standard}} (x/2)$.} which occur in the
$su(2)$ fusion rules,
\begin{equation}
F_{1} (x;t) = \sum_{n=0}^{\infty} U_{n} (x) t^{n} =
\frac{1}{1-xt+t^{2}}\ .
\end{equation}
Indeed, $\tilde{F}_{2}$ can be rewritten as
\begin{align}
\tilde{F}_{2} (y_{1},y_{2};t_{1},t_{2}) & = F_{1}
\big( \tfrac{y_{1}}{\sqrt{y_{2}}};t_{1}\sqrt{y_{2}}\big) 
\frac{1}{1-t_{2}y_{2}}\\
& = \sum_{\Lambda_{1},\Lambda_{2}} U_{\Lambda_{1}}
\big( \tfrac{y_{1}}{\sqrt{y_{2}}}\big)
y_{2}^{\frac{\Lambda_{1}}{2}+\Lambda_{2}}
t_{1}^{\Lambda_{1}}t_{2}^{\Lambda_{2}} \ ,
\end{align}
which provides us with an expression for
$\tilde{U}_{(\Lambda_{1},\Lambda_{2})}$,
\begin{equation}
\tilde{U}_{(\Lambda_{1},\Lambda_{2})} (y_{1},y_{2}) =
(\sqrt{y_{2}})^{\Lambda_{1}+2\Lambda_{2}} U_{\Lambda_{1}}
\big( \tfrac{y_{1}}{\sqrt{y_{2}}}\big) \ .
\end{equation}
By using a standard expression for the Chebyshev polynomials of the
second kind, we get
\begin{equation}
\tilde{U}_{(\Lambda_{1},\Lambda_{2})} (y_{1},y_{2}) = 
\sum_{r=0}^{\lfloor \Lambda_{1}/2\rfloor}
(-1)^{r}\binom{\Lambda_{1}-r}{r}
y_{1}^{\Lambda_{1}-2r}y_{2}^{\Lambda_{2}+r}\ .
\end{equation}

\subsection{Matrix factorisations and boundary conditions}

We now want to introduce a boundary in our Landau-Ginzburg model, and
discuss supersymmetric boundary conditions that preserve a B-type
combination of left- and right-moving supersymmetries. To preserve
this supersymmetry, one has to introduce boundary fermions together
with a boundary potential. This construction is always possible if one
finds a factorisation of the superpotential $W (x_{i})$ in terms of
matrices~\cite{Kontsevich:unpublished,Kapustin:2002bi,Orlov:2003yp,Brunner:2003dc,Kapustin:2003ga},
\begin{equation}\label{factorisation}
\EE (x_{i}) \JJ (x_{i}) = \JJ (x_{i}) \EE ( x_{i}) = W (x_{i}) \mathbf{1} \ .
\end{equation}
The matrices $\EE,\JJ$ can be combined into one matrix
\begin{equation}\label{Q-matrix}
Q (x_{i}) =\begin{pmatrix}
0 & \JJ  (x_{i}) \\
\EE (x_{i}) & 0
\end{pmatrix} \ ,
\end{equation}
such that the condition~\eqref{factorisation} above turns into $Q^{2}
(x_{i})=W (x_{i}) \mathbf{1}$. We also introduce an involution $\sigma$ as
\begin{equation}
\sigma =\begin{pmatrix}
\mathbf{1} & 0 \\
0 & -\mathbf{1}
\end{pmatrix}  \ ,
\end{equation} 
which anti-commutes with $Q$, $\sigma Q + Q\sigma =0$.
We saw that in the infrared, the bulk superpotential $W (x_{i})$ turns
into a quasi-homogeneous function, and there is a similar property for
matrix factorisations that correspond to superconformal boundary
conditions (see e.g.\ \cite{Walcher:2004tx}), namely
\begin{equation}\label{quasihomogeneous}
Q (e^{i\lambda q_{i}}x_{i}) = e^{i\lambda} \rho (x_{i},\lambda)^{-1} Q (x_{i}) \rho 
(x_{i},\lambda ) \quad ,\quad \lambda \in \mathbb{C}\ .
\end{equation}
For this to be consistent for iterated transformations, the invertible
matrices $\rho$ have to satisfy a certain composition rule; in the
case of $x$-independent $\rho$'s, this is just the representation
property,
\begin{equation}
\rho  (\lambda + \lambda ') = \rho (\lambda) \rho (\lambda ')\ .
\end{equation}
It can sometimes be useful to consider just the
infinitesimal version of the scaling behaviour. Differentiation
of~\eqref{quasihomogeneous} with respect to $\lambda$ at $\lambda =0$
yields
\begin{equation}\label{eq:Rdef}
EQ +\left[R,Q\right] = Q\ ,
\end{equation}
where
\begin{equation}
E \equiv \sum_{i=1}^n q_iy_i\frac{\partial}{\partial y_i}\quad
(\text{Euler vectorfield})\quad \text{and} \quad R \equiv
-i(\partial_{\lambda}\rho)\rho^{-1}\Big|_{\lambda =0} \ .
\end{equation}

\noindent The spectrum of chiral primary open string states can be obtained by
solving a cohomology problem. The matrix $Q$ acts linearly on the
space $N_{Q}=\mathbb{C}^{n}[x_{i}]$ of vectors with polynomial
entries, where $n$ is the size of the square matrix $Q$. Open strings
between branes given by factorisations $Q,Q'$ correspond to
homomorphisms from $N_{Q}$ to $N_{Q'}$. The space of chiral primary open string
states corresponds to the cohomology of the operator $D_{QQ'}$ defined
on $\text{Hom} (N_{Q},N_{Q'})$ by
\begin{equation}
D_{QQ'}\Phi = Q' \Phi -\sigma_{Q'}\Phi \sigma_{Q} Q \ .
\end{equation} 
Obviously, there is a $\mathbb{Z}_{2}$ action on the spectrum by
\begin{equation}
\Phi \mapsto \sigma_{Q'} \Phi \sigma_{Q} \ ,
\end{equation}
and we can split the spectrum into the part with eigenvalue $+1$ under
this operation, the bosonic spectrum, and the part with eigenvalue
$-1$, the fermionic spectrum. 

In the case of quasi-homogeneous factorisations, one also has a
$\mathbb{C}^{*}$ action on the spectrum, and we can decompose the
spectrum into eigenvectors with respect to this action,
\begin{equation}
\rho_{Q'} (\lambda) \Phi (e^{i\lambda q_{i}}x_{i}) \rho_{Q}^{-1}
(\lambda) =
e^{i\lambda q_{\Phi}} \Phi (x_{i}) \ .
\end{equation}
We call $q_{\Phi}$ the $U (1)_{R}$-charge of $\Phi$. It corresponds to
the eigenvalue of the $u (1)$-generator in the $\mathcal{N}=2$
superconformal algebra at the infrared fixed point.
In the infinitesimal version, the action on the spectrum reads
\begin{equation} \label{morphismcharge}
E\Phi  +R'\Phi-\Phi R = q_{\Phi } \Phi\ .  
\end{equation}
\smallskip

\noindent Not all different matrix factorisations correspond to
different boundary conditions. In particular, two matrix
factorisations $(Q,\sigma_{Q},\rho_{Q})$ and
$(Q',\sigma_{Q'},\rho_{Q'})$ of size $r$ that are related by a
similarity transformation
\begin{equation}
\mathcal{U}Q\mathcal{U}^{-1}=Q'  \quad \text{and}\quad
\mathcal{U}\sigma_{Q}\mathcal{U}^{-1}=\sigma_{Q'} \quad \text{and}
\quad \mathcal{U}\rho_{Q}\mathcal{U}^{-1} = \rho_{Q'}  \ ,  
\end{equation} 
with an invertible matrix $\mathcal{U} \in GL(2r,\mathbb{C}[x_{i}])$,
have the same spectra with all other branes, and are called equivalent.

\noindent Matrix factorisations can also be added (corresponding to 
superpositions of branes),
\begin{equation}
Q\oplus Q' \equiv \begin{pmatrix}
Q & 0\\
0 & Q'
\end{pmatrix} \ .
\end{equation}
We identify matrix factorisations that differ only by
direct sums of trivial matrix factorisations,
$\big(\begin{smallmatrix}0&1\\ W & 0\end{smallmatrix} \big)$ or 
$\big(\begin{smallmatrix}0&W\\ 1 & 0\end{smallmatrix} \big)$, which
have trivial spectra with all other factorisations.

There is an operation on the matrix factorisations that physically
corresponds to the map that exchanges branes and anti-branes, namely 
we can swap $\JJ$ and $\EE$, 
\begin{equation}
Q=\pmat{0 & \JJ\\ \EE & 0} \mapsto \overline{Q}= \pmat{0 & \EE\\ \JJ &
0} \ .
\end{equation}
We call $\overline{Q}$ the anti-factorisation to $Q$.
\smallskip

The spectrum of chiral primary fields can be directly compared to the
CFT description. In addition one can compare the coupling to bulk
fields (the RR charges), and the operator multiplication (for open
strings from one brane to itself, this defines a ring structure).
After the analysis of factorisations in the case of the $SU (3)/U
(2)$-model in the following section, we shall discuss their RR charges
in section~\ref{sec:LG_RR}. The multiplicative structures will not be
considered in this paper.

\subsection{Factorisations in the \texorpdfstring{$SU(3)/U(2)$}{SU(3)/SU(2)} model}
\label{sec:LG_factorisations}

We can now discuss factorisations in the Landau-Ginzburg description
of the $SU (3)/U (2)$ Kazama-Suzuki model. The superpotential is
\begin{equation}
W_{k} (y_{1},y_{2}) = \sum_{i=0}^{\lfloor \frac{k+3}{2}
\rfloor}y_{1}^{k+3-2i}y_{2}^{i}
(-1)^{i}\frac{k+3}{k+3-i}\binom{k+3-i}{i} = x_{1}^{k+3}
+ x_{2}^{k+3}  \ ,
\end{equation}
where $y_{1}=x_{1}+x_{2}$ and $y_{2}=x_{1}x_{2}$. We have
rescaled the superpotential to $W_{k}= (k+3)W_{k,2}$ ($W_{k,2}$
was given in~\eqref{superpotsu3}) to avoid disturbing prefactors in
the factorisations that we are about to discuss. 

\begin{figure}[htp]
\begin{center}
\includegraphics[width=0.9\textwidth]{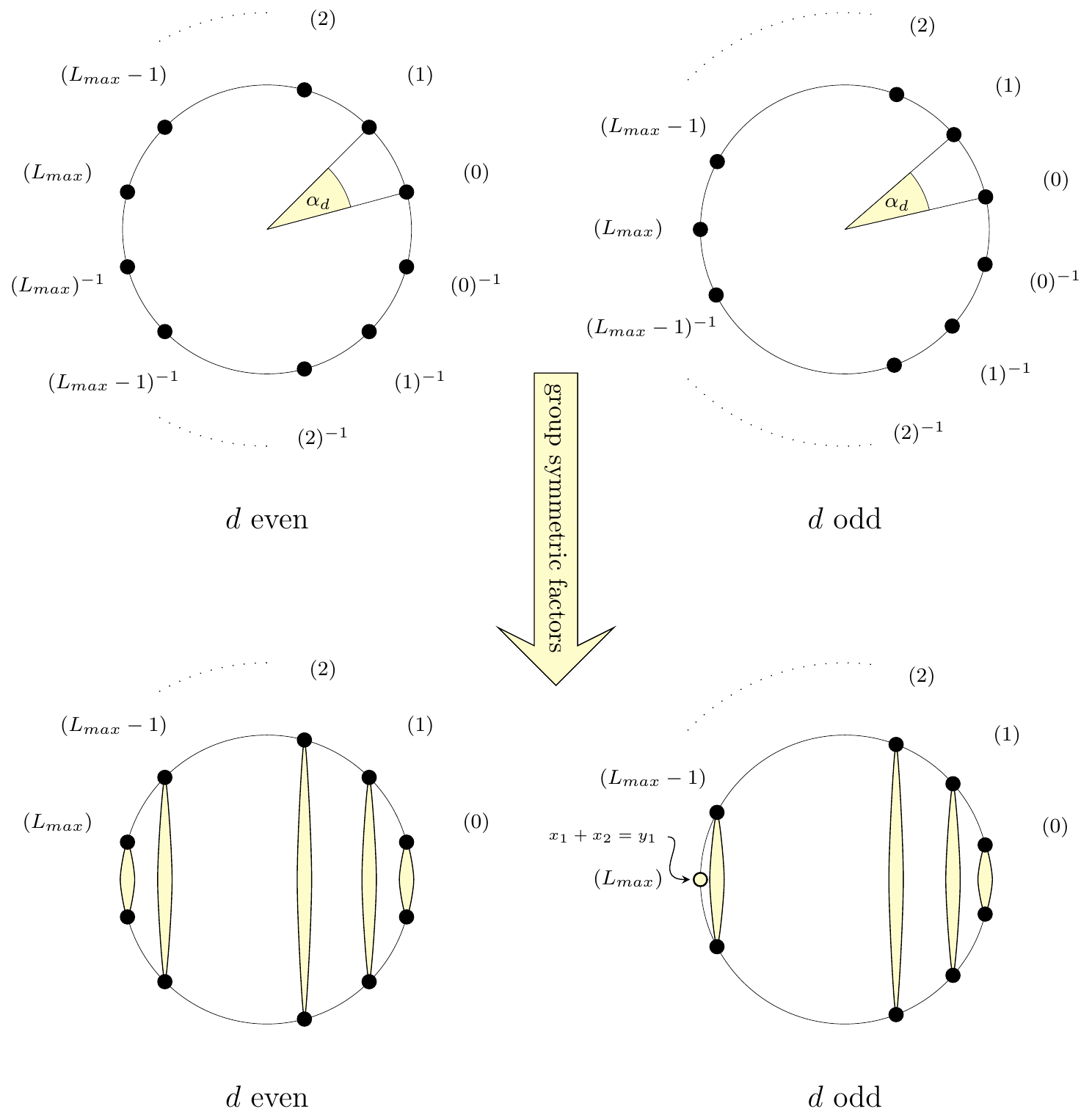}
\caption{\label{diag:factIllustr}Illustration of the polynomial
factorisations for the potential $W_{k}=x_1^d+x_2^d$ (upper
row) and for the same potential expressed in symmetric coordinates
$y_1=x_1+x_2$ and $y_2=x_1 x_2$ (lower row) with $\alpha_d =
\tfrac{2\pi}{d}$ and $L_{max}=\lfloor \tfrac{d-1}{2}\rfloor$. Each
node in the upper row corresponds to a polynomial factorisation
$(L)\hat{=} \left(x_1-e^{i\alpha_L}x_2\right)$, where
$\alpha_L=L\alpha_d+\alpha_d/2$. In the lower diagram, pairs of nodes
$(L)$ and $(L)^{-1}$ (corresponding to
$\left(x_1-e^{-i\alpha_L}x_2\right)$) are grouped together (indicated
by the shape connecting them), and we express the resulting matrix
factorisations in $y$-variables as
$\left(x_1-e^{i\alpha_L}x_2\right)\left(x_1-e^{-i\alpha_L}x_2\right)
=y_1^2-\beta_L y_2$.}
\end{center}
\end{figure}

\noindent In the variables $x_{1},x_{2}$ the superpotential is very simple, and
it can be factorised as
\begin{equation}
W_{k}= \prod_{\eta^{d}=-1} (x_{1} -\eta x_{2}) \ ,
\end{equation}
where we have set $d=k+3$. This is the factorisation that appears in
the description of permutation branes in the product of two minimal
models~\cite{Ashok:2004zb,Ashok:2004xq,Brunner:2005fv}. Let us label the
$d^{\text{th}}$ roots of $-1$ by $\eta_{j}=e^{\pi i\frac{2j+1}{d}}$,
$j=0,\dotsc ,d-1$. A factorisation in the $y$-variables
is easily obtained by noting that
\begin{equation}
(x_{1}-\eta x_{2}) (x_{1}-\eta^{-1}x_{2}) = y_{1}^{2} -
(2+\eta+\eta^{-1})y_{2} \ .
\end{equation}
This leads to a polynomial factorisation of $W_{k} (y_{i})$ in
$\lfloor \frac{d+1}{2}\rfloor$ factors (for odd $d$,
$y_{1}=x_{1}+x_{2}$ appears in the factorisation),  
\begin{equation}
W_{k} (y_{1},y_{2}) = \prod_{j=0}^{\lfloor
\frac{d-2}{2}\rfloor} (y_{1}^{2} - \beta_{j} y_{2})
\cdot \left\{\begin{array}{ll} 
y_{1} & \text{for $d$ odd}\\
1 & \text{for $d$ even.}
\end{array} \right. \ ,
\end{equation}
where
\begin{equation}
\beta_{j} = 2 +\eta_{j}+\eta_{j}^{-1} = 2 \big( 1+\cos \big( \pi
\tfrac{2j+1}{d}\big) \big)  \ .
\end{equation}
We have illustrated this arrangement of factors in
figure~\ref{diag:factIllustr}.

\begin{figure}[htbp]
\begin{center}
\includegraphics[width=0.9\textwidth]{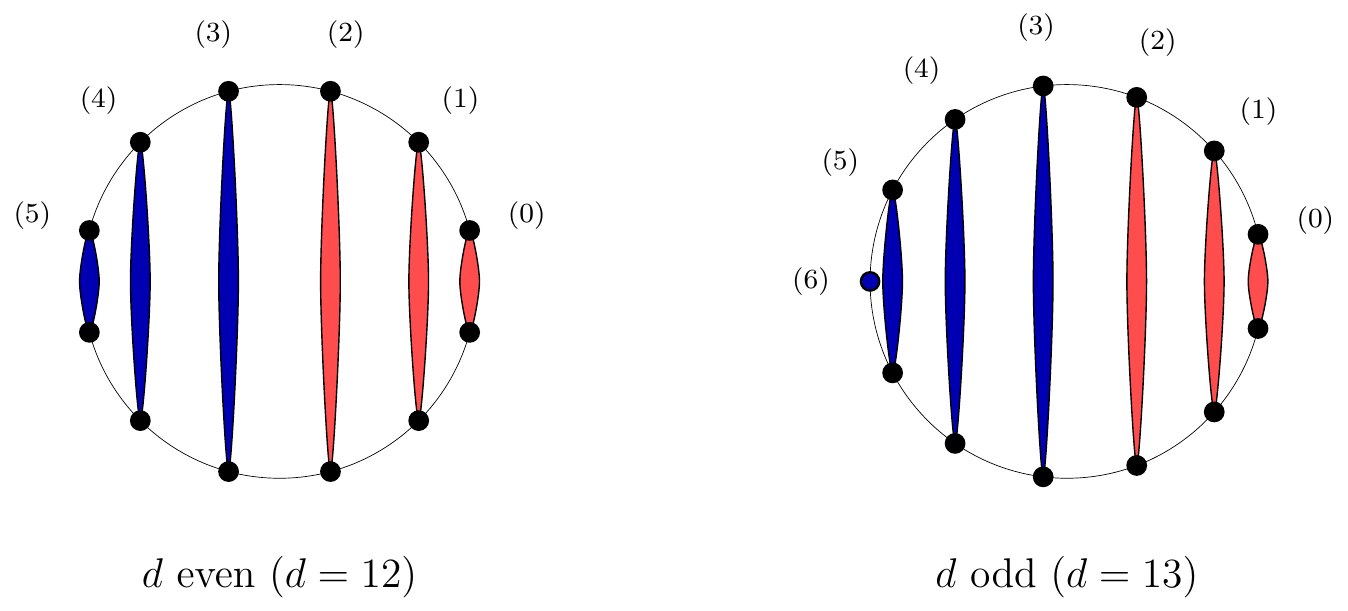}
\caption{\label{diag:LzeroIllustr}Illustration for the form of the
polynomial factorisation $Q_{[0,1,2]}$ corresponding to the CFT
boundary condition $|2,0\rangle$. The $\mathcal{J}$ part (containing
the roots in $[0,1,2]$) is colored in red (light grey in
black-and-white printouts), the $\mathcal{E}$ part (containing the other
roots) in blue (dark grey).}
\end{center}
\end{figure}

We can now easily write down matrix factorisations of the
superpotential by grouping the product formula above into two
polynomial factors $\JJ ,\EE$. It is very convenient to keep the
description in terms of the $x$-variables (indeed there is a faithful
functor of the category of matrix factorisations of $W_{k} (y_{i})$
into the category of matrix factorisations of $\tilde{W}_{k}
(x_{i})=W_{k}(x_{1}+x_{2},x_{1}x_{2})$ -- this will be discussed in
appendix~\ref{sec:functor}). Then, factorisations of $W_{k}(y_{i})$ can
be described as
\begin{equation}
\JJ_{\cI}=\prod_{\eta \in \cI} (x_{1}-\eta x_{2}) 
\quad ,\quad  
\EE_{\cI} =\prod_{\eta \in \cI^{c}} (x_{1}-\eta x_{2}) \ , 
\end{equation}
where $\mathcal{D}$ is the set of all $d^{\text{th}}$ roots of $-1$,
and $\cI\subset \mathcal{D}$ is a subset of roots that is invariant
under the map $\eta \mapsto \eta^{-1}$. The complement of $\cI$ in
$\mathcal{D}$ is denoted by $\cI^{c}=\mathcal{D}\setminus \cI$ (cf.\
figure~\ref{diag:LzeroIllustr}). These factorisations are
quasi-homogeneous in the sense of~\eqref{quasihomogeneous}. The
corresponding matrices $R_{\cI}$ are given by
\begin{equation}
R_{\cI}=\pmat{(1-q_{\cI})/2 &0\\
0 & (q_{\cI}-1)/2}\ ,
\end{equation}
where $q_{\cI}=|\cI|\frac{2}{d}$ (see~\eqref{app:UoneRrk1}).

The open string spectrum can be obtained from the open string spectra
of permutation factorisations in the product of two minimal
models~\cite{Brunner:2005fv} by a suitable projection onto open string
states that are symmetric under the exchange of $x_{1}$ and $x_{2}$
(see the discussion in appendix~\ref{sec:functor}). Essentially, by
the projection we get just half the spectrum of the corresponding
permutation factorisations, namely the number of bosonic and fermionic
fields in the spectrum between two factorisations given by $\cI$ and
$\cI'$ is
\begin{align}
\text{number of bosons} & = \frac{1}{2} |\cI \cap
\cI'|\cdot | \cI^{c}\cap \cI '^{c}|  \\
\text{number of fermions} & = \frac{1}{2} | \cI^{c}\cap \cI'|
\cdot |\cI \cap \cI '^{c} |\ .
\label{numberoffermions}
\end{align}
The detailed computations are done in
appendix~\ref{sec:PolySpectra}. Let us state here only the form of the
fermions~(see~\eqref{app:fossRk1}),
\begin{equation}
\label{fermion}
\psi_{p} = p \pmat{0& \JJ_{\cI\cap \cI'}\\ 
-\JJ_{\cI^{c}\cap \cI'^{c}} & 0} \quad 
\text{with}\  p \in \polyRing{\JJ_{\cI \cap
\cI'^{c}}}{\JJ_{\cI' \cap \cI^{c}}}\ .
\end{equation}
The $U(1)$ charge of a fermion $\psi_{p}$ with a quasi-homogeneous
polynomial $p$ is given by 
\begin{equation}
q_{\psi_{p}} = \frac{1}{d}\big(2\deg (p) + |\cI \cap
\cI'|+|\cI^{c}\cap \cI '^{c}| \big)\ . 
\end{equation}
The spectrum containing the information on $U(1)$ charges is described
by the bosonic and fermionic boundary partition functions
(see~\eqref{app:bosongenfunc} and~\eqref{app:fermiongenfunc})
\begin{align}
\label{bosongenfunc}
B_{\cI\cI'} (z) & = \frac{1-z^{2 |\cI\cap \cI'|}}{1-z^{2}}
\frac{1-z^{2|\cI^{c}\cap \cI'^{c}|}}{1-z^{4}} 
z^{|\cI^{c}\cap
\cI'|+|\cI\cap \cI'^{c}|}\\
F_{\cI\cI'} (z) & = \frac{1-z^{2 |\cI\cap
\cI'^{c}|}}{1-z^{2}} \frac{1-z^{2|\cI^{c}\cap
\cI'|}}{1-z^{4}} z^{|\cI\cap
\cI'|+|\cI^{c}\cap \cI'^{c}|}\ .
\end{align}
These are generating polynomials for the data of the spectrum -- 
the coefficient of a term $z^{n}$ gives the number of morphisms
of charge $n/d$.
\smallskip

There are $2^{\lfloor\frac{d+1}{2} \rfloor}-2$ ways of combining the
$\lfloor \frac{d+1}{2}\rfloor$ factors into two factors $\JJ$ and
$\EE$ (the $-2$ is because we ignore the trivial factorisations where
$\JJ$ or $\EE$ are constant). The common feature of these
factorisations is that they do not have any fermions in their
self-spectrum. As we shall see shortly, these factorisations can only
correspond to a subset of the boundary states that we found before. It
will therefore be necessary to find other factorisations with higher
rank matrices $\JJ,\EE$. Some of those will be constructed in
section~\ref{sec:morefact} by the technique of tachyon condensation.

\subsection{RR charges}
\label{sec:LG_RR}

To determine RR charges we have to compute one-point functions of bulk
fields in the presence of a boundary. By spectral flow, the fields
corresponding to RR ground states can be labelled by elements of the
chiral ring. For such an element $\phi$ we calculate the charge by the
Kapustin-Li formula~\cite{Kapustin:2003ga} (see also~\cite{Walcher:2004tx}),
\begin{equation}
\text{ch}_{\phi} (Q) = \frac{1}{\sqrt{2}}
\text{Res}_{W_{k}} \Big( \phi \text{Str}\big(
\partial_{y_{1}}Q\partial_{y_{2}}Q\big)\Big) \ . 
\end{equation}
Note that we have to insert a factor $1/\sqrt{2}$ if we want to
compare the results to the charges of the full boundary states in the
CFT description. (This rescaling of the RR charge also occurs e.g.\
in~\cite{Lerche:2000iv}).
The residue is formally defined as 
\begin{equation}
\text{Res}_{W_{k}} (f) = \frac{1}{(2\pi i)^{2}}\oint \oint
\frac{f}{\partial_{y_{1}}W_{k} \partial_{y_{2}}W_{k}}dy_{1}dy_{2} \ .
\end{equation}
It can be evaluated by noting that (see~\cite{Vafa:1990mu})
\begin{equation}
\text{Res}_{W_{k}} (f \partial_{y_{i}}W_{k}) = 0 \quad \text{for all}\ f\
\text{and all}\ y_{i} \ .
\end{equation}
This fixes the residue up to a normalisation which is given by the
requirement that the Hessian determinant $H$, 
\begin{equation}
H_{k}=\det (\partial_{y_{i}}\partial_{y_{j}}W_{k}) = d^{2}\; 
y_{2}^{k}\Big((U'_{k+1}(z))^{2}-U'_{k+2} (z)U'_{k} (z) \Big) \ ,
\end{equation}
($z=y_{1}/\sqrt{y_{2}}$) has as residue the number of chiral primary fields,
\begin{equation}
\text{Res}_{W_{k}} (H_{k}) = \frac{(k+1) (k+2)}{2} \ .
\end{equation}
It defines a pairing on the chiral primary fields $\tilde{U}_{(\Lambda
_{1},\Lambda_{2})} (y_{1},y_{2})$,
\begin{equation}\label{pairing}
\text{Res}_{W_{k}} \big(\tilde{U}_{(\Lambda_{1},\Lambda_{2})}
\tilde{U}_{(\Lambda_{1}',\Lambda_{2}')} \big) =
d^{2}\; \delta_{\Lambda_{1},\Lambda_{1}'}
\delta_{k-\Lambda_{1}-\Lambda_{2},\Lambda_{2}'}\ .
\end{equation}
Let us now evaluate the RR charge.
For a factorisation with a simple factor 
$\JJ_{j}=y_{1}^{2}-\beta_{j} y_{2}$ we find
\begin{equation}
\text{Str} \partial_{y_{1}}Q_{j}\partial_{y_{2}}Q_{j} =
\frac{d}{z^{2}-\beta_{j}}\Big(\beta_{j} U_{k+2} (z)-2zU_{k+1} (z)
\Big)y_{2}^{k/2} \ .
\end{equation}
To determine the charge we need to expand this polynomial in
combinations of Chebyshev polynomials in $z$, and we claim
\begin{equation}\label{claimChebyshev}
\frac{1}{z^{2}-\beta_{j}} \left(\beta_{j}U_{k+2} (z)-2z U_{k+1} (z)
\right) = 2 \sum_{i=0}^{\lfloor \frac{k}{2} \rfloor} \cos
\big(\tfrac{\pi}{d} (2j+1) (i+1) \big) U_{k-2i} (z) \ .
\end{equation}
To prove this we write $z=2\cos t$, and use an alternative expression
for the Chebyshev polynomials,
\begin{equation}
U_{n} (2\cos t) = \frac{\sin \big((n+1)t \big)}{\sin t} \ .
\end{equation}
This transforms~\eqref{claimChebyshev} into a trigonometric identity, 
\begin{multline}
\beta_{j}\sin \big((k+3)t \big) - 4\cos t \sin \big((k+2) t \big) \\
= 2 (4\cos^{2} t -\beta_{j}) \sum_{i=0}^{\lfloor \frac{k}{2}\rfloor}
\Big(\cos \big(\tfrac{\pi}{d} (2j+1) (i+1) \big) \sin \big((k-2i+1)t
\big) \Big) \ ,
\end{multline}
which can be proved straightforwardly by rewriting the trigonometric
functions in terms of exponentials and evaluating the geometric sum on
the right hand side.

Using~\eqref{claimChebyshev} and the property~\eqref{pairing} of the
residue, we can evaluate the charge corresponding to the normalised fields
$\phi_{i}=d\; U_{k-2i} (z)y_{2}^{k/2}$, and we find
\begin{equation}\label{MFRRcharges}
\text{ch}_{\phi_{i}} (Q_{j}) = 
\sqrt{2}\cos \tfrac{\pi}{d} (2j+1) (i+1) \ .
\end{equation}
This describes the charge for any factorisation $Q_{j}$ with a simple
factor $\mathcal{J}_{j}=y_{1}^{2}-\beta_{j}y_{2}$. As we will see
later in section~\ref{sec:tachyoncond}, all other polynomial
factorisations $Q_{\mathcal{I}}$ can be obtained by taking tachyon
condensates of those with a single factor in $\mathcal{J}$. The
charges add up in this process, so that the charge of
$Q_{\mathcal{I}}$ is given by
\begin{equation}
\text{ch}_{\phi_{i}} (Q_{\mathcal{I}}) = \frac{1}{\sqrt{2}} \sum_{\eta
\in \mathcal{I}} \eta^{i+1}\ ,
\end{equation}
where we made use of the formula $\eta_{j}=e^{i\pi \frac{2j+1}{d}}$
for the $d^{\text{th}}$ roots of unity and understand the sum as being
taken over those roots $\eta$ appearing in the index set $\mathcal{I}$
of the factorisation $Q_{\mathcal{I}}$ formulated in $x_{i}$
variables.  
\smallskip

This ends our discussion of the polynomial factorisations and their
properties. Let us now see how these results are related to the CFT analysis.

\section{Comparison of factorisations and boundary states}

In this section we will finally address the comparison between the boundary
states and the matrix factorisations for the $SU (3)/U (2)$-model. We
shall first identify the boundary states that correspond to polynomial
factorisations -- these already form a basis of the vector space of RR
charges. We shall then discuss tachyon condensation and RG flows, and
show how further boundary states can be identified as matrix
factorisations.

\subsection{Polynomial factorisations}

The simplest factorisations of $W_{k} (y_{1},y_{2})$ are the
polynomial factorisations that were identified in
section~\ref{sec:LG_factorisations}. One of their properties is that
they do not have fermions in their self-spectra. To do the comparison,
we first identify the boundary states that lead to fermion-free
spectra.

The fermions in the self spectrum of a brane with boundary state
$|L,l\rangle$ correspond to chiral primaries in the overlap between
$|L,l\rangle$ and $\overline{|L,l\rangle}=|L,k+1-l\rangle$. A chiral
primary $((l_{1},l_{2}),0;l_{1},l_{1}+2l_{2})$ appears there with
multiplicity $n_{(l_{1},l_{2})L}{}^{L}N^{(k+1)}_{l_{1}l}{}^{k+1-l}$.
The second factor describing the fusion rules of $su(2)$ is obviously
$0$ when $l=0$ or $l=k+1$ because $l_{1}\leq k$, thus the branes with
boundary states $|L,0\rangle$ have fermion-free open string
spectra. It turns out that for odd $k$, there are no further boundary
states with fermion-free self-spectra; for even $k$ there are in
addition the boundary states $|\frac{k}{2},l\rangle$. The detailed
analysis can be found in
appendix~\ref{sec:CFT_fermionfreeSS}.

Let us concentrate on the boundary states $|L,0\rangle$. To
characterise them further, we can compute their bosonic spectra. We
can show (see appendix~\ref{sec:CFT_l_0_spectra}) that they have
$(L+1)(k+1-2L)$ bosons in their self-spectrum. This matches with the
number of bosons for polynomial factorisations with $L+1$ elementary
factors in $\JJ$ or $\EE$. The boundary state $|0,0\rangle$ therefore
seems to correspond to a factorisation with $\JJ\sim y_{1}^{2}-\beta_{j}
y_{2}$ for some $\beta_{j}$. To determine which $\beta_{j}$ is the
correct one, we compare the RR charges. The RR charge of the boundary
state $|0,0\rangle$ is given by (see~\eqref{RRcharge})
\begin{equation}
\text{ch}_{i} (|0,0\rangle) = \frac{1}{\sqrt{2}}\frac{\sin \frac{2\pi}{d}
(i+1)}{\sin \frac{\pi}{d} (i+1)} = \sqrt{2}\cos
\tfrac{\pi}{d} (i+1)\ ,
\end{equation}
and by comparison with the RR
charges~\eqref{MFRRcharges} of the elementary factorisations, we see
that we find agreement for $j=0$. Hence we conclude that $\beta_{0}=2
(1+\cos (\frac{\pi}{d}))$ is the 
correct choice, so that 
\begin{equation}
|0,0\rangle  \leftrightarrow Q_{|0,0\rangle} = \begin{pmatrix}
0 & (y_{1}^{2}-\beta_{0}y_{2})\\
\frac{W_{k}}{y_{1}^{2}-\beta_{0}y_{2}} & 0
\end{pmatrix} \ .
\end{equation}  
The same reasoning applies to the remaining boundary states
$|L,0\rangle$ with $L\not= 0$ that should correspond to factorisations
where $\JJ$ consists of $L+1$ factors. By evaluating the RR charges we
can determine which factors appear, namely we find
\begin{align}
\text{ch}_{i} (|L,0\rangle) & = \frac{1}{\sqrt{2}} \frac{\sin \frac{2\pi
}{d} (L+1) (i+1)}{\sin \frac{\pi}{d} (i+1)}\nonumber \\
& = \sqrt{2} \sum_{j=0}^{L} \cos \tfrac{\pi}{d} (2j+1) (i+1) \nonumber\\
& = \sum_{j=0}^{L} \text{ch}_{i} (Q_{\beta_{j}}) \ .
\end{align}
We conclude that we have the following correspondence,
\begin{equation}
|L,0\rangle  \leftrightarrow Q_{|L,0\rangle} = \begin{pmatrix}
0 & \prod_{j=0}^{L} (y_{1}^{2}-\beta_{j}y_{2})\\
\frac{W_{k}}{\prod_{j=0}^{L} (y_{1}^{2}-\beta_{j}y_{2})} & 0
\end{pmatrix}  \ .
\end{equation}
To simplify notation, we define
\begin{equation}
[n_{1},\dotsc ,n_{r}]:=
\bigcup_{i=1}^{r}\{\eta_{n_{i}},\eta_{n_{i}}^{-1} \}\ , 
\end{equation}
so that $Q_{|L,0\rangle}=Q_{\cI_{|L,0\rangle}}$ with the set of roots
given by
\begin{equation}
\cI_{|L,0\rangle} = [0,\dotsc ,L] \ .
\end{equation}

It remains to check the relative spectra. Consider the factorisations
$Q_{\cI_{|L,0\rangle}}$ and
$Q_{\cI_{|L',0\rangle}}$, and assume $L'\geq L$. Then
$\cI_{|L,0\rangle}\subset \cI_{|L',0\rangle}$, and
from~\eqref{numberoffermions} we see that the spectrum does not
contain any fermions.  The bosonic spectrum is encoded in the
generating polynomial
$B_{\cI_{|L,0\rangle}\cI_{|L',0\rangle}} (z)$ given
in~\eqref{bosongenfunc}. Using
\begin{align}
|\cI_{|L,0\rangle}\cap \cI_{|L',0\rangle}| &= 2L+2 \\
|\cI_{|L,0\rangle}^{c}\cap\cI_{|L',0\rangle}^{c}| & = k+3- (2L'+2)\\
|\cI_{|L,0\rangle}^{c} \cap \cI_{|L',0\rangle}| & = 2 (L'-L)\\
|\cI_{|L,0\rangle} \cap \cI_{|L',0\rangle}^{c}| & = 0\ ,
\end{align}
the generating polynomial takes the form 
\begin{align}
B_{\cI_{|L,0\rangle}\cI_{|L',0\rangle}} (z) &= 
\sum_{\alpha_1=0}^L\sum_{\alpha_2=0}^{k-2L'}
z^{4\alpha_1 + 2\alpha_2 +2(L'-L)}\\
& = \frac{1-z^{2 (2L+2)}}{1-z^{4}}
\frac{1-z^{2 (k+3- (2L'+2))}}{1-z^{2}} 
z^{2 (L'-L)}\ .
\label{eq:LzeroSpecs}
\end{align}
This coincides precisely with the generating polynomial
$B_{|L,0\rangle ,|L',0\rangle} (z)$ in~\eqref{app:CFTbosongenfunc} of
the CFT computation. This analysis thus confirms the consistency of the
correspondence
\begin{equation}
|L,0\rangle  \leftrightarrow Q_{\cI_{|L,0\rangle}} \ .
\end{equation}
Recall that the boundary states $|L,0\rangle$ already form a basis of
the charge lattice that is spanned by the maximally symmetric boundary
states. 

For $k$ odd, these are all boundary states that can be associated to
polynomial factorisations of the superpotential.  For even $k$,
however, we also found the series $|\frac{k}{2},l\rangle$ with
fermion-free self-spectra. The analysis of RR charges leads to the
identification
\begin{equation}
|\tfrac{k}{2},l\rangle \leftrightarrow
Q_{\cI_{|\frac{k}{2},l\rangle}}
\quad  , \quad  
\cI_{|\frac{k}{2},l\rangle} = \{\eta_{-\frac{k}{2}+l+2m}:\
m\in \{0,\dotsc ,k-l+1 \} \} \ .
\end{equation}
This identification is also consistent with the spectra, which can be
verified by comparing~\eqref{bosongenfunc} and~\eqref{app:khalfspectrum}.

We conclude that all boundary states with fermion-free self-spectra
can be matched to polynomial matrix factorisations. There are,
however, other boundary states with fermions in their spectra, and 
also there are polynomial factorisations that do not
correspond to any of the maximally symmetric boundary states.

\subsection{Tachyon condensation}
\label{sec:tachyoncond}

Our aim is to identify matrix factorisations for the remaining
boundary states. We have already seen that the factorisations
$Q_{|L,0\rangle}$ form a basis of the space of RR charges. It is therefore
conceivable that we can generate all other factorisations from these
elementary ones. In this subsection we shall explain the general
mechanism of tachyon condensation for matrix factorisations that
enables us to construct new factorisations. As an example we shall
demonstrate how for even $k$ the factorisations $Q_{|\frac{k}{2},l\rangle}$
can be generated from the generating set $\{Q_{|L,0\rangle} \}$.

Let us first briefly explain how tachyon condensation works in the
matrix factorisation description. Suppose we start with the
superposition of two boundary conditions corresponding to the direct
sum $Q$ of matrix factorisations $Q_{1}$ and $Q_{2}$,
\begin{equation}
Q = \begin{pmatrix}
Q_{1} & 0\\
0& Q_{2}
\end{pmatrix} \qquad \text{with}\ \sigma = \begin{pmatrix}
\sigma_{1} & 0 \\
0 & \sigma_{2}
\end{pmatrix}\ .
\end{equation}
A fermion $\psi=\psi_{1,2}$ in the spectrum between $Q_{1}$ and $Q_{2}$
corresponds to a fermion $\Psi$ in the self-spectrum of $Q$ of the
form
\begin{equation}
\Psi = \begin{pmatrix}
0& 0 \\
\psi & 0  
\end{pmatrix} \ .
\end{equation}
It is now easy to check that $Q_{\psi}:=Q+\Psi$ is again a matrix
factorisation of $W$. We interpret the corresponding boundary
condition as the result of the condensation of the fermionic field
$\Psi$, and denote this tachyon condensate by 
\begin{equation}
\label{generalcondensate}
(Q_{1} \xrightarrow{\psi}  Q_{2}) \equiv  Q_{\psi } \equiv  \begin{pmatrix}
Q_{1} & 0 \\
\psi & Q_{2} 
\end{pmatrix} \ .
\end{equation}
In mathematics this procedure is known as \emph{cone construction},
and the object $Q_{\psi}$ fits into what is called a
\emph{distinguished triangle} (see e.g.\
\cite{Aspinwall:2004jr,govindarajan-2007-765}),
\begin{equation}\label{eq:triang}
Q_{1}[1]\xrightarrow{\psi[1]}Q_{2}\rightarrow Q_{\psi}\rightarrow
Q_{1} \ .
\end{equation}
It is understood that the first and the last term of the above
sequence are identified (therefore the name triangle) up to the action
of the shift functor $[1]$ that maps a factorisation $Q$ to its
anti-factorisation $Q[1]=\bar{Q}$. In particular, any cyclic shift of
objects in~(\ref{eq:triang}) will yield another valid distinguished
triangle. For example, shifting all objects in~(\ref{eq:triang}) one
position to the left will yield a triangle
\begin{equation}\label{shiftedtriangle}
Q_{2}\xrightarrow{\tilde{\psi}[1]}Q_{\psi}
\rightarrow Q_{1}\rightarrow Q_{2}[1] \ ,
\end{equation}
thus we learn that the object $Q_{1}$ can be obtained as a condensate
from $Q_{2}[1]$ and $Q_{\psi}$ with some morphism $\tilde{\psi}$.
This will be useful in section~\ref{sec:morefact}.

Let us exemplify this by studying condensates of two polynomial
factorisations $Q_{\cI}$ and $Q_{\cI'}$ that at least have one fermion
in their relative spectrum. From~\eqref{numberoffermions} we see that
this implies that $\cI\not \subset \cI'$ and $\cI'\not\subset
\cI$. Turning on a fermion $\psi_{p}$ (see~\eqref{fermion}) leads to
the factorisation
\begin{equation}
( Q_{\cI}  \xrightarrow{p} Q_{\cI '}) \equiv 
( Q_{\cI} \xrightarrow{\psi_{p}}Q_{\cI'}) = \begin{pmatrix}
0 & \JJ_{\cI} & 0 & 0 \\
\JJ_{\cI^{c}} & 0 & 0 & 0 \\
0 & p\JJ_{\cI\cap \cI'} & 0 & \JJ_{\cI'}\\
-p\JJ_{\cI^{c}\cap \cI'^{c}} & 0 &
\JJ_{\cI'^{c}} & 0
\end{pmatrix} \ .
\end{equation}
Consider now the fermion of lowest charge ($p=1$). By doing some
elementary transformations $Q\to \mathcal{U}Q\mathcal{U}^{-1}$, one
can verify that this factorisation is equivalent to a direct sum,
\begin{equation}\label{constantcondensation} 
(Q_{\cI} \xrightarrow{1}Q_{\cI'}) \cong
Q_{\cI\cap \cI'} \oplus Q_{\cI\cup
\cI'} \ .
\end{equation}
In case $\cI\cap \cI'=\emptyset $ or $\cI\cup \cI'=\mathcal{D}$, one
of the summands is trivial and the condensate is equivalent to a
single polynomial factorisation.\footnote{What we have described here
is very similar to the condensation processes among polynomial
factorisations in the theory of two minimal models discussed
in~\cite{Brunner:2005fv}. In fact, their arguments are directly
applicable here by applying the functor described in
appendix~\ref{sec:functor}.}
\smallskip

For even $k$, we can use the above tachyon condensations to show how
we can obtain the polynomial factorisations
$Q_{|\frac{k}{2},l\rangle}$ from our generating set $\{Q_{|L,0\rangle}
\}$. Of course $Q_{|\frac{k}{2},0\rangle}$ is already contained in the
set, so the first non-trivial example is
\begin{equation}
Q_{|\frac{k}{2},1\rangle}=Q_{[0,\dotsc ,\frac{k}{2}-1,\frac{k}{2}+1]} \ .
\end{equation}
From the condensation formula~\eqref{constantcondensation} we
see that
\begin{equation}
Q_{|\frac{k}{2},1\rangle} \cong \left(Q_{|\frac{k}{2}-1,0\rangle}
\xrightarrow{1} \overline{Q_{|\frac{k}{2},0\rangle}} \right) \ .
\end{equation}
To simplify notations, we shall denote the factorisation
$Q_{|L,0\rangle}$ by a rectangular box with label $L$, and the
factorisation $Q_{\overline{|L,0\rangle}}$ by a rounded box with label
$L$, so that the above condensation process reads
\begin{equation}
\raisebox{-1ex}{\includegraphics{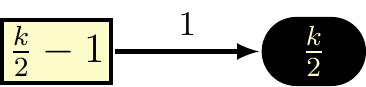}}\ .
\end{equation}
It is easy to see how this generalises: for $l\leq \frac{k}{2}$ one
has
\begin{equation}\label{khalfbranes}
\includegraphics{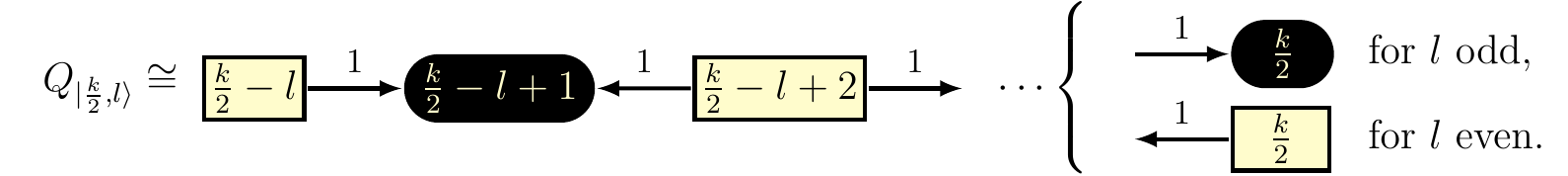}
\end{equation}
The case $l\geq \frac{k}{2}+1$ is also covered by noting that
$Q_{|\frac{k}{2},l\rangle}=\overline{Q_{|\frac{k}{2},k+1-l\rangle}}$.
Note that although multiple arrows appear, the result can still be
written as a condensate in the form~\eqref{generalcondensate} by
grouping the factorisations in rectangular boxes into $Q_{1}$ and the
ones in rounded boxes into $Q_{2}$.\footnote{That is, $Q_1=\oplus_i
Q_{1,i}$ with $Q_{1,i}=Q_{|\frac{k}{2}-l+2i,0\rangle}$ and
$Q_2=\oplus_j Q_{2,j}$ with
$Q_{2,j}=Q_{\overline{|\frac{k}{2}-l+2j+1,0\rangle}}$, such that the
individual fermions in (\ref{khalfbranes}) combine into an element of
$H^1(Q_{1},Q_{2})=\oplus_{i,j}H^1(Q_{1,i},Q_{2,j})$ (where $H^{1}(\cdot,\cdot)$ denotes the space of fermionic morphisms); this justifies
viewing the tachyon condensate as the outcome of a \emph{single}
condensation process.}  
\smallskip

Before we now go on to construct factorisations for other boundary
states, we shall first discuss the analogue of tachyon condensation on
the CFT side.

\subsection{RG flows}\label{sec:RGflows}

On the CFT side we also have some information on tachyon condensation,
which here corresponds to boundary RG flows. There is a general rule
for flows in coset models~\cite{Fredenhagen:2002qn,Fredenhagen:2003xf}
that we can apply in our setup. This rule is based on a conjecture that
certain flows that are visible for large coset levels can be extrapolated
down to arbitrary levels.

The content of the rule in our case is the following. Choose a
representation $\Lambda$ of $su (3)_{k}$, and labels $L,l$ that
parameterise boundary states. Then the rule predicts a
flow\footnote{Note the difference in notation for RG flows (denoted by
$\leadsto$) and fermionic morphisms as part of tachyon condensation
processes (denoted by $\includegraphics{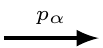}$).}
\begin{equation}\label{flowrule}
\sum_{\lambda ,l'} b^{\Lambda}_{\lambda} N^{(k+1)}_{\lambda l}{}^{l'}
|L,l'\rangle  \leadsto \sum_{L'} n_{\Lambda L}{}^{L'}
|L',l\rangle \ . 
\end{equation}
Here, $b^{\Lambda}_{\lambda}$ denotes the branching coefficient of the
regular embedding $su (2)\subset su (3)$ at embedding index $1$, $N^{(k+1)}_{\lambda l}{}^{l'}$ is the fusion coefficient of $su(2)$ at level $k+1$ and $n_{\Lambda L}{}^{L'}$ the twisted $su(3)$ fusion coefficient at level $k$. 

There is a lot of evidence that this rule correctly describes boundary
RG flows~\cite{Fredenhagen:2003xf} (see also~\cite{Bachas:2009mc}). As
a simple consistency check in our case, we can compare the RR charges
of the initial and the final configuration.  The charge of the left hand
side of equation~\eqref{flowrule} can be evaluated using~\eqref{RRcharges},
\begin{align}
\text{ch} (\text{LHS}) & = \sum_{l',\lambda ,L''} b^{\Lambda}_{\lambda}
N^{(k+1)}_{\lambda l}{}^{l'}
(N^{(k+1)}_{LL''}{}^{l'}-N^{(k+1)}_{LL''}{}^{k+1-l'}) \text{ch}
(|L'',0\rangle) \nonumber\\
 & = \sum_{l',\lambda ,L''} b^{\Lambda}_{\lambda} N^{(k+1)}_{\lambda
L}{}^{l'} (N^{(k+1)}_{l'L''}{}^{l}-N^{(k+1)}_{l'L''}{}^{k+1-l})
\text{ch} (|L'',0\rangle) \nonumber\\
 & = \sum_{L',\lambda ,L''} b^{\Lambda}_{\lambda}
\big(N^{(k+1)}_{\lambda L}{}^{L'}-N^{(k+1)}_{\lambda L}{}^{k+1-L'}
\big)\big(N^{(k+1)}_{L'L''}{}^{l}-N^{(k+1)}_{L'L''}{}^{k+1-l} \big) \text{ch}
(|L'',0\rangle) \label{LHS}\ .
\end{align}
In the last step we split the sum over $l'=0,\dotsc , k+1$ into two
parts; we introduced the new summation variable $L'=0,\dotsc , \lfloor
\frac{k}{2}\rfloor$, and replaced $l'=L'$ in the first part of the
sum, and $l'=k+1-L'$ in the second part. Now let us look at the charge
of the right hand side of equation~\eqref{flowrule},
\begin{equation}\label{}
\text{ch} (\text{RHS}) = \sum_{L',L''} n_{\Lambda L}{}^{L'}
\big(N^{(k+1)}_{L'L''}{}^{l}-N^{(k+1)}_{L'L''}{}^{k+1-l} \big)
\text{ch} (|L'',0\rangle) \ .
\end{equation}
By using formula~\eqref{twistedfusion} for the twisted $su (3)$
fusion coefficients we find precise agreement with the
result~\eqref{LHS} for the left hand side. This shows that the
suggested flows are consistent on the level of RR charges.

Let us work out one class of flows described by the rule above, where
we set $\Lambda = (1,0)$. The branching is $(1,0)\to
(0)\oplus (1)$, and so from~\eqref{flowrule} we find for $k>1$ the flows
\begin{equation}\label{RGFlows}
|L,l-1\rangle + |L,l\rangle + | L,l+1\rangle \leadsto \left\{
\begin{array}{ll}
|L-1,l\rangle +|L,l\rangle + |L+1,l\rangle 
& \text{for}\ L\not= \tfrac{k}{2}\\[1mm]
|L-1,l\rangle & \text{for}\ L=\tfrac{k}{2} \ .
\end{array} \right.
\end{equation}
Here, labels outside of the allowed range are ignored (e.g.\ if $l=0$,
then the boundary state $|L,l-1\rangle$ on the left hand side does not
appear).

The field that triggers these flows is determined as follows: consider
the adjoint representation $(1,1)$ of $SU(3)$ and decompose it into
the irreducible representations $(l,m)$ of $SU(2)\times U(1)$,
\begin{equation}
(1,1) \longrightarrow (0,0)\oplus (1,3)\oplus (1,-3) \oplus (2,0)\ .
\end{equation}
The perturbing field then has coset label $((0,0),0;l,m)$ with $l,m$
from the list above. As explained in~\cite{Fredenhagen:2002qn}, the
adjoint representation $(2,0)$ of $SU (2)\times U (1)$ has to be
removed from the list, and choosing the trivial representation $(0,0)$ would
correspond to take the identity field for the perturbation. That means
that the field $\psi $ responsible for the flows could come from two
sectors
\begin{equation}\label{pertfield}
\psi \in \mathcal{H}_{((0,0),0;1,3)}\oplus
\mathcal{H}_{((0,0),0;1,-3)} \ .
\end{equation}
The field $\psi_{((0,0),0;1,-3)}$ is the superpartner to the chiral
primary field $\psi_{((k,0),0;k,k)}$ of charge $q=\frac{k}{k+3}$.
This is the charge that we expect to see in the corresponding tachyon
condensation processes of matrix factorisations\footnote{The other
field belongs to an anti-chiral field that we do not see in the matrix
factorisation description.}. 

\subsection{Constructing more factorisations}
\label{sec:morefact}

Having identified the elementary factorisations $Q_{|L,0\rangle}$, we
can now use the information on RG flows from the CFT description to
obtain new matrix factorisations. 

Let us start with a simple example. For $L=\frac{k}{2}$ ($k$ even) and
$l=1$, the flow rule~\eqref{RGFlows} reads
\begin{equation}
|\tfrac{k}{2},0\rangle + |\tfrac{k}{2},1\rangle +
|\tfrac{k}{2},2\rangle \leadsto |\tfrac{k}{2}-1,1\rangle \ .
\end{equation}
This gives us the prescription how to build the matrix factorisation
corresponding to the boundary state
$|\frac{k}{2}-1,1\rangle$.The fermion that has to be
switched on is also uniquely fixed in this case, because a fermion
with charge $\frac{k}{k+3}$ is only found once between
$|\frac{k}{2},0\rangle$ and $|\frac{k}{2},1\rangle$, and once between
$|\frac{k}{2},2\rangle$ and $|\frac{k}{2},1\rangle$, and both have to
be turned on, because otherwise we would end up with a superposition
in the condensate. The prescription therefore is
\begin{equation}\label{condensateforkhalfone}
Q_{|\frac{k}{2}-1,1\rangle} \cong \big( Q_{|\frac{k}{2},0\rangle}
\xleftarrow{q_{\psi}=\frac{k}{k+3}} Q_{|\frac{k}{2},1\rangle }
\xrightarrow{q_{\psi}=\frac{k}{k+3}} Q_{|\frac{k}{2},2\rangle } \big) \ .
\end{equation}
A comment is in order about the directions of the arrows. We chose the
arrows such that we can write the process as a single condensation: we
can view it as turning on a single fermion between
$Q_{|\frac{k}{2},1\rangle}$ and the superposition of
$Q_{|\frac{k}{2},0\rangle}$ and $Q_{|\frac{k}{2},2\rangle}$. It is not
difficult to see that reversing both arrows leads to an equivalent
factorisation. If we only reverse one arrow, we have to view it as a
two-step condensation. If we first condense the right arrow (in
whatever direction), we find that there is only one fermion between
$Q_{|\frac{k}{2},0\rangle}$ and the condensate, and the corresponding
condensate is again equivalent to our first choice of arrows. If we first
condense the left arrow (in whatever direction), there are two
fermions left. One of them corresponds to the original fermion
corresponding to the right arrow, and the condensate is again
equivalent to our original choice of arrows. (The other fermion would
also lead to a polynomial factorisation, which could not be correct.)
Thus, although we do not have a general understanding of how to choose
the arrows to reproduce the CFT flows, we see that in our case any
choice will lead to the same result.

Let us analyse the condensate in more detail. Identifying the
morphisms of the proper charge between the polynomial factorisations
on the right hand side of~\eqref{condensateforkhalfone}, we obtain
\begin{equation}
Q_{|\frac{k}{2}-1,1\rangle} \cong \big( Q_{[0,\dotsc ,\frac{k}{2}]}
\xleftarrow{1} Q_{[0,\dotsc ,\frac{k}{2}-1,\frac{k}{2}+1]}
\xrightarrow{y_{1}} Q_{[0,\dotsc ,\frac{k}{2}-2,\frac{k}{2}]} \big) \ .
\end{equation}
The left arrow is a fermion of lowest charge, so we can condense the
factorisations according to~\eqref{constantcondensation} and find
\begin{equation}\label{polyfactforkhalfminusoneone}
Q_{|\frac{k}{2}-1,1\rangle} \cong \big( Q_{[0,\dotsc ,\frac{k}{2}-1]}
\xrightarrow{y_{1}} Q_{[0,\dotsc ,\frac{k}{2}-2,\frac{k}{2}]} \big) \ .
\end{equation}
We can rewrite the result in terms of elementary constituents
$\{Q_{|L,0\rangle} \}$ by writing the polynomial factorisation on the
right of the arrow as a condensate, which leads to
\begin{equation}\label{proposalforkhalfminusoneone}
\raisebox{-10ex}{\includegraphics{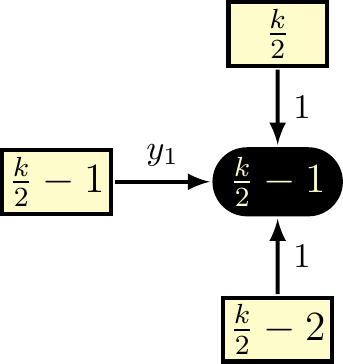}}\ .
\end{equation}
Thus we have obtained a precise proposal for the matrix factorisation
corresponding to $Q_{|\frac{k}{2}-1,1\rangle}$ from the flow rule.
\smallskip

\noindent Let us now evaluate the flow rule~\eqref{RGFlows} for $l=0$. It then
reads for $L<k/2$
\begin{equation}\label{RGflowforl=0}
|L,0\rangle + |L,1\rangle \leadsto |L-1,0\rangle + |L,0\rangle
+ |L+1,0\rangle \ ,
\end{equation}
where again boundary states are left out if the label leaves the
allowed range. This can be translated into a tachyon condensation in
terms of matrix factorisations,
\begin{equation}\label{flowforl=0}
\raisebox{-1ex}{\includegraphics{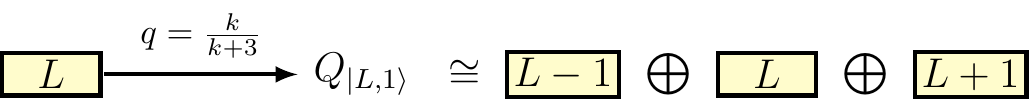}}\ .
\end{equation}
This tachyon condensate fits into the distinguished triangle
(see~\eqref{eq:triang})
\begin{equation}\label{eq:LzTriang}
Q_{|L,0\rangle}[1]\xrightarrow{\psi^{*}[1]}Q_{|L,1\rangle}\xrightarrow{}
\left( Q_{|L-1,0\rangle}\oplus Q_{|L,0\rangle}\oplus Q_{|L+1,0\rangle}
\right) \xrightarrow{} Q_{|L,0\rangle}\ ,
\end{equation}
where we write $\psi^{*}$ to denote the fermionic morphism of charge
$q=\tfrac{k}{k+3}$. By shifting the triangle
(see~\eqref{shiftedtriangle}) we see that 
$Q_{|L,1\rangle}$ can be obtained as a condensate from
$Q_{|L-1,0\rangle}\oplus Q_{|L,0\rangle}\oplus Q_{|L+1,0\rangle}$ and 
$Q_{|L,0\rangle}[1]$, i.e.\ we can invert the tachyon
condensation~\eqref{flowforl=0} to get
$Q_{|L,1\rangle}$ alone on the left hand side,
\begin{equation}
\raisebox{-9ex}{\includegraphics{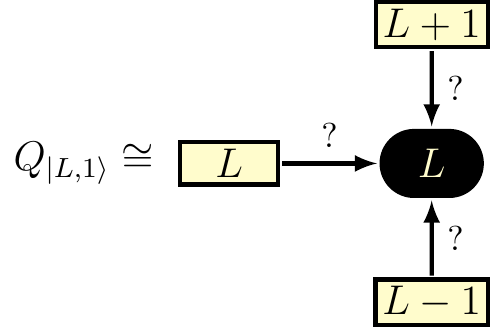}}
\end{equation}
We do not have much control over the morphisms that appear in the
condensate, but we see immediately that for $L=\frac{k}{2}-1$ we find
the same structure as in~\eqref{proposalforkhalfminusoneone}, so the
morphisms are fixed for this value of $L$. Assuming that the
morphisms will be the same for other values of $L$, we arrive at a
proposal for the factorisation corresponding to $Q_{|L,1\rangle}$
($L<\frac{k}{2}$),
\begin{equation}\label{proposalforLone}
\raisebox{-9ex}{\includegraphics{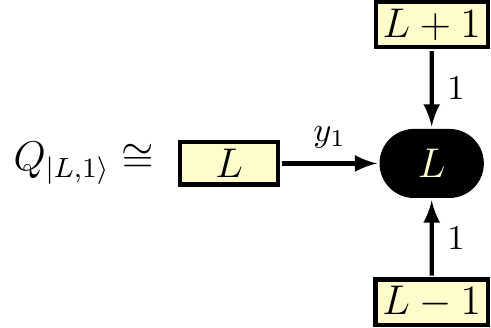}}\ .
\end{equation}
This representation of $Q_{|L,1\rangle}$ has the advantage that it
gives a description directly in terms of the basic constituents
$Q_{|L,0\rangle}$. For computations, however, it is more useful to
condense the right column of the diagram in~\eqref{proposalforLone}
into the polynomial factorisation $Q_{[0,1,\dotsc ,L-1,L+1]}$, so that
we find (similarly to~\eqref{polyfactforkhalfminusoneone})
\begin{equation}
\label{polyfactLone}
Q_{|L,1\rangle} \cong \big( Q_{[0,1,\dotsc ,L]} \xrightarrow{y_{1}}
Q_{[0,1,\dotsc ,L-1,L+1]}\big) \ .
\end{equation}
In appendix~\ref{sec:condensatespectra}, the fermionic spectrum (including the
$U(1)$ charges) of such condensates is investigated, and it agrees
with the spectrum of the $|L,1\rangle$ boundary states. Also the
relative fermionic spectra among the $Q_{|L,1\rangle}$ and between
$Q_{|L,1\rangle}$ and $Q_{|L',0\rangle }$ is determined there and shown to
be consistent with the CFT results.

Another requirement for our maximally symmetric boundary states is
that they are invariant under the exchange of left- and right-movers,
because we are considering B-type boundary states in a diagonal
theory.\footnote{Note that this requirement does not hold for general
B-type boundary states in these models, because the theory is not
diagonal with respect to the $\mathcal{N}=2$ superconformal algebra, but only
with respect to the larger coset algebra. Thus this provides a
non-trivial, necessary (but not sufficient) condition for maximally
symmetric boundary states.}  On the matrix factorisation side this
means to transpose the matrices
(see~\cite{Hori:2006ic,Brunner:2008bi}), or, in the above condensation
pictures, to reverse the arrows. Let us briefly discuss why reversing
the arrow in~\eqref{polyfactLone} leads to an equivalent
factorisation,
\begin{equation}\label{reversedarrows}
Q_{\rightarrow} =  \big( Q_{[0,1,\dotsc ,L]} \xrightarrow{y_{1}}
Q_{[0,1,\dotsc ,L-1,L+1]}\big) \cong \big( Q_{[0,1,\dotsc ,L]} \xleftarrow{y_{1}}
Q_{[0,1,\dotsc ,L-1,L+1]}\big) = Q_{\leftarrow} \ .
\end{equation}
Displaying only the $\JJ$-part of the factorisations, we have
\begin{align}
Q_{\leftarrow}\big|_{\JJ} &=
\pmat{\JJ_{[0,1,\dotsc ,L]} & y_{1}
\JJ_{[0,1,\dotsc ,L-1]}\\
0 & \JJ_{[0,1,\dotsc ,L-1,L+1]}}\\
Q_{\rightarrow}\big|_{\JJ} &= 
\pmat{\JJ_{[0,1,\dotsc ,L]} & 0\\
y_{1}\JJ_{[0,1,\dotsc ,L-1]} & \JJ_{[0,1,\dotsc ,L-1,L+1]}} \ .
\end{align}
By just exchanging columns and rows, we can bring $Q_{\leftarrow}\big|_{\JJ
}$ to the form
\begin{equation}
Q_{\leftarrow}\big|_{\JJ} \cong 
\JJ_{[0,1,\dotsc, L-1]}\pmat{\JJ_{[L+1]} & 0\\
y_{1} & \JJ_{[L]}} \ .
\end{equation}
We can now add the second row to the first with a suitable factor to
change the $(1,1)$-entry to $\JJ_{L}$. Similarly, we use the first
column to change the $(2,2)$-entry from $\JJ_{L}$ to $\JJ_{L+1}$. Under
the combined transformation the $(1,2)$-entry remains zero, and one
sees the equivalence to $Q_{\rightarrow}$.  

There is a further check that we can perform, namely we can see
whether we can reproduce the condensate in~\eqref{flowforl=0} with a
morphism that carries the right charge $q=\frac{k}{k+3}$. Indeed, as
discussed in appendix~\ref{sec:Reproducingflows} this is true, so that the RG
flow~\eqref{RGflowforl=0} is consistent with the
identification~\eqref{proposalforLone} of $|L,1\rangle$.
\smallskip

Having found the factorisations for $|L,1\rangle$, we could now try to
go further and construct factorisations for $|L,2\rangle$ by using the
RG flow rules. This is possible in principle, but in doing that one
encounters the problem that the morphisms that have to be turned on in
the condensation are in general not determined uniquely by their
$U(1)$ charge. Therefore we have a lot of freedom in the Ansatz for
the boundary states with higher label $l$, and it is not clear to us
how to determine the right choice. This is related to the fact that
for $|L,l\rangle$ with $l\geq 2$ there can be marginal fields in the
boundary spectrum, which means that these boundary conditions can be
continuously deformed.

Further progress is expected by using topological defect lines that
generate the whole spectrum of boundary states. This will be reported
elsewhere~\cite{Behr:inpreparation}.

\section{Low level examples}
For the first two levels $k=1$ and $k=2$, the $SU (3)/U (2)$ model corresponds
to a minimal model, where all matrix factorisations and boundary states
are known. At the next level $k=3$, the $SU (3)/U (2)$ model describes
a torus orbifold. In this section we want to compare our results to
known results for these low level examples.

\subsection{\texorpdfstring{$k=1$}{k=1}}

Let us start with $k=1$. The central charge is then $c_{1}=3/2$, which
is the central charge of the minimal model $SU (2)/U (1)$ at level
$2$. The superpotential is
\begin{equation}
W_{1} (y_{1},y_{2}) = y_{1}^{4} -4
y_{1}^{2}y_{2}+2y_{2}^{2} \ ,
\end{equation}
and by replacing $z=\sqrt{2}(y_{2}-y_{1}^{2})$ we obtain
\begin{equation}
\hat{W}_{1} (y_{1},z) = -y_{1}^{4} +z^{2} \ .
\end{equation}
This is the superpotential of the minimal model of level $2$ with the
$0B$ GSO projection. As discussed
in~\cite{Kapustin:2003rc,Brunner:2005pq}, the boundary states in this
model are labelled by a $su(2)_{2}$ label $\mathcal{L}=0,1,2$, but
with the identification rule $|\mathcal{L}\rangle
=|2-\mathcal{L}\rangle$. The boundary state $|1\rangle$ that is fixed
under this identification is not elementary, but decomposes into two
resolved boundary states $|1^{+}\rangle$ and $|1^{-}\rangle$. So there
are three boundary states in this model, in concordance with the
boundary states $|0,0\rangle$, $\overline{|0,0\rangle}$ and
$|0,1\rangle$ that we identified in the $SU (3)/U (2)$ model at level
$1$. According to~\cite{Brunner:2005pq} the resolved boundary states
$|1^{\pm}\rangle$ correspond to the two polynomial factorisations that
exist in this case, which we associated to the boundary states
$|0,0\rangle$ and $\overline{|0,0\rangle}$. The remaining boundary
state $|0\rangle$ (which corresponds to $|0,1\rangle$ in the $SU (3)/U
(2)$ description) then is associated~\cite{Brunner:2005pq} to the
factorisation
\begin{equation}
\begin{pmatrix}
y_{1} & z\\
-z & -y_{1}^{3}
\end{pmatrix} \begin{pmatrix}
-y_{1}^{3} & -z \\
z & y_{1} 
\end{pmatrix} = \hat{W}_{1} (y_{1},z)\cdot \mathbf{1} \ .
\end{equation}
By a similarity transformation this is equivalent to the factorisation
that we found to correspond to $|0,1\rangle$,
\begin{equation}
\begin{pmatrix}
y_{1}^{2}-\beta_{0}y_{2} & 0 \\
y_{1} & y_{1}^{2} - \beta_{1} y_{2} 
\end{pmatrix} \begin{pmatrix}
y_{1}^{2}-\beta_{1}y_{2}  & 0 \\
-y_{1}  & y_{1}^{2}-\beta_{0}y_{2}
\end{pmatrix}  = W_{1}(y_{1},y_{2}) \cdot \mathbf{1} \ ,
\end{equation}
when expressed in $y_{1},z$. Thus our general findings for the
$SU(3)/U(2)$ series agree with the minimal model analysis for $k=1$.

\subsection{\texorpdfstring{$k=2$}{k=2}}

Let us now look at the $SU (3)/U (2)$ model at level $k=2$ with
central charge $c_{2}=12/5$. The central charge is that of a minimal
model $SU (2)/U (1)$ at level $8$. The superpotential is
\begin{align}
W_{2} & = y_{1}^{5} -5y_{1}^{3}y_{2} + 5 y_{1}y_{2}^{2}
\nonumber \\
& = -\frac{1}{4} (y_{1}^{5} - y_{1}z^{2}) \ ,
\end{align}
where we changed variables by
$z=\sqrt{20}(y_{2}-\frac{1}{2}y_{1}^{2})$. This is the superpotential
of the D-type minimal model with $0B$ projection.  The boundary states
are labelled by $|\mathcal{L}\rangle$ and
$\overline{|\mathcal{L}\rangle}$, where $\mathcal{L}=0,\dotsc ,8$ and
we have the identification $|\mathcal{L}\rangle
=|8-\mathcal{L}\rangle$, and similarly for the
$\overline{|\mathcal{L}\rangle}$. The boundary state $|4\rangle$ is
fixed under this identification and can be decomposed into two
resolved boundary states $|4^{\pm}\rangle$ (similarly for
$\overline{|4\rangle}$). Thus in total we obtain $12$ boundary
states. In contrast we only find $8$ boundary states in the $SU (3)/U
(2)$ model, and these correspond precisely to the ones with $\mathcal{L}$
even. The reason why we find less is that the $SU (3)/U (2)$ coset
algebra is slightly larger than the bosonic subalgebra of the
superconformal algebra as it is the chiral algebra of the D-model,
which is obtained by a simple-current extension. The boundary states
with $\mathcal{L}$ odd correspond to twisted boundary conditions from
the point of view of the Kazama-Suzuki model. Indeed, the model at
$k=2$ has an additional automorphism. By level-rank duality, we have
the equivalence
\begin{equation}
\frac{SU(3)_{2}\times SO(4)_{1}}{U(2)} \cong \frac{SU(4)_{1}\times
SO(8)_{1}}{S(U(2)\times U(2))} \ .
\end{equation}
In the description on the right hand side, it is obvious that we have
an additional automorphism that permutes the $U(2)$'s
\cite{Ishikawa:2003kk}. Using this automorphism to twist the gluing
conditions, one finds the missing boundary states. This extra twist is
a peculiarity at $k=2$, so we are not going to work out these boundary
states explicitly here.

We have listed the (untwisted) boundary states of the Kazama-Suzuki
model together with their matrix factorisations in table~\ref{tab:keq2}. The
corresponding boundary states and factorisations of the minimal models
can be found in table~\ref{tab:keq2MM}. It is straightforward to see that the
factorisations are related to those in table~\ref{tab:keq2} by similarity
transformations, where we can use that
\begin{alignat}{2}
y_{1}^{2}-\beta_{0}y_{2} & = y_{1}^{2}-\frac{5+\sqrt{5}}{2}y_{2} && =
-\frac{\sqrt{5}+1}{4} (y_{1}^{2}+z)\\
y_{1}^{2}-\beta_{1}y_{2} & = y_{1}^{2} -\frac{5-\sqrt{5}}{2} y_{2} && = 
\frac{\sqrt{5}-1}{4} (y_{1}^{2}-z) \ .
\end{alignat} 

\begin{table}
\centering
\begin{tabular}{cc}
\toprule
$|L,l\rangle$ & $Q_{|L,l\rangle}$\\ 
\midrule\\
$|0,0\rangle$ & $ \begin{pmatrix}
0 & y_{1}^{2}-\beta_{0}y_{2} \\
y_{1}(y_{1}^{2}-\beta_{1}y_{2}) & 0
\end{pmatrix} $ \\[2em]
$|1,0\rangle $ & $\begin{pmatrix}
0 & (y_{1}^{2}-\beta_{0}y_{2})(y_{1}^{2}-\beta_{1}y_{2}) \\
y_{1} & 0 
\end{pmatrix}$ \\[2em]
$|0,1\rangle $ & $ \begin{pmatrix}
0 & 0 & y_{1}^{2}-\beta_{0}y_{2}& 0\\
0 & 0 & y_{1} & y_{1}^{2}-\beta_{1}y_{2}\\
y_{1}( y_{1}^{2}-\beta_{1}y_{2}) & 0 & 0 & 0\\
-y_{1}^{2} & y_{1}( y_{1}^{2}-\beta_{0}y_{2})
\end{pmatrix}  $ \\[3em]
$|1,1\rangle $ & $ \begin{pmatrix}
0& y_{1}(y_{1}^{2}-\beta_{0}y_{2})\\
( y_{1}^{2}-\beta_{1}y_{2}) & 0
\end{pmatrix} $ \\[2em]
\bottomrule
\end{tabular} 
\caption{\label{tab:keq2}List of boundary states and their
factorisations in the $SU(3)/U(2)$ model at level $k=2$. The other
four boundary states are just anti-branes of the ones listed here.}
\end{table}

\begin{table}
\centering
\begin{tabular}{ccc}
\toprule
$|\mathcal{L}\rangle $ &  $ Q^{|\mathcal{L}\rangle}$ & $|L,l\rangle $ \\
\midrule
$|0\rangle$ & $\begin{pmatrix}
0 & -\frac{\sqrt{5}+1}{4} (y_{1}^{2}+z)\\
\frac{\sqrt{5}-1}{4} y_{1}(y_{1}^{2}-z) & 0
\end{pmatrix}$ & $|1,0\rangle$\\[2em]
$|2\rangle$ & $\begin{pmatrix}
0 & 0 & y_{1} & \frac{\sqrt{5}-1}{4}z\\
0 & 0 & -\frac{\sqrt{5}+1}{4}z & -\frac{1}{4}y_{1}^{3}\\
-\frac{1}{4}y_{1}^{4} & -\frac{\sqrt{5}-1}{4}y_{1}z & 0 & 0\\
\frac{\sqrt{5}+1}{4}y_{1}z & y_{1}^{2} & 0 & 0 
\end{pmatrix} $& $|0,1\rangle$\\[3em]
$|4^{+}\rangle$ & $\begin{pmatrix}
0 & -\frac{\sqrt{5}+1}{4} (y_{1}^{2}+z) \\
\frac{\sqrt{5}-1}{4} y_{1}(y_{1}^{2}-z) & 0
\end{pmatrix}$& $|0,0\rangle$\\[2em]
$|4^{-}\rangle$ & $\begin{pmatrix}
0 & -\frac{\sqrt{5}+1}{4} y_{1}(y_{1}^{2}+z) \\
\frac{\sqrt{5}-1}{4} (y_{1}^{2}-z) & 0
\end{pmatrix}$& $|1,1\rangle$\\[2em]
\bottomrule
\end{tabular}
\caption{\label{tab:keq2MM}List of boundary states and matrix
factorisations in the D-type minimal model $\hat{W}_{2} (y_{1},z)
=-\frac{1}{4}(y_{1}^{5}-y_{1}z^{2})$, and the corresponding boundary
state in the $SU (3)/U (2)$ Kazama-Suzuki description.}
\end{table}

\subsection{\texorpdfstring{$k=3$}{k=3}} 

For $k=3$, the $SU (3)/U (2)$ Kazama-Suzuki model has central charge
$c=3$, and the model describes a $\mathbb{Z}_{6}$ orbifold of a torus
with complex structure $\tau =\frac{1}{2} (1+i\sqrt{3})$ and size
$R=1/\sqrt{3}$ (in units where $\alpha' =2$) without a B-field
(see~\cite{Lerche:1989cs}). It is a marginal deformation of the product
of two minimal models with superpotential $W=v^{6}+w^{3}$. The
relation of boundary states and factorisations in that model to the
torus orbifold branes has been analysed
in~\cite{Fredenhagen:2006qw}. The factorisations we found in the
Kazama-Suzuki model can be deformed to factorisations in the product
of minimal models. In particular, the deformation of the polynomial
factorisations lead to generalised permutation branes in the minimal
model description~\cite{Fredenhagen:2006qw}.

\section{Outlook}

In this article we have explored the connection between boundary
states and matrix factorisations in the $SU (3)/U (2)$ Kazama-Suzuki
model. We have identified matrix factorisations for the series
$|L,0\rangle$ of boundary states, which form a basis for the
RR charges. By using information on boundary RG flows, we have
constructed matrix factorisations also for the series $|L,1\rangle$ as
condensates of superpositions of $|L,0\rangle$ branes.
This demonstrates the power of tachyon condensation to obtain new
factorisations, and it points towards a way of how to obtain all the
factorisations corresponding to boundary states $|L,l\rangle$.

The difficulty one faces when extending the analysis to $l\geq 2$ is
that these boundary states generically possess marginal boundary
fields, and by just looking at the spectrum one will not be able to
distinguish those factorisations that are connected by marginal
deformations. For this, one needs further information, like the
boundary chiral ring.

Another way to single out the factorisations corresponding to the
boundary states that are maximally symmetric with respect to the coset
W-algebra would be to identify the W-algebra structure on the matrix
factorisation side. The Kazama-Suzuki models $SU (n)/U (n-1)$ have an
$\mathcal{N}=2$ $W_{n}$
algebra~\cite{Ito:1990ac,Ito:1991wb,Nemeschansky:1991an,Romans:1991wi}.
The construction of the corresponding currents\footnote{The
$W_{3}$-current in the coset model has been constructed
in~\cite{Ahn:1994yw}.} in the bulk LG model for $SU (3)/U (2)$ was
done in~\cite{Lerche:1994gn,Nemeschansky:1994is}, similarly to the
construction of the currents $T$ and $J$ of the $\mathcal{N}=2$ superconformal
algebra in~\cite{Fre:1992hp,Witten:1993jg}. It would be interesting to
extend this analysis to boundary theories to understand how the
symmetry of the boundary states translates into conditions on matrix
factorisations.

A promising way of how to systematically obtain the other factorisations
is by employing topological defect lines~\cite{Petkova:2000ip} to
generate factorisations for the whole set of boundary
states. Topological defects are labelled by representations of the
coset algebra. A defect $D_{((0,0),0;1,m)}$ can generate all boundary
states $|L,l\rangle$ from the elementary set $\{|L,0\rangle \}$ by
fusing the defect onto the boundary. If we identify this defect on the
LG side as a factorisation of the difference of two copies of the
superpotential~\cite{Brunner:2007qu}, we would be in the position to
iteratively obtain all relevant matrix factorisations. Of course, one
faces also here the problem that one has to fix the freedom of
marginally deforming the defect, but once a defect is fixed, there is
no further ambiguity for the matrix factorisations corresponding to
the maximally symmetric boundary states. This is currently under
investigation~\cite{Behr:inpreparation}.  
\smallskip

The methods of RG flows, tachyon condensations and topological defect
lines should also be useful when one approaches the higher rank $SU
(n)/U (n-1)$ models for $n\geq 4$. The first task would be to find an
elementary set of factorisations corresponding to boundary states
$|L,0\rangle$. The connection to a product of $n-1$ minimal models
might again be useful, and maybe it is also for this more general
setup the permutation factorisations~\cite{Enger:2005jk} that lead to
the relevant factorisations for the Kazama-Suzuki models. Another
promising approach would be to use defect lines between the product of
minimal models and the Kazama-Suzuki model, similar to the defect
corresponding to the functor described in appendix~\ref{sec:functor},
to relate factorisations on both sides.

Another interesting aspect of the relation of boundary states and
factorisations is the charge group. We have different notions of
charges in this context, and it would be interesting to compare
them. Firstly, we can compute RR charges as one-point functions of the
RR fields, which we have used to identify the correct factorisations
corresponding to the $|L,0\rangle$ boundary states. This, however, will not be
sensible to torsion charges. On the other hand, we can define charges
as dynamical invariants. On the CFT side, this means to look for
invariants under boundary RG flows like
in~\cite{Fredenhagen:2000ei}. For coset models the so defined charge
groups~\cite{Fredenhagen:2003xf, Fredenhagen:inpreparation} are given by
equivariant twisted topological
K-theory~\cite{SchaferNameki:2003aj,Schafer-Nameki:2004yr}. On the
matrix factorisation side, the group of dynamical invariants is the
Grothendieck K-group. It would be interesting to understand its
connection to the topological K-theory for the Kazama-Suzuki cosets.

Finally, the Kazama-Suzuki models can also be used to construct
Gepner-like models~\cite{Kazama:1989qp} for Calabi-Yau compactifications. It would be
interesting to repeat the charge analysis of~\cite{Caviezel:2005th} to
see whether tensor products of the factorisations that we identified
already provide a basis for the charge lattice. Furthermore, it is
known that some of the Kazama-Suzuki models can be marginally deformed
to obtain other rational models~\cite{Font:1989qc}, e.g.\  
\begin{equation}
\frac{SU (3)_{k}}{U (2)}  \rightsquigarrow \frac{SU (2)_{k+1}}{U (1)} \times
\frac{SU (2)_{\frac{k-1}{2}}}{U (1)}\quad \text{for}\ k\ \text{odd.}
\end{equation}
In the example above, under the deformation the polynomial
factorisations of the Kazama-Suzuki model go over into generalised
permutation factorisations~\cite{Caviezel:2005th} of the minimal
models. In a full Gepner-like model, it would be interesting to
investigate what happens to the properties of the branes (like their
mass) when doing complex structure deformations to go one from
Gepner-like model to another.

\section*{Acknowledgements}

It is a great pleasure to thank Ilka Brunner, Nils Carqueville,
Matthias Gaberdiel, Ilarion Melnikov and Ingo Runkel for useful
discussions. In particular we want to acknowledge Nils Carqueville's
help with setting up the SINGULAR code to test and find Ans{\"a}tze
for matrix factorisations with the help of a computer. 

\appendix

\section{CFT open string spectra}

\subsection{Explicit formula for boundary states}
\label{sec:boundarystates}

The B-type boundary states $|L,S;l\rangle$ in the $SU (3)_{k}/U (2)$
Kazama-Suzuki model are given by (see equation~\eqref{boundarystates})
\begin{equation}
|L,S;l\rangle = \sqrt[4]{\frac{2(k+3)}{3}}
\sum_{\Lambda_{1}=0}^{\lfloor \frac{k}{2}\rfloor} \sum_{\lambda
=0}^{k+1}\sum_{\Sigma} \frac{\psi_{L
(\Lambda_{1},\Lambda_{1})}^{(3)}S^{(so)}_{S\Sigma}S^{(2)}_{l\lambda
}}{\sqrt{S^{(3)}_{0
(\Lambda_{1},\Lambda_{1})}S^{(so)}_{0\Sigma}S^{(2)}_{0\lambda}}} 
| (\Lambda_{1},\Lambda_{1}),\Sigma ;\lambda ,0\dra \ .
\end{equation}
Note that in the numerator the untwisted S-matrix $S^{(2)}$ of $SU(2)$
appears because charge conjugation is an inner automorphism for
$SU(2)$. The S-matrices for $su(2)_{k+1}$ and $su(3)_{k}$ are given by
(see e.g.\ \cite{FrancescoCFT})
\begin{align}
S^{(2)}_{ll'} & = \sqrt{\frac{2}{k+3}} \sin \tfrac{\pi (l+1) (l'+1)}{k+3} \\
S^{(3)}_{0 \Lambda} & = \frac{8}{\sqrt{3} (k+3)} \sin \tfrac{\pi
(\Lambda_{1}+1)}{k+3} \sin \tfrac{\pi (\Lambda_{2}+1)}{k+3} \sin
\tfrac{\pi (\Lambda_{1}+\Lambda_{2}+2)}{k+3} \ . 
\end{align} 
For $su(3)$ we only need the S-matrix with one entry $0$.

\noindent For $so (4)_{1}$, the modular S-matrix is  
\begin{equation}
S^{(so)} = \frac{1}{2}\begin{pmatrix}
1 & 1 & 1 & 1 \\
1 & 1 & -1 & -1 \\
1 & -1 & -1 & 1 \\
1 & -1 & 1 & -1 
\end{pmatrix}  \ ,
\end{equation}
where the rows and columns are indexed by $\Sigma =0,v,s,\bar{s}$.

\noindent An expression for the twisted S-matrix $\psi^{(3)}_{L
(\Lambda_{1},\Lambda_{1})}$ for $su (3)_{k}$ can be found
in~\cite{Gaberdiel:2002qa},
\begin{equation}
\psi^{(3)}_{L (\Lambda_{1},\Lambda_{1})} = \frac{2}{\sqrt{k+3}}
\sin \frac{2\pi (L+1) (\Lambda_{1}+1)}{k+3} \ .
\end{equation}
In the computation of the spectrum we also need the modular S-matrix
of $U(1)_{6(k+3)}$,
\begin{equation}\label{app:SmatrixU1}
S^{6(k+3)}_{\mu \mu'}=\frac{1}{\sqrt{6(k+3)}}
\exp\left(-\frac{\pi i}{3(k+3)}\mu \mu' \right) \ ,
\end{equation}
the fusion rules of $su(2)_{k+1}$,
\begin{equation}\label{}
N^{(k+1)}_{l_{1}l_{2}}{}^{l_{3}} = \left\{\begin{array}{ll}
1 & \text{for}\ |l_{1}-l_{2}|\leq l_{3}\leq \min
(l_{1}+l_{2},2k+2-l_{1}-l_{2}) \ \text{and}\ l_{1}+l_{2}+l_{3}\ \text{even}\\
0 & \text{otherwise,}
\end{array} \right.
\end{equation}
and the twisted fusion rules of $su(3)_{k}$, which we discuss next.

\subsection{Twisted fusion rules of \texorpdfstring{$su(3)$}{su(3)}}
\label{sec:twisted_su3_fusion}

The formula for the open string spectra for the B-type branes in the
$SU (3)/U (2)$ model contains the twisted fusion rules for $su(3)_{k}$
that are given by
\begin{equation}
n_{\Lambda L}{}^{L'} = \sum_{\Lambda'= (\Lambda_{1}',\Lambda_{1}')}
\frac{\psi^{(3)}_{L\Lambda '}\psi^{(3)}_{L'\Lambda '}S^{(3)}_{\Lambda \Lambda
'}}{S^{(3)}_{0\Lambda '}} \ .
\end{equation}
Explicit expressions for these coefficients have been determined
in~\cite{Gaberdiel:2002qa} in terms of fusion rules of $su (2)$ at
level $2k+4$ (together with an alternative formula involving $su (2)$
fusion at level $(k-1)/2$ for odd $k$). In a similar way one can
obtain a formula involving the $su (2)$ fusion rules at level $k+1$
which is the form that is most convenient for our purposes. It is
given by (see~\eqref{twistedfusion})
\begin{equation}\label{app:twistedfusion}
n_{\Lambda L}{}^{L'} = \sum_{\gamma} b^{\Lambda}_{\gamma} \big(
N^{(k+1)}_{\gamma \ L}{}^{L'} - N^{(k+1)}_{k+1-\gamma
\ L}{}^{L'}\big) \ ,
\end{equation}
and it involves the branching coefficients $b$ of the regular embedding of
$su(2)$ in $su(3)$ (embedding index $1$).

In the following we shall prove this formula. Let us first note that
one can express the twisted S-matrix in terms of the $su (2)$ S-matrix
at level $k+1$,
\begin{equation}
\psi^{(3)}_{L (\Lambda_{1},\Lambda_{1})} = \sqrt{2}
S^{(2)}_{L,2\Lambda_{1}+1} \ .
\end{equation}
The ratio $S^{(3)}_{\Lambda \Lambda '}/S^{(3)}_{0\Lambda '}$ is given by
a character $\chi_{\Lambda}$ of the finite dimensional Lie algebra
$su(3)$ evaluated on the weight $-\frac{2\pi i}{k+3} (\Lambda
'+\rho_{su(3)})$, where $\rho_{su(3)}$ is the Weyl vector of $su
(3)$ \cite[eq.(14.247)]{FrancescoCFT}. So we get
\begin{align}
\frac{S^{(3)}_{\Lambda \Lambda '}}{S^{(3)}_{0\Lambda '}}
&= \chi_{\Lambda} (-\tfrac{2\pi i}{k+3} (\Lambda '+\rho))\nonumber\\
&= \sum_{\gamma} b^{\Lambda}_{\gamma} \chi_{\gamma} (-\tfrac{2\pi
i}{k+3} (2\Lambda_{1}'+1+\rho_{su (2)}))\nonumber\\
&= \sum_{\gamma} b^{\Lambda}_{\gamma}
\frac{S^{(2)}_{\gamma,2\Lambda_{1}'+1}}{S^{(2)}_{0,2\Lambda_{1}'+1}}\ .
\end{align}
Here, we expressed the $su(3)$-character $\chi_{\Lambda}$ as a sum
of characters $\chi_{\gamma}$ of representations
$\gamma$ of $su(2)$ that appear in the decomposition of $\Lambda$.
Inserting this into the formula~\eqref{app:twistedfusion} for $n$, we
obtain  
\begin{align}
n_{\Lambda L}{}^{L'} & = \sum_{\gamma}b^{\Lambda}_{\gamma} 
\sum_{\Lambda_{1}'=0}^{\lfloor \frac{k}{2}\rfloor}
\frac{2S^{(2)}_{L,2\Lambda_{1}'+1}S^{(2)}_{L',2\Lambda_{1}'+1}S^{(2)}_{\gamma
,2\Lambda_{1}'+1}}{S^{(2)}_{0,2\Lambda_{1}'+1}}\nonumber\\
&= \sum_{\gamma}\sum_{\mu =0}^{k+1} \frac{S^{(2)}_{L\mu}S^{(2)}_{L'\mu
}}{S^{(2)}_{0\mu }} \big( S^{(2)}_{\gamma \mu} -S^{(2)}_{k+1-\gamma
,\mu }\big) \nonumber\\
&= \sum_{\gamma} b^{\Lambda}_{\gamma} \big(N^{(k+1)}_{LL'}{}^{\gamma}
- N^{(k+1)}_{LL'}{}^{k+1-\gamma} \big) \ .
\label{app:twistedfusioncoeff}
\end{align} 
Here we used that
\begin{equation}
S^{(2)}_{\gamma \mu} -S^{(2)}_{k+1-\gamma ,\mu} 
= \left\{\begin{array}{ll}
2 S^{(2)}_{\gamma \mu} & \text{for $\mu$ odd}\\
0 & \text{for $\mu$ even} \ , 
\end{array} \right.
\end{equation}
and in the last step we used the Verlinde formula for
$su(2)_{k+1}$. This concludes the proof
of~\eqref{app:twistedfusion}. Notice that in the formula one could
replace the $su (2)_{k+1}$ fusion rules by the untruncated tensor
product coefficients of the finite dimensional $su(2)$, because
$L+L'\leq k$, so no truncation appears.

\subsection{Spectra}
\label{sec:CFTspectra}

For the comparison to the matrix factorisation results we are
interested in the chiral primaries that appear in the open string
spectra. Inserting the result~\eqref{app:twistedfusion} into the
formula~\eqref{spectrum} for the spectrum, and restricting to chiral
primaries, we obtain
\begin{align}
\langle L_{1},l_{1}|\tilde{q}^{\frac{1}{2}
(L_{0}+\bar{L}_{0}-\frac{c}{12})}|L_{2},l_{2}\rangle_{\text{ch.prim.}} 
& = \sum_{\Lambda = (\Lambda_{1},\Lambda_{2})} n_{\Lambda
L_{2}}{}^{L_{1}} N^{(k+1)}_{\Lambda_{1}l_{2}}{}^{l_{1}} \chi_{\Lambda
,0;\Lambda_{1},\Lambda_{1}+2\Lambda_{2}} (q)
\label{app:spectrum}\\
& = \sum_{\Lambda =(\Lambda_{1},\Lambda_{2})} \sum_{\gamma}
b^{\Lambda}_{\gamma} \big(N^{(k+1)}_{\gamma L_{2}}{}^{L_{1}} -
N^{(k+1)}_{k +1-\gamma,L_{2}}{}^{L_{1}} \big)\nonumber \\
& \qquad \qquad \qquad \times N^{(k+1)}_{\Lambda_{1}l_{2}}{}^{l_{1}} 
\chi_{\Lambda,0;\Lambda_{1},\Lambda_{1}+2\Lambda_{2}} (q) \ .
\end{align}
Here we have used that each chiral primary has a representative
$((\Lambda_{1},\Lambda_{2}),0;\Lambda_{1},\Lambda_{1}+2\Lambda_{2})$
(see~\eqref{chiral_primary}). 

To evaluate these expressions we also need a formula for the branching
coefficients. The representation $\Lambda = (\Lambda_{1},\Lambda_{2})$
decomposes into $su (2)$-representations according to 
\begin{equation}\label{app:branching}
(\Lambda) \to
\bigoplus_{\gamma_{1}=0}^{\Lambda_{1}}\bigoplus_{\gamma_{2}=0}^{\Lambda_{2}}
(\gamma_{1}+\gamma_{2}) \ .
\end{equation}
From this we can read off the branching coefficient
$b^{\Lambda}_{\gamma}$ that counts how often a representation $\gamma
=\gamma_{1}+\gamma_{2}$ appears in the decomposition.

It is sometimes convenient to write the branching coefficients in
terms of (untruncated) $su(2)$ fusion rules, 
\begin{equation}\label{app:branchingcoefficients}
b^{(\Lambda_{1},\Lambda_{2})}_{\gamma} = \sum_{\mu}
N_{\mu\Lambda_{1}}{}^{\Lambda_{2}} N_{\mu\,
\Lambda_{1}+\Lambda_{2}}{}^{2\gamma} \ .
\end{equation}

\subsubsection{Fermions in self spectra}
\label{sec:CFT_fermionfreeSS}

The polynomial matrix factorisations that we found in
section~\ref{sec:LG_factorisations} have the
common feature they do not have fermions in their self spectra. In the
CFT language this means that there are no chiral primaries in the spectrum
between the brane and its anti-brane. We now want to analyse which
boundary states satisfy this property. A chiral primary corresponding
to the $su(3)_{k}$-representation $\Lambda= (\Lambda_{1},\Lambda_{2})$
appears in the spectrum between $|L,l\rangle$ and
$|\overline{L,l}\rangle$ with multiplicity $n_{\Lambda
L}{}^{L}N^{(k+1)}_{\Lambda_{1}l}{}^{k+1-l}$. Obviously this is $0$ for
$l=0$ or $l=k+1$
because $\Lambda_{1}+\Lambda_{2}\leq k$, so $\Lambda_{1}< k+1$. For
$0<l<k+1$, the $su(2)$-fusion coefficient allows all $\Lambda_{1}\geq
|k+1-2l|$ that satisfy $\Lambda_{1}+k$ odd. In particular, we can
look at the multiplicity of $\Lambda = (k-1,0)$. Evaluating the
branching coefficient by formula~\eqref{app:branchingcoefficients} as 
\begin{equation}
b^{(k-1,0)}_{\gamma} = N_{k-1,k-1}{}^{2\gamma} \ ,
\end{equation}
we obtain
\begin{align}
n_{(k-1,0)L}{}^{L} & = \sum_{\gamma =0}^{k-1}\big(N_{\gamma
L}{}^{L}-N_{k+1-\gamma,L}{}^{L} \big)\nonumber\\
& = \sum_{\gamma =0}^{k-1} N_{\gamma L}{}^{L} - \sum_{\gamma =2}^{k+1}
N_{\gamma L}{}^{L}\nonumber \\
& = N_{0L}{}^{L}-N_{kL}{}^{L} \ .
\label{app:multiplicity}
\end{align}
In the last step we used that $2L\leq k$. We see that for $L<k/2$, the
multiplicity is always one, so only for $L=k/2$ (and thus only for
even $k$), there could be further boundary states with fermion-free
self-spectra. 

Let us now analyse this remaining possibility $L=k/2$ (assuming that
$k$ is even). The twisted fusion coefficient $n$ is then
\begin{align}
n_{(\Lambda_{1},\Lambda_{2})\frac{k}{2}}{}^{\frac{k}{2}} & =
\sum_{\gamma} b^{\Lambda}_{\gamma} \big(N_{\gamma
\frac{k}{2}}{}^{\frac{k}{2}} - N_{k+1-\gamma
,\frac{k}{2}}{}^{\frac{k}{2}} \big) \nonumber\\
& = \sum_{\gamma} b^{\Lambda}_{\gamma} (-1)^{\gamma} \nonumber\\
& = \sum_{\gamma_{1}=0}^{\Lambda_{1}}\sum_{\gamma_{2}=0}^{\Lambda_{2}}
(-1)^{\gamma_{1}+\gamma_{2}}\nonumber\\
& = \left\{ \begin{array}{ll} 1 & \text{for $\Lambda_{1},\Lambda_{2}$ even}\\
0 & \text{else} \ .
\end{array} \right. 
\label{app:khalffusion} 
\end{align}
In the second step we used the expression~\eqref{app:branching} for
the branching rules. The twisted fusion rules for $L=k/2$ thus allow
for all $\Lambda$ with even labels $\Lambda_{1},\Lambda_{2}$. On the
other hand the $su (2)$-fusion coefficient
$N^{(k+1)}_{\Lambda_{1}l}{}^{k+1-l}$ vanishes for even $\Lambda_{1}$ if
$k$ is even, so there are no chiral primaries in the spectrum between
$|\frac{k}{2},l\rangle$ and $|\overline{\frac{k}{2},l}\rangle$.

To summarise, we have identified two series of boundary states that
lead to fermion-free self spectra. On the one hand the series
$|L,0\rangle$ (and their anti-branes $|L,k+1\rangle$), on the
other hand, the series $|\frac{k}{2},l\rangle$, which only exists for
even $k$.

\subsubsection{Relative spectra of the $l=0$ series}
\label{sec:CFT_l_0_spectra}

For the series of boundary states $|L,0\rangle$, we shall now
determine the spectrum of chiral primaries. We encode the bosonic
spectrum (including information on the $U (1)$ charges $q_{i}$)
between a boundary state $|L,0\rangle$ and a boundary state
$|L',0\rangle$ in a generating polynomial,
\begin{equation}
B_{|L,0\rangle ,|L',0\rangle} (z) = \sum_{\text{chiral primaries}\
\phi_{i}} z^{q_{i}d} \ .
\end{equation}
Using formula~\eqref{app:spectrum} for the spectrum, we obtain
\begin{align}
B_{|L,0\rangle ,|L',0\rangle} (z) & = \sum_{\Lambda} n_{\Lambda
L}{}^{L'} N^{(k+1)}_{\Lambda_{1}0}{}^{0} z^{\Lambda_{1}+2\Lambda_{2}}\nonumber\\
& = \sum_{\Lambda_{2}=0}^{k} n_{(0,\Lambda_{2})L}{}^{L'} z^{2\Lambda_{2}}\nonumber\\
&= \sum_{\Lambda_{2}=0}^{k} \sum_{\gamma} b_{\gamma}^{(0,\Lambda_{2})}
\big(N_{\gamma L}{}^{L'} - N_{k+1-\gamma ,L}{}^{L'} \big) z^{2\Lambda_{2}}\nonumber\\
&= \sum_{\Lambda_{2}=0}^{k} \sum_{\gamma =0}^{\Lambda_{2}}
\big(N_{\gamma L}{}^{L'} - N_{k+1-\gamma ,L}{}^{L'} \big)
z^{2\Lambda_{2}} \nonumber\\
&= \sum_{\gamma =0}^{k} \sum_{\Lambda_{2} =\gamma }^{k}
\big(N_{\gamma L}{}^{L'} - N_{k+1-\gamma ,L}{}^{L'} \big)
z^{2\Lambda_{2}} \ .
\end{align}
Now assume that $L'\geq L$. Then the representations $\gamma$ that
appear in the fusion of $L$ and $L'$ can be parameterised as 
\begin{equation}
\gamma =L'-L+2m \quad \text{with}\ m=0,\dotsc ,L \ .
\end{equation}
So finally we obtain
\begin{align}
B_{|L,0\rangle ,|L',0\rangle} (z) & = \sum_{m=0}^{L}
\Bigg(\sum_{\Lambda_{2}=L'-L+2m}^{k} z^{2\Lambda_{2}} -
\sum_{\Lambda_{2}=k+1-L'+2m-L}^{k} z^{2\Lambda_{2}} \Bigg)\nonumber\\
&= \sum_{m=0}^{L} \sum_{\Lambda_{2}=L'-L+2m}^{k-L'-L+2m}z^{2\Lambda_{2}}\nonumber\\
& = z^{2 (L'-L)} \frac{1-z^{2 (2L+2)}}{1-z^{4}}\frac{1-z^{2
(k+3- (2L'+2))}}{1-z^{2}} \ .
\label{app:CFTbosongenfunc}
\end{align} 
By sending $z\to 1$, we obtain the total number of chiral
primaries in the relative spectrum,
\begin{equation}
B_{|L,0\rangle ,|L',0\rangle} (1) = \frac{1}{2} (2L+2) (k+3- (2L'+2)) \ .
\end{equation}

\noindent The spectrum encoded in the generating function $B_{|L,0\rangle
,|L',0\rangle} (z)$ can be compared to the bosonic spectrum between
two matrix factorisations. The fermionic spectrum, on the other hand,
corresponds to the spectrum of chiral
primaries between $|L,0\rangle$ and $|\overline{L',0}\rangle$. We have
already seen in appendix~\ref{sec:CFT_fermionfreeSS} that there are no
such states for $L'=L$, and similarly we find
\begin{equation}
F_{|L,0\rangle,|L',0\rangle} (1) = \sum_{\Lambda} n_{\Lambda
L}{}^{L'}N^{(k+1)}_{\Lambda_{1}0}{}^{k+1} = 0 \ ,
\end{equation}
because $\Lambda_{1}\leq k$ and so the $su(2)$ fusion rule gives
$0$. The $l=0$ series of boundary states should thus correspond to
matrix factorisations that do not have any fermions in their relative
spectra.

\subsubsection{The $L=k/2$ series}\label{sec:khalfseries}

For even $k$, we want to analyse the spectra of the boundary states
$|\frac{k}{2},l\rangle$. The bosonic partition function
$B_{|\frac{k}{2},l\rangle,|\frac{k}{2},l'\rangle}(z)$ of chiral primaries is
given by
\begin{align}
B_{|\frac{k}{2},l\rangle,|\frac{k}{2},l'\rangle}(z) 
& = \sum_{\Lambda} n_{\Lambda
\frac{k}{2}}{}^{\frac{k}{2}} 
N^{(k+1)}_{\Lambda_{1}l}{}^{l'} z^{\Lambda_{1}+2\Lambda_{2}} \nonumber\\
& =
\sum_{\Lambda_{1}=0,\text{even}}^{k}\sum_{\Lambda_{2}=0,\text{even}}^{k-\Lambda
_{1}} N^{(k+1)}_{\Lambda_{1}l}{}^{l'} z^{\Lambda_{1}+2\Lambda_{2}} \nonumber\\
&= \sum_{\mu=0}^{k/2} N^{(k+1)}_{2\mu,l}{}^{l'}
\frac{z^{2\mu}-z^{2k+4-2\mu}}{1-z^{4}} \ .
\label{app:tempresult}
\end{align}
In the second step we have used~\eqref{app:khalffusion} to evaluate the
twisted fusion rules. 
From the $su (2)$ fusion coefficient it is immediately clear that the
partition function vanishes if $l+l'$ is odd. Let us assume thus that $l+l'$ is
even and that $l'\geq l$. For the given range of the labels $\mu
,l,l'$, we can replace the $su (2)_{k+1}$ fusion rules by the
untruncated $su (2)$ tensor product coefficients,
\begin{equation}
N^{(k+1)}_{2\mu l}{}^{l'} = N_{2\mu l}{}^{l'} - N_{2k+4-2\mu
,l}{}^{l'} \quad \text{for}\ 2\mu \leq k+2,\ l,l'\leq k+1 \ .
\end{equation}
Inserting this into~\eqref{app:tempresult} we arrive at
\begin{align}
B_{|\frac{k}{2},l\rangle ,|\frac{k}{2},l'\rangle} (z)& =
\sum_{\mu=0}^{k/2} \frac{z^{2\mu}-z^{2k+4-2\mu}}{1-z^{4}}\big( N_{2\mu,l}{}^{l'} - N_{2k+4-2\mu ,l}{}^{l'}\big)\nonumber\\
&= \sum_{\mu =0}^{k/2} \frac{z^{2\mu}-z^{2k+4-2\mu}}{1-z^{4}} N_{2\mu
l}{}^{l'} - \sum_{\mu =\frac{k}{2}+2}^{k+2} 
\frac{z^{2k+4-2\mu}-z^{2\mu}}{1-z^{4}} N_{2\mu l}{}^{l'}\nonumber\\
&=\sum_{\mu\geq 0} \frac{z^{2\mu}-z^{2k+4-2\mu}}{1-z^{4}} 
N_{2\mu l}{}^{l'} \nonumber\\
&=\sum_{\mu = (l'-l)/2}^{(l+l')/2} 
\frac{z^{2\mu}-z^{2k+4-2\mu}}{1-z^{4}} \nonumber\\
&= z^{l'-l}\frac{(1-z^{2(l+1)})(1-z^{2(k+2-l')})}{(1-z^{2})(1-z^{4})}
\label{app:khalfspectrum}\ .
\end{align}
For the total number of bosons we then have (again assuming $l'\geq l$ and $l+l'$ even)
\begin{equation}
B_{|\frac{k}{2},l\rangle ,|\frac{k}{2},l'\rangle} (1)= 
\frac{1}{2} (l +1) \big( k+3- (l'+1) \big) \ .
\end{equation}

\subsubsection{The $|L,1\rangle$ series}
\label{sec:Loneseries}

For comparison with the matrix factorisation results we want to
determine the fermionic spectra of the $|L,1\rangle$ boundary states,
more precisely the self-spectrum as well as the relative spectra among
each other and with the $|L,0\rangle$ series.

The fermionic spectrum between $|L_{1},1\rangle$ and $|L_{2},1\rangle$ is the
same as the bosonic spectrum between $|L_{1},1\rangle$ and $|L_{2},k\rangle$. The
spectrum of chiral primaries is then given by~\eqref{app:spectrum},
\begin{align}
\langle L_{1},1|\tilde{q}^{\frac{1}{2}
(L_{0}+\bar{L}_{0}-\frac{c}{12})}|L_{2},k\rangle_{\text{ch.prim.}} 
&= \sum_{\Lambda = (\Lambda_{1},\Lambda_{2})} n_{\Lambda
L_{2}}{}^{L_{1}} N^{(k+1)}_{\Lambda_{1}1}{}^{k} \chi_{\Lambda
,0;\Lambda_{1},\Lambda_{1}+2\Lambda_{2}} (q) \\
& = \sum_{\Lambda_{2}=0,1} n_{(k-1,\Lambda_{2})L_{2}}{}^{L_{1}}
\chi_{(k-1,\Lambda_{2})
,0;k-1,k-1+2\Lambda_{2}} (q) \ .
\end{align}
Using~\eqref{app:twistedfusion} we find
\begin{equation}
n_{(k-1,0)L_{2}}{}^{L_{1}} = n_{(k-1,1)L_{2}}{}^{L_{1}} = 
N_{0L_{2}}{}^{L_{1}} + N_{1L_{2}}{}^{L_{1}} - N_{kL_{2}}{}^{L_{1}}\ .
\end{equation} 
For the self-spectrum ($L_{1}=L_{2}<\frac{k}{2}$), we therefore find
two chiral primaries with $SU (3)$ weights $(k-1,0)$ and $(k-1,1)$,
respectively.  The $U (1)$ charge of a chiral primary corresponding to
$(\Lambda_{1},\Lambda_{2})$ is
$q_{(\Lambda_{1},\Lambda_{2})}=\frac{\Lambda_{1}+2\Lambda_{2}}{d}$, in
our case the two fermions have $U (1)$ charges
\begin{equation}\label{app:twofermions}
q_{(k-1,0)}=\frac{d-4}{d} \quad \text{and}\quad  q_{(k-1,1)}=\frac{d-2}{d} \ . 
\end{equation}
For $L_{1}\not= L_{2}$, there are no fermions in the relative
spectrum, unless $|L_{1}-L_{2}|=1$. In that case we again find two
fermions with the same charges as in~\eqref{app:twofermions}.

The relative fermionic spectrum of $|L_{1},0\rangle$ and
$|L_{2},1\rangle$ is the same as the bosonic spectrum between
$|L_{1},0\rangle$ and $|L_{2},k\rangle$, which is given by 
\begin{align}
\langle L_{1},0|\tilde{q}^{\frac{1}{2}
(L_{0}+\bar{L}_{0}-\frac{c}{12})}|L_{2},k\rangle_{\text{ch.prim.}} 
&= \sum_{\Lambda = (\Lambda_{1},\Lambda_{2})} n_{\Lambda
L_{2}}{}^{L_{1}} N^{(k+1)}_{\Lambda_{1}0}{}^{k} \chi_{\Lambda
,0;\Lambda_{1},\Lambda_{1}+2\Lambda_{2}} (q)\\
&=  n_{(k,0)L_{2}}{}^{L_{1}}
\chi_{(k,0),0;k,k} (q) \ .
\end{align}
From~\eqref{app:twistedfusion} we conclude that
\begin{equation}
n_{(k,0)L_{2}}{}^{L_{1}} = N_{0L_{2}}{}^{L_{1}} = \delta_{L_{2}L_{1}}
\ .
\end{equation}
Hence, the fermionic spectrum is empty for $L_{1}\not= L_{2}$. For
$L_{1}=L_{2}$, there is precisely one fermion in the spectrum of
charge
\begin{equation}\label{app:onefermion}
q_{(k,0)} = \frac{d-3}{d}\ .
\end{equation}

\section{Landau-Ginzburg open string spectra}

In this appendix we shall provide details of the spectrum calculation
in the Landau-Ginzburg description of the $SU (3)/U (2)$ Kazama-Suzuki models.
We explicitly perform the calculation for the polynomial factorisations,
and for the first series of size~2 matrix factorisations corresponding
to the boundary states $|L,1\rangle$. We shall then discuss the
tachyon condensation that reproduces the RG flow of
section~\ref{sec:RGflows}. The final subsection explains the functor
from matrix factorisations in the Kazama-Suzuki model to those in the
product of minimal models.

\subsection{Spectra of polynomial factorisations}
\label{sec:PolySpectra}
The Landau-Ginzburg model corresponding to the $SU (3)/U (2)$
Kazama-Suzuki model at $su(3)$ level $d-3$ has the following
superpotential in the variables $y_1= x_1+x_2$ and $y_2=
x_1x_2$,
\begin{equation}
W_{k}(y_1,y_2) =  \prod_{p=0}^{d-1}(x_1-\eta_p x_2)  =
\prod_{j=0}^{\lfloor
\frac{d-2}{2}\rfloor} (y_{1}^{2} - \beta_{j} y_{2})
\cdot \left\{\begin{array}{ll} 
y_{1} & \text{for $d$ odd}\\
1 & \text{for $d$ even}
\end{array} \right.\ ,
\end{equation}
where 
\begin{equation}
\eta_{p}=\exp (i\pi(2p+1)/d) \quad \text{and}\quad
\beta_j=2+\eta_j+\eta_j^{-1}\ .
\end{equation} 
The simplest matrix factorisations of $W_{k}$ are of size $1$, i.e.\
polynomial factorisations. Let $\mathcal{D}=\{\eta | \eta^{d}=-1 \}$
be the set of $d^{\text{th}}$ roots of $-1$. Now we can choose a
subset $\cI\subset \mathcal{D}$ of roots and form the
factorisation
\begin{equation}
Q_{\cI} = \pmat{0 & \prod_{\eta \in \cI} (x_{1}-\eta x_{2})\\
\prod_{\eta \in \cI^{c}} (x_{1}-\eta x_{2}) & 0} \ ,
\end{equation}
where $\cI^{c}=\mathcal{D}\setminus \cI$ denotes the
complement of $\cI$ in $\mathcal{D}$. This describes a factorisation
in the $y$-variables provided $\cI$ is invariant under $\eta \to
\eta^{-1}$. 

To determine the infinitesimal $U(1)$ R-charge representation associated
to such a matrix factorisation $Q_\cI$, we make a diagonal Ansatz
$R_\cI=\text{diag}(R_{1},R_{2})$ and plug it into
eq.\ (\ref{eq:Rdef}), finding
\begin{equation}
R_{1}-R_{2}=1-q_{\cI}\ ,
\end{equation}
where $q_{\cI}=2 |\cI |/d$. We want $R_{\cI}$ to be
traceless~\cite{Walcher:2004tx}, so we find
\begin{equation} 
R_{\cI}=\pmat{(1-q_{\cI})/2 &0\\
0 & (q_{\cI}-1)/2}\ .
\label{app:UoneRrk1} 
\end{equation}

\noindent We are now ready to explicitly determine the spectra. Let us start
with the fermions. For a
fermionic morphism $\psi$,
\begin{equation}
\psi =\pmat{0 & p_{2}\\ p_{1} & 0}\ ,
\end{equation} 
in the spectrum between $Q_\cI$ and $Q_{\cI'}$, the closedness condition reads
\begin{equation}
\pmat{0& \JJ_{\cI'}\\ \JJ_{\cI'^{c}} &0}
\pmat{0 & p_2\\ p_1 &0} 
+ \pmat{0 & p_2\\ p_1 &0}
\pmat{0& \JJ_{\cI}\\ \JJ_{\cI^{c}} &0}=0\ ,
\end{equation}
which is equivalent to
\begin{align}
\JJ_{\cI'}p_1+p_2\JJ_{\cI^{c}}
=\JJ_{\cI'\cap \cI^{c}} (\JJ_{\cI\cap
\cI'}p_1 +\JJ_{\cI^{c}\cap \cI'^{c}}p_2) &=0\\
\JJ_{\cI'^{c}} p_2+p_1 \JJ_{\cI} 
=\JJ_{\cI\cap \cI'^{c}}(\JJ_{\cI\cap
\cI'}p_1 +\JJ_{\cI^{c}\cap
\cI'^{c}}p_2)&=0\ . 
\end{align}
The closed fermionic morphisms thus read
\begin{equation}\label{app:closedfermions}
\psi_{p} = p (y_{1},y_{2}) \pmat{0 & \JJ_{\cI\cap \cI'}\\ 
- \JJ_{\cI^{c}\cap \cI'^{c}} &0} \ ,
\end{equation}
with some polynomial $p (y_{1},y_{2})$. If $p$ is quasi-homogeneous ,
the charge of the corresponding fermion $\psi_{p}$ is according
to~\eqref{morphismcharge} given by
\begin{equation}\label{app:fermioncharge}
q_{\psi_{p}} = \frac{1}{d}  \big( 2\text{deg} (p) + \big| \cI\cap
\cI'\big| + \big| \cI^{c}\cap \cI'^{c}\big| \big)
\ .
\end{equation} 
The possible choices for $p$ and correspondingly the set of fermions are
determined by dividing out exact fermionic morphisms,
\begin{align}
\widetilde{\psi} = D_{\cI\cI'} \phi & = 
\pmat{0&\JJ_{\cI'} \\ \JJ_{\cI'^{c}} &0}
\pmat{v_1 & 0\\ 0 & v_2}
-\pmat{v_1 & 0\\ 0 & v_2}
\pmat{0&\JJ_{\cI}\\ \JJ_{\cI^{c}} &0}\nonumber\\
&=\pmat{0& \JJ_{\cI'} v_2-v_1\JJ_{\cI}\\ 
\JJ_{\cI'^{c}}v_1-v_2\JJ_{\cI^{c}} & 0}\nonumber\\
& =(\JJ_{\cI'\cap
\cI^{c}}v_2-\JJ_{\cI\cap \cI'^{c}}v_1)
\pmat{0 &\JJ_{\cI\cap \cI'}\\
-\JJ_{\cI^{c}\cap \cI'^{c}} & 0} \ .
\end{align}
Comparing with formula~\eqref{app:closedfermions} for the most
general closed fermionic morphism labelled by a polynomial $p$, we see
that the elements in the cohomology are of the form  
\begin{equation}
\label{app:fossRk1}
\psi_{p} = p \pmat{0& \JJ_{\cI\cap \cI'}\\ 
-\JJ_{\cI^{c}\cap \cI'^{c}} & 0} \quad 
\text{with}\  p \in \polyRing{\JJ_{\cI \cap
\cI'^{c}}}{\JJ_{\cI' \cap \cI^{c}}}\ ,
\end{equation}
where $\langle \dotsb \rangle$ denotes the ideal generated by the
polynomials between the angle brackets.  The number of fermionic open string
states is given by the dimension of the quotient ring in which $p$
takes it values. According to a generalised B{\'e}zout formula
\cite[Chapter 1, \S 3.4]{Arnold:book}, the dimension is given by the
products of the degrees of the two polynomials defining the ideal,
divided by the products of the weights of the variables $y_{i}$. In
the case at hand we find
\begin{equation}
n_{\text{fermions}} = \frac{1}{2}\big| \cI\cap
\cI'^{c}\big| \big| \cI' \cap
\cI^{c}\big|\ .
\end{equation}
Note that at least one of the sets appearing here must have even
cardinality: the roots in the sets $\cI,\cI'$ appear
as pairs $\eta,\eta^{-1}$, and the only single root $\eta =-1$ (that
could occur for odd $d$) can only be in either $\cI\cap
\cI'^{c}$ or $\cI'\cap \cI^{c}$ because they
are disjoint. Let us denote the cardinalities by $n_{1},n_{2}$, where
we choose $n_{2}$ to be even. A basis for the quotient ring is then
given by monomials $p_{\alpha}=y_{1}^{\alpha_{1}}y_{2}^{\alpha_{2}}$
where $\alpha_{1}=0, \dotsc , n_{1}-1$ and $\alpha_{2}=0,\dotsc ,
\frac{n_{2}}{2}-1$.

We can go further by not just determining the total number of
fermions, but also their $U(1)$ charges given
by~\eqref{app:fermioncharge}. We encode the spectrum in a generating
polynomial, the fermionic partition function,
\begin{equation}
F_{\cI\cI'} (z) := \sum_{\text{fermions}\ \Psi^{p}}
z^{d\, q_{\Psi^{p}}}\ ,
\end{equation}
which can straightforwardly be evaluated,
\begin{align}
F_{\cI\cI'} (z) & =
\sum_{\alpha_{1}=0}^{n_{1}-1}\sum_{\alpha_{2}=0}^{\frac{n_{2}}{2}-1}
z^{ 2\alpha_{1}+4\alpha_{2} +|\cI\cap
\cI'|+|\cI^{c}\cap \cI'^{c}|}\nonumber\\
& = \frac{1-z^{2 n_{1}}}{1-z^{2}} \frac{1-z^{2n_{2}}}{1-z^{4}} 
z^{|\cI\cap
\cI'|+|\cI^{c}\cap \cI'^{c}|} \nonumber \\
& = \frac{1-z^{2 |\cI\cap \cI'^{c}|}}{1-z^{2}}
\frac{1-z^{2|\cI^{c}\cap \cI'|}}{1-z^{4}} 
z^{|\cI\cap
\cI'|+|\cI^{c}\cap \cI'^{c}|}\ .
\label{app:fermiongenfunc}
\end{align}

\noindent The analysis for the bosonic morphisms is completely
analogous. Note that the bosonic morphisms between $Q_{\cI}$
and $Q_{\cI'}$ are in one-to-one correspondence with the
fermions between $Q_{\cI}$ and
$\overline{Q}_{\cI'}=Q_{\cI'^{c}}$; in particular the
number of bosons is given by
\begin{equation}
n_{\text{bosons}} = \frac{1}{2} \big|\cI\cap \cI'
\big| \big| \cI^{c}\cap \cI'^{c} \big| \ ,
\end{equation}
and the generating polynomial for the bosonic spectrum is
\begin{equation}\label{app:bosongenfunc}
B_{\cI\cI'} (z) = \frac{1-z^{2 |\cI\cap \cI'|}}{1-z^{2}}
\frac{1-z^{2|\cI^{c}\cap \cI'^{c}|}}{1-z^{4}} 
z^{|\cI^{c}\cap
\cI'|+|\cI\cap \cI'^{c}|}\ .
\end{equation}
Explicitly, the bosons between $Q_{\cI}$ and $Q_{\cI '}$ are given by
\begin{equation}
\phi = v \cdot 
\begin{pmatrix}\JJ_{\cI^c\cap \cI '} & 0\\ 0 &
\JJ_{\cI \cap \cI '^c}\end{pmatrix} 
\quad \text{with}\quad v
\in\polyRing{\JJ_{\cI \cap \cI '}}{\JJ_{\cI ^c\cap \cI '^c}} \ .
\end{equation}

\subsection{Tachyon condensates of two polynomial factorisations and
their spectra}
\label{sec:condensatespectra}

Having identified all polynomial factorisations, a natural procedure
to obtain more factorisations is by tachyon condensation. In this
section we discuss the situation where we superpose two polynomial
factorisations $Q_{\cI}$ and $Q_{\cI'}$ to build the
size $2$ factorisation 
\begin{equation}
Q_{\cI \cI '} = \pmat{Q_{\cI}&0\\ 0&Q_{\cI '}}\ ,
\end{equation}
and then turn on a fermion $\psi_{p^{\tau}}$ between $Q_{\cI}$ and
$Q_{\cI '}$ to obtain the tachyon condensate
\begin{equation}
Q_{\cI \cI '}^{\tau}:=
(Q_{\cI} \xrightarrow{p^{\tau}} Q_{\cI'})  
= \pmat{Q_{\cI}&0\\
\psi_{p^{\tau}} & Q_{\cI'}}\ .
\end{equation}
The fermion $\psi_{p^{\tau}}$ is of the form~\eqref{app:fossRk1} with some polynomial
$p^{\tau}$. For a generic condensate, the $U(1)$ R-charge matrix is given by
\begin{equation}\label{app:Rofcondensate}
R =\pmat{R_{\cI}+\frac{q_{\psi_{p^{\tau}}}-1}{2}
\one_{2}&  0\\
0 & R_{\cI'}-\frac{q_{\psi_{p^{\tau}}}-1}{2}
\one_{2}} \ ,
\end{equation}               
where the charge of the tachyon is given by~\eqref{app:fermioncharge}.
If the condensate matrix $Q^{\tau}_{\cI \cI '}$ is reducible, i.e.\
if it can be written as a direct sum of smaller factorisations, then the
R-charge matrix might have to be modified~\cite[Section 4.4]{Walcher:2004tx}.
In the case at hand, this only happens for the fermion with lowest
charge, $p^{\tau}=1$, in all other cases we can
employ~\eqref{app:Rofcondensate} for the condensate. 

\subsubsection{Self-spectrum} 

We first want to determine the self-spectrum of such a factorisation,
and we shall restrict the discussion to the fermions. Before the
condensation, in the superposition $Q_{\cI \cI '}$ we only have
fermions that come from the relative spectra of the constituents,
because the polynomial factorisations do not have any fermions in
their self-spectra. A basis for the fermionic spectrum is then given by
\begin{equation}\label{app:superpositionspectrum}
\pmat{0&\psi_{\cI'\cI} \\ 0&0} \quad \text{and} \quad 
\pmat{0&0\\
\psi_{\cI\cI'} & 0} \quad \text{with}\
\psi_{\cI\cI'}=p_{\cI\cI'} \pmat{0 &
\JJ_{\cI\cap \cI'}\\-\JJ_{\cI^{c}\cap
\cI'^{c}}&0} \ ,
\end{equation}
where $p_{\cI\cI'}\in \frac{\mathbb{C}[y_{1},y_{2}]}{\langle
\JJ_{\cI\cap \cI'^{c}},\JJ_{\cI'\cap
\cI^{c}}\rangle}$ (and according expressions for
$\psi_{\cI'\cI}$). 

We now want to investigate how the spectrum changes when we turn on
the fermion $\psi_{p^{\tau}}$ and form the condensate 
$Q_{\cI \cI '}^{\tau}$. A fermionic morphism $\Psi$,
\begin{equation}\label{app:closedfermionincondensate}
\Psi = \pmat{\psi_{\cI \cI} & \psi_{\cI '\cI}\\
\psi_{\cI \cI '} & \psi_{\cI '\cI '}}\ ,
\end{equation} 
in the self-spectrum of $Q^{\tau}_{\cI \cI '}$ is closed with respect
to $D^{\tau}$, precisely if the matrix blocks
in~\eqref{app:closedfermionincondensate} are of the form
\begin{subequations}
\label{app:closedfermionconditions}
\begin{align}
\psi_{\cI \cI} & = \pmat{0&p_{1} \JJ_{\cI \cap \cI '}\\
p_{2}\JJ_{\cI^{c}\cap \cI '^{c}} & 0}\\
\psi_{\cI '\cI} &= p \pmat{0& \JJ_{\cI \cap \cI '}\\
-\JJ_{\cI^{c}\cap \cI '^{c}}& 0}\\
\psi_{\cI \cI '} &= p'\pmat{0&\JJ_{\cI \cap \cI '}\\
-\JJ_{\cI^{c}\cap \cI '^{c}} & 0} - p_{3} \pmat{0 & 0\\
p^{\tau} \JJ_{\cI^{c}\cap \cI '^{c}}&0}\\
\psi_{\cI '\cI '} &= \pmat{ 0 & p_{2}\JJ_{\cI \cap \cI '}\\
p_{1}\JJ_{\cI^{c}\cap \cI '^{c}} & 0} + p_{3}\pmat{0& -\JJ_{\cI '}\\
\JJ_{\cI '^{c}} & 0} \ ,
\end{align}
\end{subequations}
where $p,p',p_{1},p_{2},p_{3}$ are polynomials satisfying
\begin{equation}\label{app:conditionforp}
p\, p^{\tau} = p_{2}\JJ_{\cI \cap \cI '^{c}} + p_{1}\JJ_{\cI^{c}\cap
\cI '} \ .
\end{equation}
The space of closed homomorphisms has to be divided by the space of
exact homomorphisms. First we observe that
\begin{equation}
D^{\tau} \pmat{0 & 0 & 0 & 0\\
0 & 0 & 0&0 \\ 0&0 &0&0\\
0&0& 0 & p_{3}} = \pmat{0&0&0&0\\
0&0&0&0\\
0&0&0&p_{3}\JJ_{\cI '}\\
p^{\tau}p_{3}\JJ_{\cI^{c}\cap \cI '^{c}}& 0& -p_{3}\JJ_{\cI '^{c}}&0}\ ,
\end{equation}
so by adding this exact homomorphism we can always remove the terms
in~\eqref{app:closedfermionconditions} involving $p_{3}$. Next we want to
show that given $p$ and $p'$, the cohomology class of the homomorphism
is already fixed. With fixed $p$ and $p'$, the only freedom
we have is to change $p_{1},p_{2}$ to some new $p_{1}'$ and
$p_{2}'$. These have to satisfy~\eqref{app:conditionforp}, and from that we
conclude that
\begin{equation}
p_{1}'=p_{1} + p_4 \JJ_{\cI \cap \cI '^{c}} \qquad p_{2}'=p_{2} -
p_{4} \JJ_{\cI^{c}\cap \cI '} \ ,
\end{equation}
with some polynomial $p_{4}$. The difference between the corresponding
homomorphisms $\Psi '$ and $\Psi$ is exact,
\begin{equation}\label{app:exactmorphism}
\Psi ' -\Psi = \pmat{0 & p_{4} \JJ_{\cI} & 0 & 0 \\
-p_{4} \JJ_{\cI^{c}}& 0&0&0\\
0&0&0& -p_{4} \JJ_{\cI '}\\
0&0&p_{4}\JJ_{\cI '^{c}}& 0} = 
D^{\tau} \pmat{0 &0 & 0 & 0\\
0 & p_{4} &0 &0\\
0&0&p_{4}&0\\
0&0&0&0}\ .
\end{equation}
Thus we conclude that the diagonal blocks $\psi_{\cI \cI}$ and
$\psi_{\cI '\cI '}$, which are specified by $p_{1}$ and $p_{2}$, do not
carry any additional information on the cohomology class that is not
contained already in the off-diagonal blocks $\psi_{\cI '\cI}$ and
$\psi_{\cI \cI '}$.

Disregarding exact homomorphisms of the form~\eqref{app:exactmorphism} or
those that would change $p_{3}$, we are now left with exact homomorphisms
of the form
\begin{equation}\label{app:exactremaining}
D^{\tau}
\pmat{v_{1}\one_{2} & \phi_{\cI '\cI}\\
\phi_{\cI \cI '} & v_{2}\one_{2}}
= \pmat{ -\phi_{\cI '\cI}\psi_{p^{\tau}} & D_{\cI '\cI}\phi_{\cI '\cI}\\
D_{\cI \cI '}\phi_{\cI \cI '} + (v_{1}-v_{2})\psi_{p^{\tau}} & 
\psi_{p^{\tau}} \phi_{\cI '\cI}} \ . 
\end{equation}
As we have discussed before, we can concentrate on the off-diagonal
blocks $\psi_{\cI '\cI}$ and $\psi_{\cI \cI '}$; they label the
fermionic morphisms in the self-spectrum of the tachyon
condensate. 

From~\eqref{app:closedfermionconditions} we see that (having
set $p_{3}=0$) these blocks have the same form as fermions in the
superposition of $Q_{\cI}$ and $Q_{\cI '}$ as given
by~\eqref{app:superpositionspectrum}. The only thing that changes is the
condition on the polynomials $p$ and $p'$. Let us first look at the
upper right block $\psi_{\cI '\cI}$ and the corresponding polynomial
$p$. The exact homomorphisms~\eqref{app:exactremaining} together with the
condition~\eqref{app:conditionforp} tells us to choose
\begin{equation}\label{app:pcondition}
p \in \mathcal{R}=\frac{\mathbb{C}[y_{1},y_{2}]}{\langle \JJ_{\cI \cap \cI
'^{c}},\JJ_{\cI^{c}\cap \cI '}\rangle} \quad \text{with}\quad 
p\, p^{\tau} = 0 \in \mathcal{R}\ .
\end{equation}
For the lower left block $\psi_{\cI \cI '}$ the changed exactness
condition from~\eqref{app:exactremaining} tells us to take
\begin{equation}\label{app:pprimecondition}
p' \in \frac{\mathbb{C}[y_{1},y_{2}]}{\langle \JJ_{\cI \cap \cI
'^{c}},\JJ_{\cI^{c}\cap \cI '},p^{\tau} \rangle} \ .
\end{equation}
We conclude that some fermions that are present in the superposition
survive, while others will disappear due to the changed conditions on
$p$ and $p'$. The details depend of course crucially on the polynomial
$p^{\tau}$ that describes the condensing field. In the extreme case
when $p^{\tau}=1$, we see immediately from~\eqref{app:pcondition}
and~\eqref{app:pprimecondition} that no fermions would survive.

The condition~\eqref{app:pcondition} on $p$ and the
condition~\eqref{app:pprimecondition} on $p'$ are dual to each other. We know
that there is an exact pairing on $\mathcal{R}$ given by a residue
formula (similarly to the one in section~\ref{sec:LG_RR}). When we identify $p'$'s
whose difference is proportional to $p^{\tau}$, the dual space is
obtained by restricting $p$ to those polynomials that are orthogonal
to $p^{\tau}$ (and everything generated from it) with respect to the
pairing. As the residue formula is non-degenerate on $\mathcal{R}$,
this is equivalent to saying that $pp^{\tau}=0$ in $\mathcal{R}$.
\smallskip

We can also determine the $U (1)$ charge of these fermions. From the
formulae~\eqref{app:UoneRrk1}, \eqref{app:Rofcondensate}
and~\eqref{app:fermioncharge} we conclude that the charge
corresponding to a fermion given by the polynomial $p'$ is
\begin{equation}\label{app:pprimecharge}
q_{\Psi^{p'}} = 1+\frac{2}{d}\big(\deg p' -\deg p^{\tau} \big) \ ,
\end{equation}
the charge of a fermion corresponding to the polynomial $p$ is
\begin{equation}\label{app:pcharge}
q_{\Psi^{p}} = \frac{2}{d}\big(\deg p + \deg p^{\tau} +|\cI \cap
\cI '|+|\cI^{c}\cap \cI '^{c}|\big) -1 \ .
\end{equation}
\medskip

\noindent Let us exemplify these considerations in the case of the
tachyon condensates~\eqref{polyfactLone} discussed in
section~\ref{sec:morefact}, namely choosing $Q_{\cI},Q_{\cI'}$
with $\cI=[0,\dotsc ,L]$ and $\cI'=[0,\dotsc
,L-1,L+1]$, and $p^{\tau}=y_{1}$.

The spectrum is then obtained by evaluating~\eqref{app:pcondition}
and~\eqref{app:pprimecondition}. The ring in~\eqref{app:pprimecondition} in which $p'$
takes its values is now
\begin{equation}
p'\in \frac{\mathbb{C}[y_{1},y_{2}]}{\langle
(y_{1}^{2}-\beta_{L}y_{2}),(y_{1}^{2}-\beta_{L+1}y_{2}),y_{1}
\rangle}\ .
\end{equation}
This ring is one-dimensional, and $p'=1$ is a representative for a
non-trivial element. Similarly, $p=y_{1}$ is a representative for the
solution of $pp^{\tau}=0$ in $\mathcal{R}$ as in~\eqref{app:pcondition}.

\noindent The condensate thus has two fermions; according to~\eqref{app:pprimecharge}
and~\eqref{app:pcharge} their charges are 
\begin{equation}
q_{\Psi^{p'}} = \frac{d-2}{2} \quad \text{and}\quad q_{\Psi^{p}} =
\frac{d-4}{d} \ .
\end{equation}
This fits precisely with the CFT result in~\eqref{app:twofermions}.

\subsubsection{Relative spectra}

Now we want to study the relative fermionic spectrum between two
condensates $Q_{i}$ ($i=1,2$) of polynomial factorisations,
\begin{align}
Q_{i} & = (Q_{\cI_{i}} \xrightarrow{p^{\tau}_{i}} Q_{\cI_{i}'}) \\
& = \begin{pmatrix}
Q_{\cI_{i}} & 0\\
\psi_{i}^{\tau} & Q_{\cI_{i}'}
\end{pmatrix} \ ,
\end{align}
where
\begin{equation}
\psi^{\tau}_{i} = p_{i}^{\tau} \begin{pmatrix}
0 & \JJ_{\cI_{i} \cap \cI_{i}' }\\
-\JJ_{\cI_{i}^{c}\cap \cI_{i}'^{c}} & 0
\end{pmatrix} \ .
\end{equation}
For a fermionic morphism $\Psi_{12}$ from $Q_{1}$ to $Q_{2}$,
\begin{equation}
\Psi_{12} = \begin{pmatrix}
\psi_{\cI_{1} \cI_{2}} & \psi_{\cI_{1}' \cI_{2}} \\
\psi_{\cI_{1}\cI_{2}'} & \psi_{\cI_{1}' \cI_{2}'}
\end{pmatrix} \ ,
\end{equation}
the closedness condition reads
\begin{align}
\label{app:condition1}
D_{\cI_{1}'\cI_{2}} \psi_{\cI_{1}'\cI_{2}} & = 0 \\
\label{app:condition2}
D_{\cI_{1} \cI_{2}} \psi_{\cI_{1} \cI_{2}} & = -\psi_{\cI_{1}'\cI_{2}}
\psi_{1}^{\tau} \\
\label{app:condition3}
D_{\cI_{1}' \cI_{2}'} \psi_{\cI_{1}' \cI_{2}'} & = -\psi_{2}^{\tau} 
\psi_{\cI_{1}'\cI_{2}} \\
\label{app:condition4}
D_{\cI_{1} \cI_{2}'} \psi_{\cI_{1} \cI_{2}'} & = -\psi_{2}^{\tau} 
\psi_{\cI_{1}\cI_{2}} - \psi_{\cI_{1}'\cI_{2}'}
\psi_{1}^{\tau} \ .
\end{align}
To simplify our analysis, we now specify the precise case that we are
interested in. We want in particular to analyse the relative spectra
of the factorisations $Q_{|L,1\rangle}$, so we take 
\begin{align}
Q_{1} = Q_{|L_{1},1\rangle} &= (Q_{[0,\dotsc ,L_{1}-1,L_{1}+1]} 
\xrightarrow{y_{1}}  Q_{[0,\dotsc ,L_{1}]})\\ 
Q_{2} = Q_{|L_{2},1\rangle} &= (Q_{[0,\dotsc ,L_{2}]}\xrightarrow{y_{1}} 
Q_{[0,\dotsc ,L_{2}-1,L_{2}+1]}) \ ,
\end{align}
and we assume $L_{1}>L_{2}$. Note that we have chosen different
presentations for the two factorisations to simplify our analysis
(see~\eqref{reversedarrows}). Explicitly we then have
\begin{align}
\label{app:sets}
\cI_{1} &= [0,\dotsc ,L_{1}-1,L_{1}+1] \supset [0,\dotsc ,L_{2}] =
\cI_{2}\\
\label{app:setsprime}
\cI_{1}' &= [0,\dotsc ,L_{1}] \supset [0,\dotsc ,L_{2}-1,L_{2}+1] =
\cI_{2}' \ ,
\end{align}
and also $\cI_{1}'\supset \cI_{2}$. This last condition means that
there are no fermions in the spectrum between $Q_{\cI_{1}'}$ and
$Q_{\cI_{2}}$, i.e.\ that all closed fermionic morphisms are exact
with respect to $D_{\cI_{1}'\cI_{2}}$. On the other hand, the
closedness condition~\eqref{app:condition1} for the
$\psi_{\cI_{1}'\cI_{2}}$-part in the spectrum of the condensates 
just means that $\psi_{\cI_{1}'\cI_{2}}$ is closed and hence exact with
respect to $D_{\cI_{1}'\cI_{2}}$. Also the structure of exact
morphisms is unchanged in the $\psi_{\cI_{1}'\cI_{2}}$-sector, so that
we can always achieve 
\begin{equation}
\psi_{\cI_{1}'\cI_{2}} = 0
\end{equation}
by adding exact morphisms. This simplifies the other closedness
conditions~\eqref{app:condition2}and~\eqref{app:condition3} to the
usual closedness conditions of the constituents, and because of the
relations in~\eqref{app:sets} and~\eqref{app:setsprime}, all these
morphisms are exact and hence can be set to zero,
\begin{align}
\psi_{\cI_{1}\cI_{2}} &= 0\\
\psi_{\cI_{1}'\cI_{2}'} & = 0 \ .
\end{align}
This again simplifies the closedness condition~\eqref{app:condition4}
to the usual one, and so $\psi_{\cI_{1}\cI_{2}'}$ is of the form
\begin{equation}\label{app:specialclosed}
\psi_{\cI_{1}\cI_{2}'} = p \begin{pmatrix}
0 & \JJ_{\cI_{1}\cap \cI_{2}'}\\
-\JJ_{\cI_{1}^{c}\cap \cI_{2}'^{c}} & 0 
\end{pmatrix}  \ .
\end{equation}
What happens to the exactness condition? Here we have to distinguish
two cases. If $L_{1}> L_{2}+1$, then $\cI_{1}\supset \cI_{2}'$ and
thus there are no fermions between $Q_{\cI_{1}}$ and $Q_{\cI_{2}'}$
and we can also set $\psi_{\cI_{1}\cI_{2}'}=0$: in that case there are
no fermions in the spectrum. 

Let us therefore assume that $L_{1}=L_{2}+1$. We are now looking
for the most general exact fermionic morphism $D_{12}\Phi_{12}$ that
has all entries vanishing except $\psi_{\cI_{1}\cI_{2}'}$,
\begin{equation}
\label{app:specialexact}
D_{12}\Phi_{12} = \begin{pmatrix}
D_{\cI_{1}\cI_{2}} \phi_{\cI_{1}\cI_{2}}
-\phi_{\cI_{1}'\cI_{2}}\psi^{\tau}_{1} &
D_{\cI_{1}'\cI_{2}}\phi_{\cI_{1}'\cI_{2}} \\
D_{\cI_{1}\cI_{2}'} \phi_{\cI_{1}\cI_{2}'}
+\psi^{\tau}_{2}\phi_{\cI_{1}\cI_{2}} -
\phi_{\cI_{1}'\cI_{2}'}\psi^{\tau}_{1} & D_{\cI_{1}'\cI_{2}'}\phi_{\cI
_{1}'\cI_{2}'} -\psi^{\tau}_{2}\phi_{\cI_{1}'\cI_{2}}
\end{pmatrix}  
= \begin{pmatrix}
0 & 0\\
\psi^{\text{ex}}_{\cI_{1}\cI_{2}'} & 0
\end{pmatrix}  \ .
\end{equation}
First we note that $\phi_{\cI_{1}'\cI_{2}}$ has to be closed with
respect to $D_{\cI_{1}'\cI_{2}}$, and so it is of the form
\begin{equation}
\phi_{\cI_{1}'\cI_{2}} = v_{1} \begin{pmatrix}
\JJ_{\cI_{1}'^{c}\cap \cI_{2}} & 0\\
0 & \JJ_{\cI_{1}'\cap \cI_{2}^{c}} 
\end{pmatrix} \ ,
\end{equation}
with some arbitrary polynomial $v_{1}$. A straightforward analysis
yields the form of $\phi_{\cI_{1}\cI_{2}}$,
\begin{equation}
\phi_{\cI_{1}\cI_{2}} = \begin{pmatrix}
v_{2} \JJ_{\cI_{1}^{c}\cap \cI_{1}'^{c}\cap \cI_{2}} & 0\\
0 & v_{2}' \JJ_{\cI_{1}\cap \cI_{1}'\cap \cI_{2}^{c}}
\end{pmatrix}\ ,
\end{equation}
with 
\begin{equation}
v_{2}' \JJ_{\cI_{1}^{c}\cap \cI_{1}' \cap \cI_{2}} 
- v_{2} \JJ_{\cI_{1}\cap \cI_{1}'^{c}\cap \cI_{2}^{c}} = v_{1}
p_{1}^{\tau } \ .
\end{equation}
In our case the polynomial accompanying $v_{2}'$ is
$\JJ_{\cI_{1}^{c}\cap \cI_{1}' \cap \cI_{2}}=1$ because
$\cI_{1}\supset \cI_{2}$, so that we can express $v_{2}'$ in terms of
the other polynomials, and we get
\begin{equation}
\phi_{\cI_{1}\cI_{2}} = v_{2} \begin{pmatrix}
1 & 0\\
0 & \JJ_{\cI_{1}\cap \cI_{2}^{c}}
\end{pmatrix} + \begin{pmatrix}
0 & 0\\
0 & v_{1}p_{1}^{\tau}  \JJ_{\cI_{1}\cap \cI_{1}'\cap \cI_{2}^{c}}
\end{pmatrix} \ .
\end{equation}
Similarly we have 
\begin{equation}
\phi_{\cI_{1}'\cI_{2}'} = v_{3} \begin{pmatrix}
1 & 0 \\
0 & \JJ_{\cI_{1}'\cap\cI_{2}'^{c}} 
\end{pmatrix}
- \begin{pmatrix}
0 & 0\\
0 & v_{1}p_{2}^{\tau}  \JJ_{\cI_{1}'\cap \cI_{2}^{c}\cap \cI_{2}'^{c}}
\end{pmatrix} \ .
\end{equation}
Parameterising 
\begin{equation}
\phi_{\cI_{1}\cI_{2}'} = \begin{pmatrix}
v_{4} & 0\\
0 & v_{5} \end{pmatrix}\ ,
\end{equation}
the $\psi^{\text{ex}}_{\cI_{1}\cI_{2}'}$ component of an exact morphism of the
form~\eqref{app:specialexact} reads
\begin{equation}
\psi^{\text{ex}}_{\cI_{1}\cI_{2}'} = 
\big(v_{5} \JJ_{[L_{1}]} - v_{4} \JJ_{[L_{1}-1,L_{1}+1]} -
p^{\tau}_{1}v_{3}\JJ_{[L_{1}-1]} + p_{2}^{\tau}v_{2}\JJ_{[L_{1}+1]} + v_{1}
p^{\tau}_{1}p^{\tau}_{2} \big) \begin{pmatrix}
0 & \JJ_{\cI_{1}\cap \cI_{2}'}\\
-\JJ_{\cI_{1}^{c}\cap \cI_{2}'^{c}} & 0
\end{pmatrix} \ .
\end{equation}
Comparing this with the form~\eqref{app:specialclosed} of a closed
morphism, we see that the fermions in the spectrum are labelled by the
quotient ring
\begin{equation}
\mathcal{R}_{12} = \frac{\mathbb{C}[y_{1},y_{2}]}{\langle
\JJ_{[L_{1}]},\JJ_{[L_{1}-1,L_{1}+1]},p^{\tau}_{1}\JJ_{[L_{1}-1]},p^{\tau
}_{2}\JJ_{[L_{1}+1]},p^{\tau}_{1}p^{\tau}_{2}\rangle} \ .
\end{equation} 
In our case $p^{\tau}_{1}=p^{\tau}_{2}=y_{1}$, and so 
\begin{equation}
\mathcal{R}_{12} = \frac{\mathbb{C}[y_{1},y_{2}]}{\langle
y_{1}^{2},y_{2}\rangle} \ , 
\end{equation}
which is two-dimensional with representatives $p=1,y_{1}$. We finally
conclude that there are two fermions in the spectrum between
$Q_{|L_{1},1\rangle}$ and $Q_{|L_{1}-1,1\rangle}$.

The charges of these two fermions can then be determined using the
charge matrix~\eqref{app:Rofcondensate} and we obtain
\begin{equation}
q = \frac{d-4}{d} +\frac{2}{d} \text{deg} (p)\ ,
\end{equation}
so the two fermions have charges $\frac{d-4}{d}$ and $\frac{d-2}{d}$.
This matches the CFT result in appendix~\ref{sec:Loneseries} (see the
discussion below~\eqref{app:twofermions}).

\subsubsection{Relative spectrum with polynomial factorisations}
\label{sec:Relspectrawithpoly}
We want to determine the fermionic spectrum between a polynomial
factorisation $Q_{1}=Q_{\cI_{1}}$ and the condensate $Q_{2}$ of two
polynomial factorisations, 
\begin{equation}
Q_{2}= (Q_{\cI_{2}}\xrightarrow{p^{\tau}}Q_{\cI_{2}'}) \ .
\end{equation}
A closed fermion,
\begin{equation}
\Psi = \begin{pmatrix}
\psi_{\cI_{1}\cI_{2}}\\
\psi_{\cI_{1}\cI_{2}'}
\end{pmatrix}  \ ,
\end{equation}
satisfies
\begin{align}
\psi_{\cI_{1}\cI_{2}} & = p_{1} \begin{pmatrix}
0 & \JJ_{\cI_{1}\cap \cI_{2}}\\
-\JJ_{\cI_{1}^{c}\cap \cI_{2}^{c}}&0
\end{pmatrix} \\
\psi_{\cI_{1}\cI_{2}'} &= \begin{pmatrix}
0 & p_{2} \JJ_{\cI_{1}\cap \cI_{2} \cap \cI_{2}'}\\
p_{2}' \JJ_{\cI_{1}^{c}\cap \cI_{2}^{c}\cap \cI_{2}'^{c}} & 0
\end{pmatrix}  \ ,
\end{align}
with
\begin{equation}
p_{2}'\JJ_{\cI_{1}\cap \cI_{2}^{c} \cap \cI_{2}'} +
p_{2}\JJ_{\cI_{1}^{c}\cap \cI_{2}\cap \cI_{2}'^{c}} = p_{1}p^{\tau} \ .
\end{equation}
In the cases we are interested in, we have either $\cI_{1}\subset
\cI_{2}$ or $\cI_{2}\subset \cI_{1}$, i.e.\ there are no fermions
between $Q_{\cI_{1}}$ and $Q_{\cI_{2}}$. By condensation, the
closedness and exactness condition do not change for the
$\psi_{\cI_{1}\cI_{2}}$ component, and we can use exact morphisms to set
$p_{1}$ to $0$. Then $\psi_{\cI_{1}\cI_{2}'}$ is
closed with respect to $D_{\cI_{1}\cI_{2}'}$,
\begin{equation}\label{app:closedfermionrelspec}
\psi_{\cI_{1}\cI_{2}'} = p \begin{pmatrix}
0 & \JJ_{\cI_{1}\cap \cI_{2}'}\\
-\JJ_{\cI_{1}^{c}\cap \cI_{2}'^{c}} & 0
\end{pmatrix} \ .
\end{equation}
The remaining exact morphisms $D_{12}\Phi$ then come from bosons
\begin{equation}
\Phi = \begin{pmatrix}{c}
\phi_{\cI_{1}\cI_{2}}\\
\phi_{\cI_{1}\cI_{2}'}
\end{pmatrix} \ ,
\end{equation}
where $\phi_{\cI_{1}\cI_{2}}$ is closed with respect to $D_{\cI_{1}\cI
_{2}}$,
\begin{equation}
\phi_{\cI_{1}\cI_{2}} = v_{1} \begin{pmatrix}
\JJ_{\cI_{1}^{c}\cap\cI_{2}} & 0\\
0 & \JJ_{\cI_{1}\cap \cI_{2}^{c}}
\end{pmatrix} \ ,
\end{equation}
with some polynomial $v_{1}$. Writing the bosonic component $\phi_{\cI_{1}\cI_{2}'}$ as
\begin{equation}
\phi_{\cI_{1}\cI_{2}'} = \begin{pmatrix}
v_{2} & 0\\
0 & v_{2}'
\end{pmatrix} \ ,
\end{equation}
the remaining exact fermionic morphisms $\Psi^{\text{ex}}$ have
$\psi^{\text{ex}}_{\cI_{1}\cI_{2}}=0$ and
\begin{align}
\psi^{\text{ex}}_{\cI_{1}\cI_{2}'} &= \begin{pmatrix}
0 & v_{2}'\JJ_{\cI_{2}'} -v_{2}\JJ_{\cI_{1}} +
p^{\tau}v_{1}\JJ_{\cI_{1}\cap \cI_{2}^{c}}\JJ_{\cI_{2}\cap \cI_{2}'}\\
v_{2}\JJ_{\cI_{2}'^{c}}-v_{2}'\JJ_{\cI_{1}^{c}} -
p^{\tau}v_{1}\JJ_{\cI_{1}^{c}\cap
\cI_{2}}\JJ_{\cI_{2}^{c}\cap\cI_{2}'^{c}} & 0
\end{pmatrix} \\
&= (v_{2}'\JJ_{\cI_{1}^{c}\cap \cI_{2}'} -v_{2}\JJ_{\cI_{1}\cap
\cI_{2}'^{c}} +p^{\tau} v_{1}\JJ_{\cI_{1} \cap \cI_{2}^{c} \cap \cI
_{2}'^{c}} \JJ_{\cI_{1}^{c}\cap \cI_{2}\cap \cI_{2}'}) \begin{pmatrix}
0 & \JJ_{\cI_{1}\cap \cI_{2}'}\\
-\JJ_{\cI_{1}^{c}\cap \cI_{2}'^{c}} & 0
\end{pmatrix} \ .
\label{app:exactfermionsrelspec}
\end{align}
Here we used again that either $\cI_{1}\subset \cI_{2}$ or vice
versa. The fermions are then labelled by polynomials $p$
(see~\eqref{app:closedfermionrelspec}) modulo identifications that
come from the exact morphisms~\eqref{app:exactfermionsrelspec}, hence we
can view $p$ as living in the quotient
\begin{equation}
p\in \frac{\mathbb{C}[y_{1},y_{2}]}{\langle \JJ_{\cI_{1}^{c}\cap
\cI_{2}'},\JJ_{\cI_{1}\cap
\cI_{2}'^{c}}, p^{\tau}\JJ_{\cI_{1} \cap \cI_{2}^{c} \cap \cI
_{2}'^{c}} \JJ_{\cI_{1}^{c}\cap \cI_{2}\cap \cI_{2}'}\rangle} \ .
\end{equation}
For explicitness we now set 
\begin{align}
\cI_{1} &= [0,\dotsc ,L_{1}]\\
\cI_{2} & = [0,\dotsc ,L_{2}]\\
\cI_{2}' &= [0,\dotsc ,L_{2}-1,L_{2}+1] \\
p^{\tau} & = y_{1}\ .
\end{align}
For $L_{1}<L_{2}$ we have $\JJ_{\cI_{1}\cap \cI_{2}'^{c}}=1$, so there
are no fermions. Similarly for $L_{1}>L_{2}$ we have
$\JJ_{\cI_{1}^{c}\cap \cI_{2}'}=1$, and no fermion remains. On the
other hand, for $L_{1}=L_{2}$, the spectrum is given by
\begin{equation}
p \in \frac{\mathbb{C}[y_{1},y_{2}]}{\langle
\JJ_{[L_{1}+1]},\JJ_{[L_{1}]},y_{1}\rangle} \ ,
\end{equation}
which is one-dimensional. In conclusion, we find precisely one fermion
in the spectrum between $Q_{|L_{1},0\rangle}$ and
$Q_{|L_{2},1\rangle}$ if $L_{1}=L_{2}$, and no fermions otherwise. The
fermion that appears for $L_{1}=L_{2}$ has charge
\begin{equation}
q=\frac{d-3}{d}\ ,
\end{equation} 
which can be determined using~\eqref{morphismcharge} and the charge
matrices~\eqref{app:UoneRrk1} and \eqref{app:Rofcondensate}.
This coincides with the CFT result in~\eqref{app:onefermion}.

\subsection{Reproducing the CFT flows}
\label{sec:Reproducingflows}

Having identified matrix factorisations for the $|L,0\rangle$ and
$|L,1\rangle$ series, we can now compare the RG flows~\eqref{RGFlows}
between boundary states to tachyon condensation in the matrix
factorisation language. Let us consider the RG flow
\begin{equation}\label{app:exampleflow}
|L,0\rangle + |L,1\rangle \leadsto |L-1,0\rangle +|L,0\rangle +
|L+1,0\rangle \ .
\end{equation}
From formula~\eqref{app:onefermion} we can see that in the relative spectrum
between $|L,0\rangle$ and $|L,1\rangle$, there is precisely one
fermion $\psi$ of charge $q_{\psi}=\frac{d-3}{d}$. Condensing this fermion
corresponds in the CFT to perturb with a field from this coset sector,
so this is compatible with the considerations that led to~\eqref{pertfield}. 
In the language of matrix factorisations this fermion is given 
by~\eqref{app:closedfermionrelspec} with $p=1$.

Let us explicitly work out the tachyon condensate. To make the
equations more readable, we only write the $\JJ$-part of the matrix in
a notation like in~\eqref{Q-matrix}. We obtain
\begin{equation}
(Q_{|L,0\rangle} \xrightarrow{\psi} Q_{|L,1\rangle})_{\JJ}= 
\begin{pmatrix}
 \JJ_{[0,\dotsc , L]} & 0    & 0\\
0     & \JJ_{[0,\dotsc ,L]}    & 0\\
\JJ_{[0,\dotsc , L-1]}    & y_{1}\JJ_{[0,\dotsc , L-1]} &
\JJ_{[0,\dotsc ,L-1,L+1]}
  \end{pmatrix} \ .
\end{equation}
We can use the term $\JJ_{[0,\dotsc, L-1]}$ to eliminate all other
entries in that row or column by elementary transformations.
Having done this one can immediately see the equivalence to the matrix
\begin{equation}
\big (Q_{|L-1,0\rangle + |L,0\rangle +|L+1,0\rangle}\big)_{\JJ} 
= \begin{pmatrix}
\JJ_{[0,\dotsc, L-1 ]} & 0 & 0\\
0 & \JJ_{[0,\dotsc, L]} & 0 \\
0 & 0 & \JJ_{[0,\dotsc ,L+1]}
\end{pmatrix} \ .
\end{equation}
This reproduces the RG flow~\eqref{app:exampleflow} in terms of matrix
factorisations. Note that the brane $|L,0\rangle$, although appearing
in both the initial and the final configuration, is not purely a
spectator brane, but is involved in the flow.

\subsection{A faithful functor}\label{sec:functor}

Given a matrix factorisation $Q (y_{1},y_{2})$ for the superpotential
$W_{k} (y_{1},y_{2})$, we can construct from it a matrix factorisation
$\tilde{Q} (x_{1},x_{2}):=Q (x_{1}+x_{2},x_{1}x_{2})$ of the
superpotential
\begin{equation}
\tilde{W}_{k} (x_{1},x_{2}) =  W_{k} (x_{1}+x_{2},x_{1}x_{2}) =
x_{1}^{k+3}+x_{2}^{k+3} \ .
\end{equation}
This map gives rise to a functor from the category of matrix
factorisations of $W_{k}$ to the category of $\tilde{W}_{k}$. It maps
a morphism $\Phi (y_{1},y_{2})$ from $Q_{1}$ to $Q_{2}$ (seen as a
matrix with polynomial entries) to 
\begin{equation}
\tilde{\Phi} (x_{1},x_{2}) = \Phi (x_{1}+x_{2},x_{1}x_{2}) \ .
\end{equation}
Obviously, $\tilde{\Phi}$ is closed if $\Phi$ is. On the other hand,
if we change $\Phi$ by an exact morphism, then obviously the
corresponding image also differs from $\tilde{\Phi}$ by an exact term.

The most interesting property of this map is that it defines a
faithful functor, i.e.\ it is injective on the morphism spaces. This
can be seen as follows: let $\Phi$ be such that $\tilde{\Phi} = D_{x}\Psi$
is exact. Decompose $\Psi (x_{1},x_{2})=\Psi_{\text{sym}}
(x_{1},x_{2})+\Psi_{\text{asym}} (x_{1},x_{2})$ into a symmetric and
an antisymmetric part with respect to the exchange of $x_{1}$ and
$x_{2}$. Since $\tilde{\Phi} (x_{1},x_{2})$ is symmetric by
construction, and also $D_{x}$ is symmetric, we know that
$D_{x}\Psi_{\text{asym}}=0$. Hence
\begin{equation}
\tilde{\Phi} = D_{x}\Psi_{\text{sym}}\ .
\end{equation}
A symmetric polynomial can be rewritten in terms of $y_{1},y_{2}$, so
that there exists a morphism $\Psi ' (y_{1},y_{2})$ such that
$\Psi_{\text{sym}} (x_{1},x_{2})=\Psi '
(x_{1}+x_{2},x_{1}x_{2})$. Therefore 
\begin{equation}
\Phi (y_{1},y_{2}) = D_{y} \Psi ' (y_{1},y_{2})\ ,   
\end{equation}
from which we conclude that the functor is indeed faithful.

This property makes it possible to use known results on factorisations
of $\tilde{W}_{k} (x_{1},x_{2})$ to obtain information on factorisations
of $W_{k}(y_{1},y_{2})$. Namely given a factorisation $Q$ of $W_{k}$, express
it in $x$-variables to get a factorisation $\tilde{Q}$ of
$\tilde{W}_{k}$. Then determine the spectrum, and decompose it into one part
that is symmetric under exchange of $x_{1}$ and $x_{2}$, and one part
that is anti-symmetric. The symmetric part is the isomorphic image of
the spectrum of the factorisation $Q$ in the variables $y_{1},y_{2}$.

The functor we have discussed here, can be realised in terms of a
defect\footnote{We thank Nils Carqueville and Ingo Runkel for
discussions on this point.} separating the theories with
superpotentials $W_{k}$ and $\tilde{W}_{k}$. This will be discussed
elsewhere~\cite{Behr:inpreparation}.

\end{document}